\documentstyle[11pt]{article}

\makeatletter
\@addtoreset{equation}{section}
\makeatother

\def\dalemb#1#2{{\vbox{\hrule height .#2pt
        \hbox{\vrule width.#2pt height#1pt \kern#1pt
                \vrule width.#2pt}
        \hrule height.#2pt}}}

\def\half{{\textstyle{1\over2}}}
\let\a=\alpha \let\b=\beta  \let\d=\delta \let\e=\epsilon
 \let\h=\eta \let\q=\theta \let\i=\iota \let\k=\kappa
  \let\n=\nu   
\let\s=\sigma     

\let\C=\Chi      
\let\la=\label  
\def\nn{\nonumber} \def\bd{\begin{document}} \def\ed{\end{document}}
\def\ds{\documentstyle} \let\fr=\frac \let\bl=\bigl \let\br=\bigr
\let\Br=\Bigr \let\Bl=\Bigl 
\let\bm=\bibitem
\let\na=\nabla
\let\pa=\partial \let\ov=\overline
\def\ie{{\it i.e.\ }} 
\newcommand{\be}{\begin{equation}} 
\newcommand{\ee}{\end{equation}} 
\def\ba{\begin{array}}
\def\ea{\end{array}}
\def\ft#1#2{{\textstyle{{\scriptstyle #1}\over {\scriptstyle #2}}}}
\def\fft#1#2{{#1 \over #2}}
\def\del{\partial}
\def\sst#1{{\scriptscriptstyle #1}}
\def\oneone{\rlap 1\mkern4mu{\rm l}}
\def\td{\tilde}
\def\wtd{\widetilde}
\def\im{{\rm i}}
\def\bog{Bogomol'nyi\ }
\def\q{{\tilde q}}
\def\hast{{\hat\ast}}
\def\0{{\sst{(0)}}}
\def\1{{\sst{(1)}}}
\def\2{{\sst{(2)}}}
\def\3{{\sst{(3)}}}
\def\4{{\sst{(4)}}}
\def\5{{\sst{(5)}}}
\def\6{{\sst{(6)}}}
\def\7{{\sst{(7)}}}
\def\8{{\sst{(8)}}}
\def\n{{\sst{(n)}}}
\newcommand{\w}[1]{\\[0.#1cm]}
\def\hA{\hat{\cal A}}
\def\ns{{\sst {\rm NS}}}
\def\rr{{\sst {\rm RR}}}
\def\tH{{\widetilde H}}
\def\tB{{\widetilde B}}
\def\cA{{\cal A}}
\def\cF{{\cal F}}
\def\tF{{\wtd F}}
\def\v{{{\cal V}}}
\def\Z{\rlap{\sf Z}\mkern3mu{\sf Z}}
\def\ep{{\epsilon}}
\def\IIA{{\rm IIA}}
\def\IIB{{\rm IIB}}
\def\ads{{\rm AdS}}
\def\R{\rlap{\rm I}\mkern3mu{\rm R}}
\def\ua{\underline{\alpha}}
\def\ub{\underline{\phantom{\alpha}}\!\!\!\beta}
\def\uc{\underline{\phantom{\alpha}}\!\!\!\gamma}
\def\um{\underline{\mu}}
\def\ud{\underline\delta}
\def\ue{\underline\epsilon}
\def\una{\underline a}
\def\unA{\underline A}
\def\unb{\underline b}
\def\unB{\underline B}
\def\unc{\underline c}
\def\unC{\underline C}
\def\und{\underline d}
\def\unD{\underline D}
\def\une{\underline e}
\def\unE{\underline E}
\def\unf{\underline{\phantom{e}}\!\!\!\! f}
\def\unF{\underline F}
\def\ung{\underline g}
\def\unm{\underline m}
\def\unM{\underline M}
\def\unn{\underline n}
\def\unN{\underline N}
\def\unp{\underline{\phantom{a}}\!\!\! p}
\def\unP{\underline P}
\def\unH{\underline{H}}
\def\unF{\underline{F}}
\def\unT{\underline{T}}
\def\ovA{\overline{A}}
\def\ovB{\overline{B}}
\def\uC{{\underline C}}
\def\ns{\normalsize}
\def\vs{\vspace{-0.25cm}}
\def\se{\;\;=\;\;}
\def\de{\;\;:=\;\;}
\def\cF{{\cal F}}
\def\cH{{\cal H}}
\def\cK{{\cal K}}

\def\atwo{\alpha_{2}}
\def\aone{\alpha_{1}}
\def\afive{\alpha_{5}}
\def\ap{\alpha_p}
\def\azero{\alpha_o}
\def\afour{\alpha_{4}}
\def\appt{\alpha_{p+2}}
\def\apmo{\alpha_{p-1}}

\def\cE{{\cal E}}
\def\tr{{\rm tr}}
\def\bC{{\bar \C}}

\newcommand{\bea}{\begin{eqnarray}} 
\newcommand{\eea}{\end{eqnarray}} 
\newcommand{\ra}{\rightarrow}
\newcommand{\Tr}{{\rm Tr} } 
\newcommand{\tamphys}{\it Randall Laboratory, Department of Physics, 
University of Michigan,\\ Ann Arbor, MI 48109-1120}

\newcommand{\auth}{M. J. Duff\footnote{mduff@umich.edu}}

\begin{document}
\begin{flushright}
\hfill{UM-TH-99-07}\\
\hfill{hep-th/9912164}\\
\end{flushright}
\vspace{24pt}

\begin{center}
{ \large 
{\bf TASI LECTURES ON BRANES, BLACK HOLES \\ AND ANTI-DE SITTER SPACE\footnote{Based on 
talks delivered at the Theoretical Advanced Study Institute, Boulder, Colorado, 
June 1999 and the Banff Summer School, Alberta, Canada, June-July 1999.
Research supported in part by NSF Grant PHY-9722090.}}.}

\vspace{36pt}

\auth

\vspace{10pt}

{\tamphys}

\vspace{44pt}

\underline{ABSTRACT}

\end{center}

In the light of the duality between physics in the 
bulk of anti-de Sitter space and a conformal field theory on the 
boundary, we review the $M2$, $D3$ and $M5$ branes and how their 
near-horizon geometry yields the compactification of $D=11$ supergravity on 
$S^{7}$, Type $IIB$ supergravity on $S^{5}$ and $D=11$ supergravity on 
$S^{4}$, respectively. We discuss the ``Membrane at the End of the 
Universe'' idea and its relation to the corresponding superconformal singleton 
theories that live on the boundary of the $AdS_{4}$, $AdS_{5}$ and 
$AdS_{7}$ vacua. The massless sectors of these compactifications 
are described by the maximally supersymmetric $D=4$, $D=5$ and $D=7$ gauged 
supergravities. We construct the non-linear 
Kaluza-Klein ans\"atze describing the embeddings of the $U(1)^4$, $U(1)^3$ and 
$U(1)^2$ truncations of these supergravities, which admit $4$-charge $AdS_{4}$, 
$3$-charge $AdS_{5}$ and $2$-charge $AdS_{7}$ black hole solutions. These 
enable us to embed the black hole solutions back in ten and eleven dimensions 
and reinterpret them 
as $M2$, $D3$ and $M5$ branes spinning in the transverse dimensions 
with the black hole charges given by the angular momenta of the branes. 
A comprehensive Appendix lists the field equations, symmetries and 
transformation rules of $D=11$ supergravity, Type $IIB$ supergravity, 
and the $M2$, $D3$ and $M5$ branes.  

{\vfill\leftline{}\vfill

\pagebreak
\setcounter{page}{1}

\tableofcontents
\addtocontents{toc}{\protect\setcounter{tocdepth}{2}}
\newpage
\section{\bf INTRODUCTION}
\la{World}

\subsection{Supergravity, supermembranes and $M$-theory}
\la{Supergravity}

A vital ingredient in the quest for a unified theory embracing all physical
phenomena is {\it supersymmetry}, 
 a symmetry which (a) unites bosons and
fermions, (b) requires the existence of gravity and (c) places an upper limit of
{\it eleven} on the dimension of spacetime.  For these reasons, in the early 1980s
many physicists looked to eleven-dimensional supergravity in the hope that it
might provide that elusive superunified theory.  Then in 1984 superunification underwent a major paradigm-shift: eleven-dimensional
supergravity was knocked off its pedestal by ten-dimensional 
superstrings ,
one-dimensional objects whose vibrational modes represent the elementary
particles.  Unlike eleven-dimensional supergravity, superstrings provided a
perturbatively finite theory of gravity which, after compactification to four
spacetime dimensions, seemed in principle capable of explaining the Standard Model
of the strong, weak and electromagnetic forces including the required {\it chiral}
representations of quarks and leptons. 

Despite these major successes, however, nagging doubts persisted about
superstrings. First, many of the most important questions in string 
theory --- How do
strings break supersymmetry? How do they choose the right vacuum state? How do
they explain the smallness of the cosmological constant? How do they resolve the
apparent paradoxes of quantum black holes? --- seemed incapable of being answered
within the framework of a weak coupling perturbation expansion. They seemed to
call for some new, {\it non-perturbative}, physics. Secondly, why did there appear
to be {\it five} different mathematically consistent superstring theories: the $E_8
\times E_8$ heterotic string, the $SO(32)$ heterotic string, the $SO(32)$ Type $I$
string, the Type $IIA$ and Type $IIB$ strings? If one is looking for a unique
{\it Theory of Everything}, this seems like an embarrassment of riches! Thirdly, if
supersymmetry permits eleven dimensions, why do superstrings stop at ten? This
question became more acute with the discoveries of the elementary {\it
supermembrane} in 1987 and its dual partner, the 
solitonic {\it superfivebrane}, in
1992. These are supersymmetric extended objects with respectively two and five
dimensions moving in an eleven-dimensional spacetime. Finally, therefore, if we
are going to generalize zero-dimensional point particles to one-dimensional
strings, why stop there? Why not two-dimensional membranes or more generally
$p$-dimensional objects (inevitably dubbed {\it $p$-branes})? Although 
this latter possibility was pursued by a small but dedicated group of 
theorists, starting in about 1986, it was largely ignored by
the mainstream physics community.

Well, the year 1995 witnessed a new paradigm-shift: perturbative
ten-dimensional superstrings have in their turn been superseded by a new {\it
non-perturbative} theory called {\it $M$-theory}, which describes, amongst other
things, supermembranes and superfivebranes, which subsumes the above five
consistent strings theories, and which has as its low-energy limit,
eleven-dimensional supergravity! According to  Fields Medalist Edward Witten ``M
stands for magical, mystery or membrane, according to taste''. New evidence in
favor of this theory is appearing daily on the internet and represents the most
exciting development in the subject since 1984 when the superstring revolution
first burst on the scene.
\be
\left.
\begin{array}{ll}
E_{8}е\times E_{8}е~~ heterotic~~ string &\\
SO(32)~~ heterotic~~ string &\\
SO(32)~~ Type~~I~~string & \\
Type~~ IIA ~~string&\\
Type~~ $IIB$ ~~string&
\end{array}
\right \}M~~theory
\la{M theory}
\ee

Thus this new framework now provides the starting point for
understanding a wealth of new non-perturbative phenomena, including
string/string duality, Seiberg-Witten theory, quark confinement, QCD,
particle physics phenomenology, quantum black holes, cosmology and, 
ultimately, their complete synthesis in a final theory of physics.

\subsection{The Kaluza-Klein idea}
\la{KK}

Cast your minds back to 1919.  Maxwell's theory of electromagnetism was well
established and Einstein had recently formulated his General Theory of
Relativity.
By contrast, the strong and weak interactions were not well understood.  In
searching for a unified
theory of the fundamental forces, therefore, it was natural to attempt to merge
gravity with
electromagnetism. This Kaluza was able to do through the
remarkable device of
postulating an extra fifth dimension for spacetime. Consider Einstein's theory
of pure gravity in
five spacetime dimensions with signature $(-,+,+,+,+)$. The line element is
given by
\begin{equation}
d{s}_{5}{}^2={g}_{{M}{N}}d{x}^{{M}}d{x}^{{N}}
\la{Einstein}
\end{equation}
where $M = 0,1,2,3,4$. 
Kaluza then made
the $4+1$ split
\begin{equation}
{g}_{{M}{N}}=e^{\phi/\sqrt{3}}
\pmatrix{g_{\mu\nu}+e^{-\sqrt{3}\phi}A_{\mu}A_{\nu}&e^{-\sqrt{3}\phi}A_{\mu}\cr
e^{-\sqrt{3}\phi}A_{\nu}&e^{-\sqrt{3}\phi}}
\la{KKansatz}
\end{equation}
where ${x}^{{M}}=(x^{\mu},y)$, $\mu = 0,1,2,3,$.  Thus the fields 
$g_{\mu\nu}(x)$, 
$A_{\mu}(x)$ and $\phi (x)$
transform
respectively as a tensor, a vector and a scalar under four-dimensional general
coordinate
transformations.  All this was at the classical level, of course, but in the
modern parlance of
quantum field theory, they would be described as the spin $2$ graviton, the
spin $1$ photon and
the spin $0$ dilaton\footnote{The dilaton was considered an embarrassment in 1919, and
was
(inconsistently) set equal to zero. However, it was later revived 
and subsequently stimulated Brans-Dicke theories of gravity.
The dilaton also plays a crucial role in $M$-theory}.
Of course it is
not enough to call $A_{\mu}$ by the name photon, one must demonstrate that it
satisfies Maxwell's
equations and here we see the Kaluza miracle at work.   After making the same
$4+1$ split of the
five-dimensional Einstein equations ${R}_{{M}{N}}=0$, we
correctly recover
not only the Einstein equations for $g_{\mu\nu}(x)$ but also the Maxwell
equation for
$A_{\mu}(x)$ and the massless Klein-Gordon equation for $\phi (x)$.  Thus
Maxwell's theory of electromagnetism is an inevitable consequence of Einstein's
general theory of
relativity, given that one is willing to buy the idea of a fifth dimension.

Attractive though Kaluza's idea was, it suffered from two obvious drawbacks.
First, although the
indices were allowed to range over $0,1,2,3,4$, for no very good reason the
dependence on the extra
coordinate $y$ was suppressed. Secondly, if there is a fifth dimension why
haven't we seen it? The
resolution of both these problems was supplied by Oskar Klein in
1926.
Klein insisted on treating the extra dimension seriously but assumed the fifth
dimension to have
circular topology so that the coordinate $y$ is periodic, $0\leq my \leq 2\pi$,
where $m$ has dimensions of mass. Thus the space has topology $R^4 \times
S^1$.
It is difficult to envisage a spacetime with this topology but a simpler
analogy is
provided by a garden hose: at large distances it looks like a line $R^1$ but
closer inspection
reveals that at every point on the line there is a little circle, and the
topology is $R^1 \times
S^1$.  So it was that Klein suggested that there is a little circle at each
point in four-dimensional
spacetime.

Let us consider Klein's proposal from a modern perspective. We start with pure
gravity in five
dimensions described by the action
\begin{equation}
I_{5}е=\frac{1}{2\kappa_{5}е{}^2}\int d^5{x}\sqrt{-{g}}{R}
\la{D5action}
\end{equation}
$I_{5}е$ is invariant under the
five-dimensional general coordinate transformations
\begin{equation}
\delta g_{MN}= \partial_{M} \xi^{P}
g_{PN} +\partial_{N} \xi^{P} g_{
MP}
+ \xi^{P}\partial_{P} g_{MN}
\la{general}
\end{equation}
The periodicity in $y$ means that the fields $g_{MN}(x,y)$, $A_{M}(x,y)$
and $\phi
(x,y)$ may be expanded in the form
\[
 g_{\mu\nu}(x,y)= \sum_{n=-\infty}^{n=\infty} g_{\mu\nu (n)}(x)e^{inmy},
\]
\[
A_{\mu}(x,y)=\sum_{n=-\infty}^{n=\infty}A_{\mu
(n)}(x)e^{inmy},
\]
\be
\phi(x,y)=\sum_{n=-\infty}^{n=\infty}\phi_{(n)}e^{inmy}
\la{fourier}
\ee
with
\begin{equation}
 g^*{}_{\mu\nu (n)}(x)=g_{\mu\nu (-n)}(x)
\end{equation}
etc. So (as one now finds in all the textbooks) a Kaluza-Klein theory describes
an infinite
number of four-dimensional {\it fields}. However (as one finds in none of the
textbooks) it also
describes an infinite number of four-dimensional {\it symmetries} since we may
also Fourier expand
the general coordinate parameter $ \xi^{ \mu}(x,y)$ as follows
\[ 
\xi^{\mu}(x,y)=\sum_{n=-\infty}^{n=\infty} \xi^{\mu}{}_{(n)}(x)e^{inmy} \]
\begin{equation}
  \xi^{4}(x,y)=\sum_{n=-\infty}^{n=\infty} \xi^{4}{}_{(n)}(x)e^{inmy}
\la{fourier2}
\end{equation}
with $ \xi^*{}^{M}{}_{(n)}= \xi^{M}{}_{-(n)}$.

Let us first focus on the $n=0$ modes in (\ref{fourier}), which are  just
Kaluza's
graviton, photon and dilaton. Substituting (\ref{KKansatz}) and  (\ref{fourier})
in the action
(\ref{D5action}), integrating over $y$ and retaining just the $n=0$ terms we
obtain (dropping the
$0$ subscripts) \begin{equation}
 I_{4}е=\frac{1}{2 \kappa_{4}^2}\int
d^4{x}\sqrt{{-g}}[R-\frac{1}{2}\partial_{\mu}\phi\partial^{\mu}\phi
-\frac{1}{4}e^{-\sqrt{3}\phi}F_{\mu\nu}F^{\mu\nu}]
\la{action2}
\end{equation}
where $F_{\mu\nu}=\partial_{\mu}A_{\nu}-\partial_{\nu}A_{\mu}$ and 
$2\pi\kappa_{4}{}^2=m\kappa_{5}^2$.  Newton's constant is given by 
$\kappa_{4}{}^{2}=8\pi G$.  This form 
of the action explains our choice of parameterization in 
(\ref{KKansatz}): we obtain the usual Einstein-Hilbert term for 
gravity, the conventional Maxwell kinetic term for electromagnetism 
and the conventional Klein-Gordon term for the dilaton.  Note, 
however, that we have normalized $\phi$ and $A_{\mu}$ so that $1/2\kappa_{4}е^{2}е$ is common to 
both gravity and matter terms.   From (\ref{general}), this action is 
invariant under general coordinate
transformations with
parameter $\xi^{\mu}{}_0$, i.e, (again dropping the $0$ subscripts) \[
\delta  g_{ \mu \nu}= \partial_{ \mu}  \xi^{ \rho}
g_{ \rho \nu} +\partial_{ \nu}  \xi^{ \rho}g_{ \mu\rho}
+\xi^{\rho}\partial_{\rho}g_{\mu\nu}
\]
\[
\delta A_{\mu}=\partial_{\mu} \xi^{\rho}A_{\rho}+\xi^{\rho}\partial_{\rho}
A_{\mu}
\]
\begin{equation}
\delta \phi=\xi^{\rho}\partial_{\rho}\phi,
\la{general2}
\end{equation}
local gauge transformations with parameter $\xi^{4}{}_0$
\begin{equation}
\delta A_{\mu}=\partial_{\mu}\xi^{4}
\la{gauge}
\end{equation}
and global scale transformations with parameter $\lambda$
\begin{equation}
\delta A_{\mu}=\lambda A_{\mu},\,\,\,
\delta \phi=2\lambda/\sqrt 3
\la{scale}
\end{equation}
The symmetry of the vacuum, determined by the VEVs
\begin{equation}
\langle g_{ \mu \nu} \rangle =\eta_{ \mu \nu},\,\,\,
\langle A_{\mu} \rangle=0,\,\,\,
\langle \phi \rangle =\phi_0
\la{vevs}
\end{equation}
is the four-dimensional Poincare group $\times
R$. Thus, the masslessness of the graviton is due to general covariance,
the masslessness of
the photon to gauge invariance, but the dilaton is massless because it is the
Goldstone boson
associated with the spontaneous breakdown of the global scale invariance. Note
that the gauge
group is $R$ rather than $U(1)$ because this truncated $n=0$ theory has lost
all memory of the
periodicity in $y$.

Now, however, let us include the $n\neq0$ modes. An important observation is
that the assumed
topology of the ground state, namely $R^4 \times S^1$, restricts us to general
coordinate
transformations periodic in $y$.  Whereas the general covariance
(\ref{general2}) and local
gauge invariance (\ref{gauge}) simply correspond to the $n=0$ modes of
(\ref{general})
respectively, the global scale invariance is no longer a symmetry because it
corresponds to a
rescaling
\begin{equation}
\delta  g_{MN}= -\frac{1}{2}\lambda  g_{MN}
\end{equation}
combined with a  general coordinate transformation
\begin{equation}
\xi^4=-\lambda y/m
\end{equation}
which is now forbidden by the periodicity requirement.  The field $\phi$ is
therefore merely a
pseudo-Goldstone boson.

Just as ordinary general covariance may be regarded as the local gauge symmetry
corresponding to
the global Poincar\'{e} algebra  and local gauge invariance as the gauge symmetry
corresponding to
the  global abelian algebra, so the infinite parameter local transformations
(\ref{fourier2})
correspond to an infinite-parameter global algebra with generators
\[
P^{\mu}{}_n=e^{inmy}\partial^{\mu}
\]
\[
M^{\mu\nu}{}_n=e^{inmy}(x^{\mu}\partial^{\nu}-x^{\nu}\partial^{\mu})
\]
\be
Q_n=ie^{inmy}\partial/\partial(my)
\ee
It is in fact a Kac-Moody-Virasoro generalization of the Poincar\'{e}/gauge algebra.
Although this larger algebra describes a symmetry of the four-dimensional
theory, the symmetry
of the vacuum determined by (\ref{vevs}) is only Poincar\'{e} $\times U(1)$.  Thus
the gauge
parameters $\xi^{\mu}{}_n$ and $\xi^{4}{}_n$ with $n\neq 0$ each correspond to
spontaneously
broken generators, and it follows that for $n\neq 0$ the fields $A_{\mu}{}_n$
and $\phi_n$ are the
corresponding Goldstone boson fields. The gauge fields $g_{\mu\nu}{}_n$, with
two degrees of
freedom, will then each acquire a mass by absorbing the two degrees of
freedom of each vector
Goldstone boson $A_{\mu}{}_n$ and the one degree of freedom of each scalar
Goldstone boson
$\phi_n$ to yield a pure spin $2$ massive particle with five degrees of
freedom.  This accords
with the observation that the massive spectrum is pure spin two.
Thus we
find an infinite tower of charged, massive spin $2$ particles with charges
$e_n$ and masses $m_n$
given by
\begin{equation}
e_n=n\sqrt 2\kappa_{4}е m,\,\,\, m_n=|n|/R
\la{charge}
\end{equation}
where $R$ is the radius of the $S^{1}$ given by $R^{2}m^{2}=
\langle g_{yy} \rangle$.
Thus Klein explained (for the first time) the quantization of electric
charge. (Note also that charge conjugation is just parity
tranformation $y \rightarrow
-y$ in the fifth dimension.)  Of course, if we identify the fundamental unit of
charge $e=\sqrt
2\kappa_{4}е m$ with the charge on the electron, then we are forced to take $m$ to
be very large: the
Planck mass $10^{19}$ $GeV$, way beyond the range of any current or forseeable
accelerator. This
answers the second question left unanswered by Kaluza because with $m$ very
large, the radius of the
circle must be very small: the Planck size $10^{-35}$ $meters$, which
satisfactorily accords with our
everyday experience of living in four spacetime dimensions.\footnote{A 
variation on the Kaluza-Klein theme is that our universe is a $3$-brane embedded 
in a higher dimensional spacetime\cite{Akama,Rubakov,Gibbonswiltshire}. This is 
particularly compelling in the context 
of the Type $IIB$ threebrane \cite{Horowitzstrominger} since the worldvolume 
fields  
necessarily include gauge fields \cite{DLgauge}. Thus the  
strong, 
weak and electromagnetic 
forces might be confined to the worldvolume of the brane while gravity 
propagates in the bulk. It has recently been suggested that, in such  
schemes, the extra dimensions might be much larger than $10^{-35}~{\rm 
meters}$ \cite{Antoniadis,Wittencalabi} and may even be a large as a 
millimeter \cite{Lykken,Arkani,Dienes,Randallsundrum1}. In yet another variation, the 
brane occupies the boundary of $AdS_{5}е$ and 
the extra fifth dimension is infinite \cite{Randallsundrum2}. 
Once again, the near horizon geometry of the $D3$-brane provides an example of 
this, as discussed in section \ref{AdS5}.}

It is interesting to note that, despite the inconsistency problems that arise in
coupling a finite number of massive spin two particles to gravity and/or
electromagnetism,
Kaluza-Klein theory is consistent by virtue of having an {\it infinite} tower
of such states.  Any
attempt to truncate to a finite non-zero number of massive modes would
reintroduce the inconsistency. We also note, however, that these massive  Kaluza-Klein modes
have the unusual
gyromagnetic ratio $g=1$.  Moreover, when
we embed the theory in a superstring theory or $M$-theory, these Kaluza-Klein states will
persist as a subset of
the full string or $M$-theory spectrum.  It is sometimes claimed that $g \neq 2$ 
leads to unacceptable
high-energy behaviour
for Compton scattering. However, although the classical value $g=2$ is 
required in QED, the Standard Model, and indeed open string theory, 
this is not a universal rule. Tree level unitarity applies only in 
the energy regime $M_{Planck}>E>m/Q$ for a particle of mass m and 
charge $Q$, and this 
range is empty for Kaluza-Klein theory.

In summary, it seems that a five-dimensional world with one of its dimensions
compactified on a
circle is operationally indistinguishable from a four-dimensional world with a
very particular
(albeit infinite) mass spectrum.

\subsection{The field content}
\la{Content}

Eleven is the maximum spacetime dimension in which one can formulate
a consistent supersymmetric theory, as was first recognized by Nahm 
\cite{Nahm}
in his classification of supersymmetry algebras.  The easiest way
to see this is to start in four dimensions and note that one supersymmetry
relates states differing by one half unit of helicity. If we now make the
reasonable assumption that there be no massless particles with spins greater than
two, then we can allow up to a maximum of $N=8$ supersymmetries taking us from
helicity $-2$ through to helicity $+2$.  Since the  minimal supersymmetry
generator is a Majorana spinor with four off-shell components, this means a total
of $32$ spinor components.  A discussion of spinors and Dirac matrices 
in $D$ spacetime dimensions may be found in the reprint volume of Salam and 
Sezgin \cite{Salamsezgin}. Now in a spacetime with $D$ dimensions and
signature $(1,D-1)$, the maximum value of $D$ admitting a $32$ component
spinor is $D=11$. (Going to $D=12$, for example, would require $64$ components.)
See Table \ref{minimal}\footnote{Conventions differ on how to count 
the supersymmetries and in later sections we follow the more usual 
convention \cite{Supergravity}  
that $N_{max}=8$ in $D=5$ and in $N_{max}=4$ $D=7$ }.  Furthermore, as we shall see 
later, $D=11$
emerges naturally as the maximum dimension admitting supersymmetric extended
objects, without the need for any  assumptions about higher spin.    
Not long after Nahm's paper, Cremmer, Julia and Scherk 
\cite{Supergravity}
realized  that supergravity not only permits up to seven extra dimensions but in
fact takes its simplest and most elegant form when written in its full
eleven-dimensional glory.  The unique $D=11,N=1$ supermultiplet is comprised
of a graviton $g_{MN}$, a gravitino $\psi_M$ and $3$-form gauge field 
$A_{MNP}$ with $44$, $128$ and $84$ physical degrees of freedom, respectively.
For a counting of on-shell degrees of freedom in higher dimensions, 
see Table \ref{Degrees}.
\begin{table}
\[
\begin{array}{llll}
&D-bein&e_{M}е{}^{A}е&D(D-3)/2\\
&gravitino&\Psi_{M}е&2^{(\alpha-1)}е(D-3)\\
&p-form&A_{M_{1}еM_{2}е\ldots M_{p}е}е&
\left(
\begin{array}{c}
D-2\\
p
\end{array}
\right)
\\
&spinor&\chi&2^{(\alpha-1)}е
\end{array}
\]
\caption{On-shell degrees of freedom in $D$ dimensions. $\alpha=D/2$ if $D$ is even, 
$\alpha=(D-1)/2$ if $D$ is odd. We assume Majorana fermions and divide by 
two if the fermion is Majorana-Weyl. Similarly, we assume real 
bosons and divide by two if the tensor field strength is self-dual. } 
\la{Degrees}
\end{table}
The theory may also be formulated in 
superspace.  Ironically,
however, these extra dimensions were not at first taken seriously but rather
regarded merely as a useful device for deriving supergravities in four
dimensions. Indeed $D=4,N=8$ supergravity was first obtained by Cremmer and Julia
 via the process of {\it dimensional reduction} i.e. by
requiring that all the fields of $D=11,N=1$ supergravity be independent of the
extra seven coordinates.  

For many years the Kaluza-Klein idea of taking extra dimensions seriously was largely
forgotten but the arrival of eleven-dimensional supergravity provided the missing
impetus.  
The kind of four-dimensional world we end up with depends on how we
{\it compactify} these extra dimensions: maybe seven of them would allow us to
give a gravitational origin, \`{a} la Kaluza-Klein, to the strong and weak forces as
well as the electromagnetic.  In a very influential paper, Witten 
\cite{Wittensearch} drew attention to the fact that in such a scheme the
four-dimensional gauge group is determined by the {\it isometry} group of the
compact manifold ${\cal K}$. Moreover, he proved (what to this day seems 
to be merely a gigantic coincidence) that seven is not only the maximum dimension
of ${\cal K}$  permitted by supersymmetry but the minimum needed for the isometry
group to coincide with the standard model gauge group $SU(3) \times SU(2)
\times U(1)$.
\begin{table}
\halign{\indent #&\qquad\hfil# \hfil&\quad\hfil
#\hfil&\quad\hfil # \hfil &
\quad \hfil # \hfil &\quad \hfil # \hfil &\quad #\hfil\cr
&&&Dimension & Minimal Spinor& Supersymmetry&\cr
&&&($D$ or $d$) & ($M$ or $m$) & ($N$ or $n$)&\cr
&&&11 & 32 & 1&\cr
&&&10 & 16 & 2, 1&\cr
&&&9 & 16 & 2, 1&\cr
&&&8 & 16 &2, 1&\cr
&&&7 & 16 & 2, 1&\cr
&&&6 & 8 & 4, 3, 2, 1&\cr
&&&5 & 8 & 4, 3, 2, 1&\cr
&&&4 & 4 & 8, $\ldots$, 1&\cr
&&&3 & 2 & 16, $\ldots$, 1&\cr
&&&2 & 1 & 32, $\ldots$, 1&\cr}
\caption{Minimal spinor components and supersymmetries.}
\la{minimal}
\end{table}

In the early 80's there was great interest in 
four-dimensional $N$-extended supergravities for which the global $SO(N)$ is 
promoted to a gauge
symmetry \cite{Das}. In these theories the underlying supersymmetry algebra is 
no longer 
Poincar\'{e} but rather anti-de Sitter ($AdS_4$) and the
Lagrangian has a non-vanishing cosmological constant $\Lambda$ proportional to
the square of the gauge coupling constant $g$:  
\be
G\Lambda \sim -g^{2}е
\la{G}
\ee
where $G$ is Newton's constant. The $N>4$ gauged supergravities were 
particularly interesting since the cosmological constant $\Lambda$ does 
not get renormalized \cite{CDGR} and hence the $SO(N)$ gauge symmetry has 
vanishing $\beta$-function\footnote{For $N\leq 4$, the 
$beta$ function (which receives a contribution from the spin $3/2$ 
gravitinos) is positive and the pure supergravity theories are not 
asymptotically free. The addition of matter supermultiplets only makes 
the $\beta$ function more positive \cite{Supergravity81} and hence 
gravitinos 
can
never be  confined. I am grateful to Karim Benakli, Rene Martinez Acosta 
and
Parid Hoxha  for discussions on this point.}.  The relation (\ref{G}) 
suggested
that there might be a Kaluza-Klein interpretation since in such theories 
the 
coupling constant of the gauge group arising from the isometries of 
the extra dimensions is given by
\be
g^{2}е \sim Gm^{2}е
\la{e}
\ee
where $m^{-1}е$ is the size of the compact space. Moreover, there is 
typically a negative cosmological constant
\be
\Lambda\sim -m^{2}е
\la{Lambda}
\ee
Combining (\ref{e}) and (\ref{Lambda}), we recover (\ref{G}).  Indeed, 
the
maximal $(D=4,N=8)$ gauged supergravity \cite{DN} was seen to correspond 
to the 
massless sector of $(D=11,N=1)$
supergravity \cite{Cremmer} compactified on an $S^7$ whose metric admits 
an $SO(8)$
isometry and $8$ Killing spinors \cite{DP}. An important ingredient in 
these 
developments that
had been insufficiently emphasized in earlier work on Kaluza-Klein
theory was that the $AdS_4 \times S^7$ geometry was not fed in by hand
but resulted from a {\it spontaneous compactification}, i.e. the
vacuum state was obtained by finding a stable solution of the
higher-dimensional field equations \cite{CS}.  The mechanism of
spontaneous compactification appropriate to the $AdS_4 \times S^7$
solution of eleven-dimensional supergravity was provided by the 
Freund-Rubin mechanism \cite{Freundrubin} in which the $4$-form field strength in
spacetime $F_{\mu\nu\rho\sigma}$ ($\mu=0,1,2,3$) is proportional to
the alternating symbol $\epsilon_{\mu\nu\rho\sigma}$ \cite{DV}:
\be
F_{\mu\nu\rho\sigma}е \sim \epsilon_{\mu\nu\rho\sigma}
\la{F_{4}е}
\ee
A summary of this $S^{7}е$ and other $X^{7}е$ compactifications of $D=11$ 
supergravity  down to $AdS_{4}$ may be found in
\cite{DNP}. By applying a similar mechanism to the $7$-form dual of this 
field
strength one could also find compactifications on $AdS_{7} \times
S^{4}$ \cite{PTV} whose massless sector describes gauged maximal
$N=4$, $SO(5)$ supergravity in $D=7$ \cite{PPV,TV}. Type IIB supergravity 
in $D=10$, with its self-dual
$5$-form field strength, also admits a Freund-Rubin compactification
on $AdS_{5}\times S^{5}$ \cite{Schwarzcompact,GM,KRV} whose massless sector 
describes
gauged maximal $N=8$ supergravity in $D=5$ \cite{PPV2,GRW}.

\begin{table}
\begin{center}
\begin{tabular}{ccc}
{\bf Compactification}&{\bf Supergroup}&{\bf Bosonic~subgroup}\\
$AdS_{4}е\times S^{7}$е&$OSp(4|8)$&$SO(3,2) \times SO(8)$\\
$AdS_{5}е\times S^{5}$е&$SU(2,2|4)$&$SO(4,2) \times SO(6)$\\
$AdS_{7}е\times S^{4}$е&$OSp(6,2|4)$&$SO(6,2) \times SO(5)$
\end{tabular}
\end{center}
\medskip
\caption{Compactifications and their symmetries.}
\label{supergroups}
\end{table}

In the three cases given above, the symmetry of the vacuum is
described by the supergroups $OSp(4|8)$, $SU(2,2|4)$ and $OSp(6,2|4)$
for the $S^7$, $S^5$ and $S^4$ compactifications respectively, as 
shown in Table \ref{supergroups}.  Each of these groups is known to admit 
the so-called singleton, doubleton or 
tripleton\footnote{Our nomenclature is based on the 
$AdS_{4}е,AdS_{5}е$ and $AdS_{7}е$ groups having ranks $2,3$ and $4$, 
respectively,  and differs from that of G\"{u}naydin.} 
supermultiplets \cite{Gunaydin1} as shown in Table \ref{fields}. 
\begin{table}
\begin{center}
\begin{tabular}{ccc}
{\bf Supergroup}&{\bf Supermultiplet}&{\bf Field~content}\\
$OSp(4|8)$&$(n=8,d=3)$~singleton&8~scalars,8~spinors\\
$SU(2,2|4)$&$(n=4,d=4)$~doubleton&1~vector,8~spinors,6~scalars\\
$OSp(6,2|4)$&$((n_{+}е,n_{-}е)=(2,0),d=6)$
~tripleton&1~chiral~2-form,8~spinors,5~scalars
\end{tabular}
\end{center}
\medskip
\caption{Superconformal groups and their singleton, doubleton 
and 
tripleton repesentations.}
\label{fields}
\end{table}
We recall that 
singletons are those strange representations of $AdS$ first identified 
by Dirac \cite{Dirac} which admit no analogue in flat spacetime. They 
have been much studied by Fronsdal and collaborators 
\cite{Fronsdal,Flato}.
  
This Kaluza-Klein approach to $D=11$ supergravity \cite{Kaluza} 
eventually fell out of favor for three reasons. First, as emphasized by 
Witten \cite{Wittenfermion}, it is impossible to derive by the conventional
Kaluza-Klein  technique of compactifying on a manifold a {\it chiral theory} in
four spacetime dimensions starting from a non-chiral theory such as
eleven-dimensional supergravity. Ironically,
Horava and Witten \cite{Horava} were to solve this problem  years later by 
compactifying on
something that is not a manifold!. Secondly, in spite of its maximal
supersymmetry and other intriguing features, eleven dimensional supergravity was,
after all, still a {\it field theory} of gravity with all the attendant problems
of non-renormalizability. For a recent discussion, see \cite{Deser}.  This problem also had to await the dawn of 
$M$-theory, since we now regard $D=11$ supergravity not as a 
fundamental theory in its own right but the effective low-energy 
Lagrangian of $M$-theory. 
Thirdly, these $AdS$ vacua necessarily have
non-vanishing cosmological constant unless cancelled by fermion
condensates \cite{Orzalesi}.} and this was deemed unacceptable at the 
time. However, as we shall now describe, $AdS$ is currently undergoing a
renaissance thanks to the $AdS/CFT$ correspondence.
 
\subsection{The AdS/CFT correspondence}
\la{AdSCFT}

A by-product of $M$-theory has been the revival of anti-de Sitter 
space  
brought about by Maldacena's conjectured duality between physics in the 
bulk of $AdS$ and a conformal field theory on the boundary \cite{Maldacena1}.
In particular, $M$-theory on $AdS_{4}е\times S^{7}$ is dual to a 
non-abelian $(n=8,d=3)$ superconformal theory, Type $IIB$ string theory on 
$AdS_{5}е\times S^{5}$ is dual to a $d=4$ $SU(N)$ super Yang-Mills theory 
theory and $M$-theory on $AdS_{7}е\times S^{4}$ is dual to a non-abelian 
$((n_+,n_-)=(2,0),d=6)$ conformal theory. In particular, as has been spelled 
out most clearly in the $d=4$ $SU(N)$ Yang-Mills case, there is seen to be a 
correspondence between the Kaluza-Klein mass spectrum in the bulk and the 
conformal dimension of operators on the boundary 
\cite{Gubserklebanovpolyakov,Wittenads}.  This 
duality thus holds promise not only of a deeper understanding of 
$M$-theory, but may also throw light on non-perturbative aspects of 
the theories that live on the boundary which can include 
four-dimensional gauge theories. Models of this kind, where a bulk 
theory with gravity is equivalent to a boundary theory without 
gravity, have also been advocated by `t Hooft \cite{thooft} and by 
Susskind \cite{Susskind1} who call them {\it holographic} theories. The 
reader may notice a striking similarity to the earlier idea of ``The 
membrane at the end of the 
universe''\cite{Fifteen,BDPS,BD,BDPS2,Sutton}, whereby the $p$-brane occupies 
the $S^{1}е\times S^{p}е$ boundary of $AdS_{p+2}$ and is described by a 
superconformal singleton theory and to the ``membrane/supergravity 
bootstrap'' \cite{Fifteen,BDPS,BD} which conjectured that the dynamics of the supergravity 
in the bulk of $AdS$ was dictated by the membrane on its boundary and 
vice-versa.   For example, one immediately recognises 
that the dimensions and supersymmetries of the three conformal theories 
in Maldacena's duality are exactly the same as the singleton, doubleton and tripleton 
supermultiplets of Table \ref{supergroups}. 
Further interconnections between the two are currently being explored. 
See \cite{Duffads,Seibergwitten} and Section \ref{Universe}.

\subsection{Plan of the lectures}
\la{Plan}

The first purpose of these lectures will be to explain an area of 
research that was very active about ten years ago, namely anti-de Sitter 
space, the {\it Membrane at the End of the Universe}, singletons, superconformal 
theories, the {\it Membrane/Supergravity Bootstrap} and all that. These 
topics have recently undergone a revival of interest thanks to the 
the $AdS/CFT$ conjecture of Maldacena \cite{Maldacena1} which suggests a duality 
between physics in the bulk of AdS and a superconformal theory on its 
boundary. We may thus regard this earlier work as a ``prequel'' 
($AdS/CFT$ Episode 1?). The second purpose is to discuss some very 
recent work on black holes in $AdS$ which are interesting in their own 
right as well as finding application in the $AdS/CFT$ correspondence.
Both these topics will first require a thorough grounding in 
Kaluza-Klein theory, $D=11$ supergravity, $D=10$ Type $IIB$ 
supergravity and $M2$, $D3$ and $M5$ branes. 
  
In Section \ref{eleven}, we discuss the bosonic sector of $D=11$ 
supergravity and those brane solutions (so called BPS branes) that preserve 
half of the supersymmetry: 
the supermembrane ($M2$-brane) and the superfivebrane ($M5$-brane). These 
solutions, 
together with the plane wave \cite{Hull} and Kaluza-Klein monopole 
\cite{Han}, which also preserve half the supersymmetry, are the progenitors of 
the lower dimensional BPS objects of $M$-theory. 

In Section \ref{IIB}, we discuss the bosonic sector of Type $IIB$ supergravity 
and the self-dual super threebrane ($D3$-brane) solution which also preserves 
one half of the supersymmetry.

These $M2$, $D3$ and $M5$ BPS branes may be regarded as the extremal mass=charge 
limit 
of more general two-parameter black brane 
solutions that exhibit event horizons. In the near-horizon limit, they 
tend respectively to the $AdS_4 \times S^{7}$, the $AdS_5 \times 
S^{5}$ and the $AdS_{7}е \times S^{4}е$ vacua of the maximally 
supersymmetric $D=4$, $D=5$ and 
$D=7$ gauged supergravities that correspond to the massless sector of the 
compactification of $D=11$ supergravity on $S^{7}$, Type $IIB$ supergravity on 
$S^{5}$ 
and $D=11$ supergravity on $S^{4}$, respectively. These are also 
treated in Sections \ref{eleven} and \ref{IIB}, as is the issue of 
consistent Kaluza-Klein truncation to the massless sector.   
 
The zero modes of these three brane solutions are described by Green-Schwarz type 
covariant brane actions whose bosonic sectors are treated in Section 
\ref{Branes}. These will be crucial for Section \ref{Singleton}, whose 
purpose will be to explain anti-de Sitter 
space, the membrane at the end of the universe, singletons, superconformal 
theories and the membrane/supergravity bootstrap. In 
particular, by focussing on the bosonic radial mode, we show in detail how the 
superconformal singleton  
action on $S^{1} \times S^{2}$ (including the scalar ``mass term'') 
follows from the action for a spherical $M2$ brane on the boundary of 
$AdS_{4}е \times S^{7}е$ 
in the limit of large radius: the membrane at the end of the universe. 
(We also indicate, following \cite{Seibergwitten} how this generalizes 
to 
generic branes with spherical topology but arbitrary geometry.)
Then we 
compare and contrast these old results with the comparatively recent 
Maldacena conjecture.  Since the $AdS/CFT$ correspondence will be the subject 
matter of 
several other lecturers at this school, we shall not dwell on the 
implications of our results for this correspondence but simply present 
them as interesting in their own right. 

As we have seen, these $AdS$ spaces emerge as the vacua of the maximally 
supersymmetric $D=4$, $D=5$ and 
$D=7$ gauged supergravities that correspond to the massless sector of the 
compactification of $D=11$ supergravity on $S^{7}$, Type $IIB$ supergravity on 
$S^{5}$ 
and $D=11$ supergravity on $S^{4}$, respectively. Recently, therefore, 
black hole solutions of {\it gauged} supergravity have attracted a good 
deal of attention and these are the subject of Section \ref{Blackhole}  

In \cite{Duffliu}, 
for example, new anti-de Sitter black hole solutions of gauged 
$N=8$, $D=4$, $SO(8)$ supergravity were presented. By focussing on the 
$U(1)^{4}$ Cartan subgroup, non-extremal $1$, $2$, $3$ and $4$ charge solutions 
were found. In the extremal limit, they may preserve up to 
$1/2$, $1/4$, $1/8$ and $1/8$ of the supersymmetry respectively. By 
contrast, the magnetic solutions preserve none. Since $N=8$, 
$D=4$ supergravity is a consistent truncation of 
$N=1$, $D=11$ supergravity, resulting from the $S^{7}е$ 
compactification, it follows that these black holes will also 
be solutions of this theory. In \cite{Duffliu}, it 
was conjectured that a subset of the extreme electric black holes
preserving $1/2$ the supersymmetry may be identified with the 
$S^{7}е$ Kaluza-Klein spectrum, with the non-abelian quantum numbers 
provided by the fermionic zero modes.

In \cite{Cvetic10} the non-linear 
Kaluza-Klein ans\"atze describing the embeddings of the $U(1)^4$, $U(1)^3$ and 
$U(1)^2$ truncations of these supergravities were presented, which admit $4$-charge AdS$_{4}$, 
$3$-charge $AdS_{5}$ and $2$-charge $AdS_{7}$ black hole solutions. These 
enable us to embed the 
black hole solutions back in ten and eleven dimensions and reinterpret them 
as $M2$, $D3$ and $M5$ branes spinning in the transverse dimensions 
with the black hole charges given by the angular momenta of the branes.
It is curious that the same $U(1)^4$, $U(1)^3$ and 
$U(1)^2$ black hole charges appear in these spherical 
compactifications as in the toroidal compactifications but for totally 
different reasons. Instead of arising from the intersection of 
different non-rotating branes, they arise from the different angular 
momenta of a single brane. This is indicative of deeper levels of 
duality yet to be uncovered.

At the end of each of these sections there are problems whose 
solutions may be found in Section \ref{Solutions}.

Much of what we need to do in these lectures involves only the 
bosonic sectors of the supergravities and the branes. For 
completeness, however, a comprehensive Appendix lists the complete 
field equations, symmetries and transformation rules of $D=11$ 
supergravity, Type $IIB$ supergravity and the $M2$, $D3$ and $M5$ branes, 
with all the fermionic terms included.

Introductory treatments of supersymmetry and supergravity may be found in 
\cite{Wess,Gates,Srivastava,West,Bailin,vanN}, of Kaluza-Klein theories 
in \cite{Wittensearch,Kaluza,Klein,Castellani,Salamsezgin}, of 
supermembranes in  
\cite{Fifteen,String,Classical,Duffsupermembranes,Stelle,
Polchinski}, of $M$-theory in \cite{Schwarzpower,M,TownsendM,Eleven} and of 
the 
$AdS/CFT$ correspondence in \cite{Petersen,Aharony1,Duffads}.     

\subsection{Problems 1}
\la{Problems1}

\begin{enumerate}

\item 

Derive the four-dimensional action (\ref{action2}) by substituting 
(\ref{KKansatz}) into the five-dimensional Einstein action 
(\ref{Einstein}) and 
assuming all fields are independent of the fifth coordinate $y$. 

\item 

Derive the particle content of $(D=4,N=8)$ supergravity starting 
from $(D=11,N=1)$ supergravity.

\end{enumerate}

\section{\bf ELEVEN DIMENSIONAL SUPERGRAVITY}
\la{eleven}

\subsection{Bosonic field equations}
\la{Bosonic}

We will frequently seek solutions to the field equations of $D=11$ 
supergravity in which all fermion fields are set equal to zero. The 
bosonic field equations following from Appendix \ref{D=11rules} are
\be
R_{MN}-{1\over 2}g_{MN} R={1\over 12}
\left(F_{MPQR}F_N{}^{PQR} -{1\over 8} g_{MN}
F_{PQRS}F^{PQRS}\right),
\la{einstein}
\ee
and
\be
\partial_{M}(е\sqrt{-g}F^{MUVW}е)
+\frac{1}{1152}\epsilon^{UVWMNOPQRST}еF_{MNOP}еF_{QRST}е=0
\la{tensor}
\ee
or, in the language of differential forms  
\be
d {}^* F+{1\over 2} F\wedge F=0
\ee
where $*$ denotes the Hodge dual.

\subsection{AdS$_{4}е \times$ S$^{7}е$}
\la{AdS4}
In this section, we examine the spontaneous compactification of $D=11$ 
supergravity. We are interested in obtaining a 
four-dimensional theory which 
admits maximal spacetime symmetry. With signature $(-+++)$, this 
means that the vacuum should be invariant under $SO(4,1)$, Poincar\'{e} 
or $SO(3,2)$ according as the cosmological constant is positive, zero 
or negative, corresponding to de Sitter, Minkowski or anti-de Sitter 
space, respectively. 

The first requirement of maximal symmetry is that the vev of any 
fermion field should vanish and accordingly we set
\be
\langle \Psi_{M}е \rangle =0
\la{vev}
\ee 
and focus our attention on vacuum solutions of the bosonic equations 
(\ref{einstein}) and (\ref{tensor}). We look for solutions of the 
direct product form $M_{4}е \times M_{7}е$ compatible with maximal 
spacetime symmetry. We denote the spacetime coordinates by 
$x^{\mu}е$, $\mu=0,1,2,3$ and the internal coordinates by $y^{m}е$, 
$m=1,\ldots,7$. The ansatz of Freund and Rubin \cite{Freundrubin} 
is to set
\be
F_{\mu\nu\rho\sigma}=\frac{3a}{2}\epsilon_{\mu\nu\rho\sigma}
\la{FRansatz2}
\ee 
with all other components vanishing, where $a$ is a real constant and 
the factor $3/2$ is chosen for future convenience. For future 
reference, it will prove convenient to rewrite this in terms of the 
dual field strength
\be
{F}_{7}е \equiv *F_4 +\frac{1}{2} A_{3} \wedge F_{4}ее= 6L^{6}е\epsilon_7
\ee
where $\epsilon_{7}$ is the volume form on the internal  manifold, 
and where $L=2/a$. At the moment,
$L$ is just an arbitrary constant of integration, but in Section 
\ref{Membrane} we 
shall relate it to 
the tension of an $M2$ brane. Substituting 
into the field equations we find that (\ref{tensor}) is trivially 
satisfied while (\ref{einstein}) yields the product of a 
four-dimensional Einstein spacetime
\be
R_{\mu\nu}=-3a^{2}g_{\mu\nu}=-\frac{12}{L^{2}е}g_{\mu\nu}
\la{einstein12}
\ee
with signature $(-+++)$ and seven-dimensional Einstein space
\be
R_{mn}е=\frac{3a^{2}}{2}g_{mn}е =\frac{6}{L^{2}е}g_{mn}
\la{einstein22}
\ee
with signature $(+++++++)$.

For future reference we also record the form taken by the 
supercovariant derivative ${\tilde D}_{M}е$ (\ref{covariantderiv})  when 
evaluated in the Freund-Rubin background geometry. First we decompose 
the $D=11$ gamma matrices $\Gamma_{A}е$
\be
 \Gamma_{A}е=(\gamma_{\alpha}е \otimes 1, \gamma_{5}е \otimes 
 \Gamma_{a}е)
\ee
where
\be
\{\gamma_{\alpha}е,\gamma_{\beta}\}= -2\eta_{\alpha\beta}е
\ee
\be
\{\Gamma_{a}е,\Gamma_{b}е\}=-2\delta_{ab}е
\ee
and where $\alpha,\beta\ldots$ are spacetime idices for the tangent 
space group $SO(1,3)$ and $a,b,\ldots$ are the extra-dimensional 
indices for the tangent space group $SO(7)$. Substituting the Freund- 
Rubin ansatz into the covariant drivative we find that
\be
\tilde D_{\mu}е =D_{\mu}е+\frac{1}{L}\gamma_{\mu}е\gamma_{5}е
\ee
\be
\tilde D_{m}е = D_{m}е-\frac{1}{2L} \Gamma_{m}е
\la{internalderiv}
\ee
where $\gamma_{\mu}е=e_{\mu}е^{\alpha}е\gamma_{\alpha}е$ and 
$\Gamma_{m}е=e_{m}е^{a}е\Gamma_{a}е$.

Some comments are now in order. The constancy of $a$ in the ansatz 
(\ref{FRansatz2}) is necessary to 
solve the field equations, but other ans\"{a}tze are possible and are  
discussed in \cite{DNP}. The maximally spacetime symmetric 
solution of (\ref{einstein12}) is, in fact, $AdS_{4}е$ since the cosmological 
constant
\be
\Lambda=-3a^{2}е
\ee
is negative. $AdS_{4}$ can be defined as the four-dimensional 
hyperboloid 
\be
\eta_{ab}y^{a}y^{b}=-\frac{1}{a^{2}}
\ee
in $R^{5}$ with Cartesian coordinates $y^{a}е$, where
\be
\eta_{ab}е =diag(-1,1,1,1,-1)
\ee
In polar coordinates $x^{\mu}е=(t,r,\theta,\phi)$ the line element 
may be written
\be
ds_{4}^{2}=g_{\mu\nu}dx^{\mu}dx^{\nu}
=-(1+a^{2}еr^{2}е)dt^{2}е+(1+a^{2}еr^{2}е)^{-1}еdr^{2}е
+r^{2}е(d\theta^{2}е+\sin^{2}е \theta d\phi^{2}е)
\la{metric2}
\ee
It is sometimes useful to employ the change of variable
\be
ar=\sinh \rho
\ee
for which
\be
ds_{4}^{2}=-\cosh^{2}е\rho 
dt^{2}е+\frac{1}{a^{2}}еd\rho^{2}е+\frac{1}{a^{2}е} \sinh^{2}е\rho 
(d\theta^{2}е+\sin^{2}е \theta d\phi^{2}е)
\la{adsmetric}
\ee

Representations of $SO(3,2)$ are denoted $D(E_{0},s)$, where $E_{0}е$ 
is the lowest energy eigenvalue (in units of $a$) and $s$ is the 
total angular momentum. The representation is unitary provided 
$E_{0}е\geq s+1/2$ for $s=0,1/2$ and $E_{0}е\geq s+1$ for $s \geq 1$. 
The representations are all infinite dimensional. Of course, in 
addition to being the $AdS$ group in four dimensions, 
$SO(3,2)$ has the interpretation as the {\it conformal} group in 
three dimensions, where the quantum number $E_{0}е$ plays the role of 
the conformal weight. This will prove significant in the $AdS/CFT$ 
correspondence.

There are infinitely many seven-dimensional Einstein spaces $M_{7}е$ 
satisfying (\ref{einstein22}) and we now turn to the question of how 
much unbroken supersymmetry survives. Recall from (\ref{vev}) that the vev of the 
gravitino has been set to zero. For a supersymmetric vacuum we 
require that it remain zero under a supersymmetry transformation 
(\ref{susytransforms})
\be
\langle \delta\Psi_{M}е \rangle = \langle {\tilde D}_{M}е\epsilon 
\rangle =0
\ee
To solve this in the Freund-Rubin background, we look for solutions of 
the form
\be
\epsilon(x,y)=\epsilon (x) \eta (y)
\ee
where $\epsilon(x)$ is an anticommuting four-component spinor in 
$D=4$ and $\eta(y)$ is a commuting eight-component spinor in $D=7$, 
satisfying
\be
{\tilde D}_{\mu}е\epsilon (x)=0
\ee
\be
{\tilde D}_{m}е\eta (y)=0
\ee
Thus the problem of counting unbroken supersymmetries is equivalent to 
the problem of counting {\it Killing spinors}. It is not dificult to 
establish that $AdS_{4}е$ admits the maximum number (i.e. four) as 
far as spacetime is concerned, and so the number $N$ of unbroken 
generators of $AdS$ supersymmetry is given by the number of Killing 
spinors on $M_{7}е$. From (\ref{internalderiv}) these are seen to satisfy the 
integrability condition \cite{ADP}
\be
[{\tilde D}_{m}е,{\tilde 
D}_{n}е]=-\frac{1}{4}C_{mn}е^{ab}е\Gamma_{ab}е\eta=0
\la{integrability}
\ee
where $C_{mn}е^{ab}е$ is the Weyl tensor. The subgroup of $Spin(7)$ (the 
double cover of the tangent space group $SO(7)$) generated by these 
linear combinations of the $Spin (7)$ generators $\Gamma_{ab}е$ 
corresponds to the {\it holonomy} group ${\cal H}$ of the generalized 
connection appearing in ${\tilde D_{m}е}$. Thus the maximum number of 
unbroken supersymmetries $N_{max}е$ is equal to the 
number of spinors left invariant by ${\cal H}$. This in turn is given 
by the number of singlets appearing in the decomposition of the $8$ of
$Spin(7)$ under ${\cal H}$. For example, the squashed $S^{7}$ of 
\cite{ADP,DNP2} has ${\cal H}=G_{2}$ and hence $N_{max}=1$. 
   
In the 
supersymmetric context, all linear irreducible representations of 
$N=1$ $AdS$ supersymmetry were classified by Heidenreich 
\cite{Heidenreich}. They fall into 4 classes:

1. $D(1/2,0) \oplus D(1,1/2)$ 
 
2. $D(E_{0}е,0) \oplus D(E_{0}е+1/2,1/2) \oplus D(E_{0}е+1,0), E_{0}е
\geq 1/2$

3. $D(s+1,s) \oplus D(s+3/2,s+1/2), s\geq 1/2$

4. $D(E_{0}е,s) \oplus D(E_{0}е+1/2,s+1/2) \oplus D(E_{0}е+1/2,s-1/2) 
\oplus D(E_{0}е+1,s).$

Class 1 is the singleton supermultiplet which has no analogue in Poincar\'{e}
supersymmetry.  Singletons are the subject of Section 
\ref{singletons}. Class 2 is the Wess-Zumino supermultiplet. Class 3 is the 
gauge supermultiplet with spins $s$ and $s+1/2$ with $s\geq1/2$. Class 4 is the
higher spin supermultiplet. The corresponding study of $OSp(4|N)$
representations was neglected in the literature until their importance in
Kaluza-Klein supergravity became apparent. For example, the
round $S^7$ leads to massive $N=8$ supermultiplets with maximum spin $2$. 
This corresponds to an $AdS$ type of multiplet shortening analogous to the shorteneing due to 
central charges in Poincar\'{e} supersymmetry \cite{FN}. Two features emerge: (1)  
$OSp(4|N)$ multiplets may be decomposed into the $OSp(4|1)$ multiplets discussed
above; (2) In the limit as $a \rightarrow 0$ and the $OSp(4|N)$ contracts 
to the $N$-extended Poincar\'{e} algebra, all short $AdS$ multiplets become massless
Poincar\'{e} multiplets.
 
In this section we shall focus on the maximally symmetric round 
$S^{7}е$, for which $C_{mn}е^{ab}=0$ yielding the maximum $N=8$ Killing 
spinors, and hence on the eleven-dimensional vacuum $AdS_{4}е\times 
S^{7}е$. The Kaluza-Klein 
mechanism will give rise to an effective $D=4$ theory with $N=8$ 
supersymmetry and local $SO(8)$ invariance, describing a massless $N=8$ 
multiplet coupled to an infinite tower of massive $N=8$ 
supermultiplets with masses quantized in units of $L^{-1}е$, the inverse 
radius of $S^{7}$. Combining the internal $SO(8)$ symmetry and $N=8$ 
supersymmetry with the $SO(2,3)$ of the $AdS_{4}е$ spacetime, it is 
readily seen that the complete symmetry is $OSp(4|8)$. It follows 
without any further calculation that the massless sector on $S^{7}е$ is given by 
the familiar massless $N=8$ supermultiplet consisting of (1 
spin 2, 8 spin 3/2, 28 spin 1, 56 spin 1/2, 70 spin zero), but a calculation is 
required to see the $D=11$ origin of these fields, which is different 
from that of $T^{7}$. See Table \ref{N=8}.  

\subsection{Consistent truncation to the massless modes}

An entirely different question is whether the massive modes can be 
consistently truncated to yield just gauged $N=8$ supergravity. A 
{\it consistent truncation} is defined to be one for which all 
solutions of the truncated theory are solutions of the original theory. 
It requires that the truncated fields must never appear linearly in 
the action, otherwise setting them to zero would result in further 
constraints on the massless modes over and above the equations of 
motion of the massless theory \cite{DNPW,Duffpope,Pope,deWitnicolai}. 
Interestingly enough, for generic Kaluza-Klein theories, truncation to 
the massless sector is {\it not} consistent. To illustrate this, 
consider the field equations of pure gravity with a positive 
cosmological constant $\Lambda$ in $D=4+k$ dimensions
\be
{\hat R}_{MN}е=\Lambda{\hat g}_{MN}е
\ee
This theory admits the ground-state solution of ($D=4$ de Sitter 
spacetime) $\times$ (compact manifold $M_{k}е$). In some earlier 
Kaluza-Klein literature, it was generally 
believed that the correct ansatz for the metric ${\hat g}_{MN}(x,y)$ 
is given by
\[
{\hat g}_{\mu\nu}(x,y)е=g_{\mu\nu}(x)е+
A_{\mu}е^{i}K^{mi}е(y)е(x)A_{\nu}е^{j}е(x)K^{nj}е(y)g_{mn}е(y) 
\]
\[
{\hat g}_{\mu n}(x,y)=A_{\mu}е^{i}е(x)K^{mi}е(y)g_{mn}е(y)
\]
\be
{\hat g}_{mn}(x,y)=g_{mn}(y)
\la{standard}
\ee
where ${\hat g}_{mn}(y)$ is the metric on $M_{k}е$. The quantities 
$K^{mi}е(y)$ are the Killing vectors corresponding to the isometries 
of this metric and $i$ runs over the dimension of the isometry group 
$G$. The claim that this is the correct ansatz was based on the 
observation that substituting this ansatz into the higher-dimensional 
Einstein action and integrating over $y$, one obtains the 
four-dimensional Einstein-Yang-Mills action with metric 
$g_{\mu\nu}(x)$ and gauge potential $A_{\mu}е^{i}е(x)$.

However, the correct Kaluza-Klein ansatz must be consistent with the 
higher dimensional field equations and, as we shall now demonstrate, 
this is not in general true. For example, the four-dimensional 
Einstein equations read
\be
R_{\mu\nu}е-\frac{1}{2}g_{\mu\nu}е+\Lambda g_{\mu\nu}=
\frac{1}{2}(F_{\mu\nu}е^{i}еF_{\nu}е^{\rho j}е-
\frac{1}{4}F_{\rho\sigma}е^{i}еF^{\rho\sigma j}е)
K_{n}е^{i}еK^{nj}е   
\ee
where $F_{\mu\nu}е^{i}$ is the Yang-Mills field strength. The 
inconsistency is now apparent. The left-hand side is independent of 
$y$ while the right-hand side in general depends on $y$ via the 
Killing vector combination $K_{n}е^{i}еK^{nj}е$. For example, when 
$M_{k}е=S^{k}е$ with its $SO(k+1)$ invariant metric.
\be
K_{n}е^{i}еK^{nj}е=\delta_{ij}е+Y^{ij}е(y)
\ee
where $Y^{ij}е(y)$ is that harmonic of the scalar Laplacian with next 
to lowest non-vanishing eigenvalue $2\Lambda(k+1)/(k-1)$ belonging to 
the $k(k+3)/2$ dimensional representation of $SO(k+1)$.

This situation changes radically when we turn to the $S^{7}$ 
compactification е$D=11$ supergravity. 
The reason for the difference is the presence of the three-index gauge 
field ${\hat A}_{MNP}е$ in addition to the metric ${\hat g}_{MN}$е.
The crucial observation is that the standard Kaluza-Klein ansatz
(\ref{standard}) must be augmented by the additional ansatz 
\cite{DP,DNPW} 
\be
{\hat F}_{\mu \nu p q}е=-\frac{1}{2a}\epsilon_{\mu\nu\rho\sigma}
F^{\rho\sigma i}е\nabla_{[p}еK_{q]}е^{i}е
\la{nonstandard} 
\ee
Substituting (\ref{standard}) and (\ref{nonstandard}) into the $D=11$ 
field equations (\ref{einstein}) and (\ref{tensor}) now yields the 
$D=4$ Einstein equation
\be
R_{\mu\nu}е-\frac{1}{2}g_{\mu\nu}е+\Lambda g_{\mu\nu}=
\frac{1}{2}(F_{\mu\nu}е^{i}еF_{\nu}е^{\rho j}е-
\frac{1}{4}F_{\rho\sigma}е^{i}еF^{\rho\sigma j}е)
(K_{n}е^{i}еK^{nj}е+
\frac{1}{a^{2}е}\nabla_{m}еK_{n}е^{i}е\nabla^{m}еK^{ni}е)
\la{consistenteinstein}   
\ee
The miracle of the $S^{7}$ compactification is that
\be
K_{n}е^{i}еK^{nj}е+
\frac{1}{a^{2}е}\nabla_{m}еK_{n}е^{i}е\nabla^{m}еK^{ni}е=\delta^{ij}е
\la{delta} 
\ee
and so the right-hand side of (\ref{consistenteinstein}) becomes the correct 
energy-momentum tensor of $SO(8)$ Yang-Mills! Indeed the round 
$S^{7}е$ is the only $M_{7}е$ solution of (\ref{einstein22}) known to 
satisfy (\ref{delta}).
 
\subsection{The supermembrane solution}
\la{Membrane}

Historically, the equations of motion of $D=11$ supergravity were 
written down in 1978 \cite{Supergravity} and 
the $D=11$ supermembrane was 
discovered as a fundamental object in its own right in 1987 \cite{BST1,BST2}. 
In 1990, however, it was realized \cite{dust} that the $D=11$ 
supermembrane arises as a 
classical solution of the supergravity field equations which 
preserves one half of the spacetime supersymmetry. 
We shall now describe this solution.

We begin by making an ansatz for the $D=11$ gauge fields 
$g_{MN}е=e_{M}е {}^{A}еe_{N}е{}^{B}е\eta_{AB}е$ and $A_{MNP}е$ 
corresponding to the most general three-eight split invariant under 
$P_{3}е\times SO(8)$, where $P_{3}е$ is the $D=3$ Poincar\'{e} group. We 
split the $D=11$ coordinates
\be
x^{M}е=(x^{\mu}е,y^{m}е)
\la{split1}
\ee
where $\mu=0,1,2$ and $m=3,4,\ldots ,10$, and write the line element 
as 
\be
ds^{2}е=e^{2A}е\eta_{\mu\nu}еdx^{\mu}еdx^{\nu}е
+e^{2B}е\delta_{mn}еdx^{m}еdx^{n}е
\la{metric}
\ee
and the $3$-form gauge field as
\be
A_{012}е= e^{C}е
\la{3form}
\ee
All other components of $A_{MNP}$ and all components of the 
gravitino $\Psi_{M}е$ are set equal to zero. $P_{3}е$ invariance 
requires that the arbitrary functions $A,B$ and $C$ depend only on 
$y^{m}е$; $SO(8)$ invariance then requires that this dependence be 
only through $y=\sqrt{\delta_{mn}еy^{m}еy^{n}е}$.

As we shall show, the three arbitrary functions $A$, $B$, and $C$ are 
reduced to one by the requirement that the field configuration 
(\ref{metric}) and (\ref{3form}) preserve some unbroken supersymmetry. 
In other words, there must exist Killing spinors $\epsilon$ satisfying
\be
{\tilde D}_{M}е \epsilon =0
\la{covariant}
\ee
where ${\tilde D}_{M}$ is the bosonic part of the supercovariant derivative appearing in 
the supersymmetry transformation rule of the gravitino 
(\ref{susytransform})
\be
{\tilde D}_{M}е=\partial_{M}е+\frac{1}{4}\omega_{M}е{}^{AB}е\Gamma^{AB}е
-\frac{1}{288}(\Gamma_{M}е{}^{PQRS}е-8\delta_{M}е{}^{P}е\Gamma^{QRS}е)
F_{PQRS}е
\la{covariantderiv}
\ee
We make the three-eight split
\be
\Gamma^{A}е=(\gamma_{\alpha}е\otimes \Gamma_{9}е, {\bf 1} \otimes 
\Sigma_{a}е)
\ee
where $\gamma_{\alpha}$ and $\Sigma_{a}е$ are the $D=3$ and $D=8$ 
Dirac matrices respectively and where
\be
\Gamma_{9}е=\Sigma_{3}е\Sigma_{4}е\ldots\Sigma_{10}е
\ee
so that $\Gamma_{9}е{}^{2}е=1$.  We also decompose the spinor field as
\be
\epsilon(x,y)=\zeta (x) \otimes \eta(y)
\la{spinordecomp}
\ee
where $\zeta$ is a constant spinor of $SO(1,2)$ and $\eta$ is an 
$SO(8)$ spinor which may further be decomposed into chiral eigenstates 
via the projection operators $(1+ \Gamma_{9}е)/2$.

In our background (\ref{metric}) and (\ref{3form}), the supercovariant 
derivative becomes:
\[
{\tilde D}_{\mu}е=\partial_{\mu}е
-\gamma_{\mu}е\frac{1}{2}е\Sigma^{m}е\partial_{m}е{A}\Gamma_{9}е
- \gamma_{\mu}еe^{-3A}е\Sigma^{m}е\partial_{m}еe^{C}е,
\]
\be
{\tilde D}_{m}е=\partial_{m}е
+\frac{1}{4}e^{-B}е(\Sigma_{m}е\Sigma^{n}е-\Sigma^{n}е\Sigma_{m}е)\partial_{n}еe^{B}е
- 
\frac{1}{24}e^{-3A}е(\Sigma_{m}е\Sigma^{n}е-\Sigma^{n}е\Sigma_{m}е)\partial_{n}е
e^{C}\Gamma_{9} + \frac{1}{6}e^{-3A}е\partial_{m}еe^{C}\Gamma_{9}
\ee
Note that the $\Gamma_{\mu}е$ and $\Sigma_{m}е$ carry world indices. 
Hence we find that (\ref{covariant}) admits two non-trivial solutions
\be
(1 + \Gamma_{9}е)\eta =0
\la{chiral}
\ee
where 
\be
\eta=e^{C/6}е\eta_{0}е
\ee
where $\eta_{0}е$ is a constant spinor and
\[
A=\frac{1}{3}C
\]
\be
B=-\frac{1}{6}C+constant
\la{AB}
\ee
In each case, (\ref{chiral}) means that one half of the maximal possible 
supersymmetry survives.

With the substitutions (\ref{metric}), (\ref{3form}) and (\ref{AB}),
the Einstein equation and the $3$-form equation reduce to the single 
equation for one unknown:
\be
\delta^{mn}е \partial_{m}е\partial_{n}еe^{-C}е=0
\la{harmonic}
\ee
and hence, imposing the boundary condition that the metric be 
asymptotically Minkowskian, we find
\be
e^{-C}е=1+\frac{b^{6}е}{y^{6}е}
\la{singlebrane}
\ee
where $b$ is a constant, at this stage arbitrary, and 
$y^{2}е=\delta_{mn}еy^{m}еy^{n}е$. Thus the 
metric is given by
\be
ds^2=(1+b^{6}е/y^6)^{-2/3}dx^{\mu}dx_{\mu}+
(1+b^{6}е/y^6)^{1/3}(dy^2+y^2d\Omega_7{}^2)
\ee
and the $7$-form field strength by
\be
{F}_{7}е \equiv *F_4 +\frac{1}{2} A_{3} \wedge F_{4}ее= 6b^{6}е\epsilon_7
\ee
Here $\epsilon_7$ is the volume form on $S^7$ and $\Omega_7$ is the volume. 
The mass per unit area of the membrane ${\cal M}_3$ is given by: 
\be
\kappa_{11}е^{2}е{\cal M}_3=3b^{6}ее\Omega_{7}е
\ee
This {\it elementary} solution is a singular solution of the supergravity 
equations coupled to a supermembrane source and carries a Noether
``electric'' charge  
\be
Q=\frac{1}{\sqrt{2}\kappa_{11}}\int_{S^7}   F_7
=\sqrt{2}\kappa_{11}{\cal M}_3 
\ee
Hence the solution saturates the Bogomol'nyi-Prasad-Sommerfield (BPS) bound
\be
\sqrt{2}\kappa_{11}{\cal M}_3\geq Q.
\ee
This is a consequence of the preservation of half the supersymmetries. 
Of course, the equation (\ref{harmonic}) displays a delta-function 
singularity at the origin and hence requires a source term. This is 
provided by adding to the supergravity action (\ref{D11action}) the action 
of the supermembrane itself \cite{dust}, discussed in Section 
\ref{M2}. We find, as expected, that the mass per unit volume ${\cal 
M}_{3}е$ is just the membrane tension $T_{3}е$
\be
{\cal M}_{3}е=T_{3}е  
\ee

The zero modes of this solution
belong to a $(d=3,n=8)$ supermultiplet consisting of eight scalars and
eight spinors $(\phi^I,\chi^I)$, with $I=1,...,8$, which correspond to the
eight Goldstone bosons and their superpartners associated with breaking of
the eight translations transverse to the membrane worldvolume. 

A straightforward generalization to exact, stable multimembrane 
configurations can be obtained by replacing the single membrane 
expression (\ref{singlebrane}) by a linear superposition
\be
e^{-C}е=1+\sum_{l}е\frac{b_{l}е^{6}}{|{\bf y}-{\bf y}_{l}|^{6}е}
\la{multibrane}
\ee
where ${\bf y}_{l}е$ corresponds to the position of each brane, each 
with charge ${b_{l}е}^{6}е$.  The 
ability to superpose solutions of this kind is a well-known 
phenomenon in soliton and instanton physics and goes by the name of 
the ``no-static-force condition''. In the present context, it means that  
the gravitational attractive force acting on each of the branes is 
exactly cancelled by an equal but repulsive force due to the 3-form.
This condition is closely related to the saturation of the BPS bound and to the 
existence of unbroken supersymmetry. In the case that $N$ branes with 
the same charge are stacked together, we have
\be
e^{-C}е=1+\frac{Nb^{6}}{y^{6}}
\ee
or in terms of the Schwarzschild-like coordinate $r$ given by
\be
r^{6}е=y^{6}+Nb^{6}е
\ee
we have
\be
e^{C}е=1-\frac{Nb^{6}е}{r^{6}е}
\ee
and the solution exhibits an event-horizon at $r=N^{1/6}еb$. Indeed 
the solution may be analytically continued down to $r=0$ where there 
is a curvature singularity, albeit hidden by the event horizon 
\cite{DGT}. Of particular interest is now the near horizon limit $y\rightarrow 0$, 
or equivalently the large $N$ limit, because then the metric reduces to
\cite{GT,DGT,GHT} the $AdS_{4}е\times S^{7}е$ vacuum of Section 
\ref{AdS4} with
\be
L^{6}е=Nb^{6}е
\ee
Thus
\be
ds^{2}=\frac{y^{4}}{L^{4}}еdx^{\mu}еdx_{\mu}е+\frac{L^{2}е}{y^{2}е}dy^{2}е
+L^{2}еd\Omega_{7}е^{2}е
\ee
which is just $AdS_{4}е\times S^{7}е$ with the $AdS$ metric written in 
horospherical coordinates.

\subsection{AdS$_{7}е \times$ S$^{4}е$}
\la{AdS7}
In this section, we are interested in obtaining a seven-dimensional 
theory which 
admits maximal $AdS$ spacetime symmetry. With signature $(-++++++)$, this 
means that the vacuum should be invariant under $SO(6,2)$. We again look 
for solutions of the 
direct product form $M_{7}е \times M_{4}е$ compatible with maximal 
spacetime symmetry. We denote the spacetime coordinates by 
$x^{\mu}е$, ($\mu=0,\ldots ,6$) and the internal coordinates by $y^{m}е$, 
($m=1,\ldots,4$). The ansatz of Freund and Rubin \cite{Freundrubin} 
is to set
\be
F_{4}е=3L^{2}е\epsilon_{4}е
\la{FRansatz5}
\ee
with all other components vanishing, where $L$ is a real constant. 
Here $\epsilon_{4}е$ is the volume form on the internal manifold.
As before,
$L$ is just an arbitrary constant of integration, but in Section 
\ref{Fivebrane} we 
shall relate it to 
the tension of an $M5$-brane. Substituting 
into the field equations we find that (\ref{tensor}) is trivially 
satisfied while (\ref{einstein}) yields the product of a 
seven-dimensional Einstein spacetime
\be
R_{\mu\nu}е=-\frac{3}{2L^{2}е}еg_{\mu\nu}е
\la{einstein15}
\ee
with signature $(-++++++)$ and four-dimensional Einstein space
\be
R_{mn}е=\frac{3}{L^{2}е}g_{mn}е 
\la{einstein25}
\ee
with signature $(++++)$.

The maximally spacetime symmetric solution of (\ref{einstein15}) is
$AdS_{7}е$. In these lectures, we shall focus also on the maximally 
symmetric round $S^{4}е$ solution of (\ref{einstein25}).  The Kaluza-Klein 
mechanism will give rise to an effective $D=7$ theory with $N=4$ 
supersymmetry and local $SO(5)$ invariance, describing a massless $N=4$ 
multiplet coupled to an infinite tower of massive $N=4$ 
supermultiplets with masses quantized in units of $2a$, the inverse 
radius of $S^{4}$. Combining the internal $SO(5)$ symmetry and $N=4$ 
supersymmetry with the $SO(6,2)$ of the $AdS_{7}е$ spacetime, it is 
readily seen that the complete symmetry is $OSp(6,2|4)$. It follows 
without any further calculation that the massless sector on $S^{7}е$ is 
given by the massless $N=4$ supergravity supermultiplet, and 
there indeed exists a consistent truncation to the massless sector 
\cite{Nastase1,Nastase2}.

\subsection{The superfivebrane solution}
\la{Fivebrane}

The superfivebrane was discovered as a soliton solution
of $D=11$  supergravity also preserving half the spacetime supersymmetry 
\cite{Gueven}. 
Now we make the six/five split $x^M=(x^{\mu},y^m)$ where
$\mu=0,1,2,3,4,5$ and $m=6,...,10$ and proceed in a way similar to the 
supermembrane solution. In particular, we again look for solutions 
preserving supersymmetry but this time we look for a non-vanishing 
{\it magnetic} charge carried by $F_{4}е$ :
\be
F_4=3b^{3}е\epsilon_4
\ee
The metric is given by
\be
ds^2=(1+b^{3}е/y^3)^{-1/3}dx^{\mu}dx_{\mu}+
(1+b^{3}е/y^3)^{2/3}(dy^2+y^2d\Omega_4{}^2)
\ee
where the fivebrane mass per unit $5$-volume ${\cal M}_6$, which we 
identify with the fivebrane tension $T_{6}$е, is related to 
the constant $b^{3}е$
by
\be
b^{3}е=\frac{2\kappa_{11}{}^2{\cal M}_6}{3\Omega_4}
\ee
Here $\epsilon_4$ is the volume form on $S^4$ and $\Omega_4$ is the volume.
This {\it solitonic} solution is a non-singular solution of the source-free
equations and carries a topological ``magnetic'' charge   
\be 
P=\frac{1}{\sqrt{2}\kappa_{11}}\int_{S^4}F_4=\sqrt{2}\kappa_{11}{\cal
M}_6 
\ee
Hence the solution saturates the Bogomol'nyi
bound 
\be
\sqrt{2}\kappa_{11}{\cal M}_6\geq P
\ee
Once again, this is a consequence of the preservation of half the 
supersymmetries. The soliton zero modes
are described by the chiral antisymmetric tensor multiplet
$(B^-{}_{\mu\nu},\lambda^I,\phi^{[IJ]})$.  Note that in addition to the five scalars
corresponding to the five translational Goldstone bosons, there is also a
$2$-form $B^-{}_{\mu\nu}$ whose $3$-form field strength is anti-self-dual and
which describes three degrees of freedom.

The electric and magnetic charges obey a Dirac quantization rule 
\be
QP=2\pi n \qquad n={\rm integer}
\ee
Or, in terms of the tensions,
\begin{equation}
2\kappa_{11}{}^2 {T}_3 { {T}}_6 =2\pi  n
\la{Dirac11}
\end{equation}
This naturally suggests a $D=11$ membrane/fivebrane duality. Note that 
this reduces the three dimensionful parameters $T_{3}е$, $T_{6}$ and 
$\kappa_{11}е$ down to two. Moreover, it can be shown
\cite{Duffliuminasian} that they are not independent.  To see this, we note
from Appendix \ref{M2rules} that $A_3$ has period $2\pi/T_3$ so that $F_4$ is quantized
according to   %
\begin{equation} \int F_4={2\pi n\over T_3}\,\,\,\,\,n=integer \label{kquant}
\end{equation}
Consistency of such $A_3$ periods with the spacetime action,
(\ref{D11action}), gives the relation
\begin{equation}
{(2\pi)^2\over\kappa_{11}{}^2T_3^3}\in 2Z
\label{eq:k11t3}
\end{equation}
From (\ref{Dirac11}), this may also be written as 
\begin{equation}
2\pi { T_6\over T_3{}^2}\in Z
\label{eq:newdirac}
\end{equation}
Thus the tension of the singly charged fivebrane is given by
\be
 T_6=\frac{1}{2\pi}T_3{}^2
\la{tension}
\ee
In problem \ref{Problems3}, you are asked to show that a  
derivation \cite{Schwarzpower} based on $M/IIB$ duality gives the 
same result
\cite{Dealwis}. 

As for the membrane, multifivebrane solutions may be obtained by 
superposition. If we again consider $N$ singly charged fivebranes stacked one 
upon the other, we find in the near horizon, or large $N$, limit the 
$AdS_{7}е\times S^{4}е$ geometry with
\be
L^{3}е=Nb^{3}е
\ee
namely
\be
ds^{2}е=\frac{y}{L}dx^{\mu}еdx_{\mu}е+\frac{L^{2}е}{y^{2}е}dy^{2}е
+L^{2}еd\Omega_{4}е^{2}е
\ee

\subsection{Problems 2}
\la{Problems2}

\begin{enumerate}

\item Derive the bosonic field equations (\ref{einstein}) and 
(\ref{tensor}) by varying the $D=11$ supergravity action (\ref{D11action}).

\item Prove that the $D=11$ superfivebrane preserves one half of the 
supersymmetry.

\end{enumerate}

\section{\bf TYPE IIB SUPERGRAVITY}
\la{IIB}

\subsection{Bosonic field equations}
\la{Bosonic2}

Next we consider Type $IIB$ supergravity in $D = 10$ 
\cite{Schwarz,Howewest} which also describes 128 + 128 degrees of freedom, 
and corresponds to the field-theory limit of the Type $IIB$ superstring. 
The spectrum of 
the supergravity theory consists of a complex scalar $B$, a complex 
spinor $\lambda$, a complex 2-form $A_{MN}е$, a complex Weyl 
gravitino $\psi_{M}е$, a real graviton $e_{M}е^{R}$ and a real 
4-form $A_{MNPQ}$ whose 5-form field strength 
$F_{MNPQR}$ obeys a self-duality condition. 
Owing to this self-duality , there exists no covariant action 
principle and
it is therefore simplest to work directly with the field equations.
Once again, it is convenient to isolate just the bosonic field 
equations following from Appendix \ref{TypeIIB} which read
\be
D^{M}еP_{M}е=\frac{1}{24}\kappa_{10}е^{2}еG_{MNP}еG^{MNP}е
\ee
\be
D^{P}еG_{MNP}е=P^{P}еG^{*}е_{MNP}е-\frac{2}{3}i\kappa_{10}еF_{MNPQR}еG^{*}е{}^{PQR}е
\ee
\be
R_{MP}е-\frac{1}{2}g_{MP}еR=
P_{M}еP^{*}е_{P}е+P^{*}е_{M}еP_{P}е-g_{MP}еP^{R}еP^{*}е_{R}е
+\frac{1}{6}\kappa_{10}е^{2}еF_{R_{1}е\ldots R_{4}еM}еF^{R_{1}е\ldots R_{4}е}е_{P}е
\ee
\be
+\frac{1}{8}\kappa_{10}е^{2}е(G_{M}е^{RS}еG^{*}{}е_{PRS}е
+G^{*}{}е_{M}е^{RS}еG_{PRS}е)
-\frac{1}{24}\kappa_{10}е^{2}еg_{MP}еG^{RST}еG^{*}е{}_{RST}
\ee
\be
F_{MNPQR}е=*F_{MNPQR}е
\la{selfdual}
\ee
where
\be 
P_{M}е=f^{2}е\partial_{M}еB
\ee
\be
G_{MNP}е=f(F_{MNP}-BF^{*}е{}_{MNP}е)
\ee
\be
F_{MNP}е=3\partial_{[M}еA_{NP]}е
\ee
\be
f=(1-B^{*}еB)^{-1/2}е
\ee
\be
F_{MNPQR}е=5\partial_{[M}еA_{NPQR]}е-
\frac{5}{4}\kappa_{10}е{\rm Im}(A_{[MN}еF^{*}е{}_{PQR]}е)
\ee
Our notation is that $X^{*}$ is the complex conjugate of $X$ while $*X$ 
is the Hodge dual of $X$.
 
\subsection{AdS$_{5}е \times$ S$^{5}е$}
\la{AdS5}
In this section, we are interested in obtaining a five-dimensional 
theory which 
admits maximal $AdS$ spacetime symmetry. With signature $(-++++)$, this 
means that the vacuum should be invariant under $SO(4,2)$. We again look 
for solutions of the 
direct product form $M_{5}е \times M'_{5}е$ compatible with maximal 
spacetime symmetry. We denote the spacetime coordinates by 
$x^{\mu}е$, ($\mu=0,\ldots, 4$) and the internal coordinates by $y^{m}е$, 
($m=1,\ldots,5$). The ansatz of Freund and Rubin \cite{Freundrubin} 
is to set
\be
F_{5}е=G_{5}е+*G_{5}е
\ee
and 
\be
G_{5}е=4L^{4}\epsilon_{5}ее
\la{FRansatz3}
\ee 
with all other components vanishing, where $L$ is a real constant. 
Here $\epsilon_{5}е$ is the volume form on the internal manifold.
As before,
$L$ is just an arbitrary constant of integration, but in Section 
\ref{Threebrane}, we 
shall relate it to 
the tension of an $D3$-brane. Substituting 
into the field equations we find that (\ref{tensor}) is trivially 
satisfied while (\ref{einstein}) yields the product of a 
five-dimensional Einstein spacetime
\be
R_{\mu\nu}е=-\frac{4}{L^{2}}g_{\mu\nu}е
\la{einstein13}
\ee
with signature $(-++++)$ and five-dimensional Einstein space
\be
R_{mn}е=\frac{4}{L^{2}}еg_{mn}е 
\la{einstein23}
\ee
with signature $(+++++)$.

The maximally spacetime symmetric solution of (\ref{einstein13}) is
$AdS_{5}е$. In these lectures, we shall focus also on the maximally 
symmetric round $S^{5}е$ solution of (\ref{einstein23}).  The Kaluza-Klein 
mechanism will give rise to an effective $D=5$ theory with $N=8$ 
supersymmetry and local $SO(6)$ invariance, describing a massless $N=8$ 
multiplet coupled to an infinite tower of massive $N=8$ 
supermultiplets with masses quantized in units of $a$, the inverse 
radius of $S^{5}$. Combining the internal $SO(6)$ symmetry and $N=8$ 
supersymmetry with the $SO(2,4)$ of the $AdS_{5}е$ spacetime, it is 
readily seen that the complete symmetry is $SU(2,2|4)$. It follows 
without any further calculation that the massless sector on $S^{5}е$ 
is given by the massless $N=8$ supergravity supermultiplet. To date, 
however, the proof of a complete consistent truncation to the massless 
sector is still lacking.

\subsection{The self-dual superthreebrane solution}
\la{Threebrane}

The resulting field equations admit as a solution the self-dual 
threebrane \cite{HS} which in the extremal limit preserves half the 
supersymmetry \cite{DLgauge} just like the extremal 2-brane and 5-brane of $D=11$ 
supergravity. We need keep only the graviton, dilaton and 4-form and 
make the four/six split $x^{M}е=(x^{\mu}е,y^{m}е)$ where $\mu=0,1,2,3$ 
and $m=5, \ldots, 10$. The solution is given by
\be
ds^{2}е=(1+b^{4}ее/y^{4}е)^{-1/2}еdx^{\mu}еdx_{\mu}е
+(1+b^{4}ее/y^{4}е)^{1/2}е(dy^{2}е+y^{2}еd\Omega_{5}е^{2}е)
\la{threebranemetric}
\ee
\be
e^{2\phi}е=e^{2\phi_{0}е}е=constant
\la{threebranedilaton}
\ee
\be
F_{5}е=G_{5}е+*G_{5}е
\la{threebrane5form}
\ee
where
\be
G_{5}е=4b^{4}ее\epsilon_{5}е
\ee
Here we have employed the string frame metric 
$g_{MN}е(string)=e^{\phi/2}еg_{MN}е(Einstein)$.
The threebrane mass per unit volume ${\cal M}_{4}$ is then related to 
the tension $T_{4}е$ by \cite{String}
\be
{\cal M}_{4}е=e^{-\phi_{0}е}еT_{4}е 
\ee
and related to 
the constant $k_{4}е$ by
\be
b^{4}=е\frac{2\kappa_{10}е^{2}е{{\cal M}_{4}е}}{\Omega_{5}е}
\ee
It is perhaps worth saying a few more words about the
self-duality of the superthreebrane. By virtue of the self-duality 
condition (\ref{selfdual}), the electric Noether charge 
\be
Q=\frac{1}{{\sqrt 2} \kappa_{10}е}\int_{S^{5}е}е*F_{5}е
\ee
coincides with the topological magnetic charge 
\be
P=\frac{1}{{\sqrt 2} \kappa_{10}е}\int_{S^{5}е}еF_{5}е
\ee
so
\be
Q=P
\ee
and hence
\be
2\kappa_{10}^{2}е{\cal M}_{4}е^{2}е=2\pi n
\ee
(Note that such a condition is possible only in theories allowing a real
self-duality condition i.e. in $D=2$ modulo $4$ dimensions, assuming Minkowski
signature. The $D=6$ self-dual string of \cite{lublack} is another example.) 

We can also count bosonic and fermionic zero modes. We know that one half of
the supersymmetries are broken, hence we have 16 fermionic zero modes. 
Regrouping these 16 fermionic zero modes, we get four Majorana spinors in $d
= 4$. Hence the $d = 4$ worldvolume supersymmetry is $N = 4$. Worldvolume
supersymmetry implies that the number of fermionic and bosonic on-shell
degrees of freedom must be equal, so we need a total of eight bosonic zero
modes. There are the usual six bosonic translation zero modes, but we are
still short of two. The two extra zero modes come from the excitation of the
complex antisymmetric field strength $G_{MNP}$ and correspond to a real vector 
field on the worldvolume \cite{DLgauge}. Together with the other zero modes,
these fields make up the $d = 4, N = 4$ gauge supermultiplet
$(A_{\mu}, \lambda^I,\phi^{[IJ]})$.

As for the membrane and fivebrane, multithreebrane solutions may be obtained by 
superposition. If we again consider $N$ singly charged threebranes stacked one 
upon the other, we find in the near horizon, or large $N$, limit the 
$AdS_{7}е\times S^{4}е$ geometry with
\be
L^{4}е=Nb^{4}е
\ee
namely
\be
ds^{2}е=\frac{y^{2}е}{L^{2}е}dx^{\mu}еdx_{\mu}е+\frac{L^{2}е}{y^{2}е}dy^{2}е
+L^{2}еd\Omega_{5}е^{2}е
\ee

\subsection{Problems 3}
\la{Problems3}

\begin{enumerate}

\item
Show, using the duality between $M$-theory on $T^2$ and Type $IIB$ theory on
$S^1$, that the $M$-theory membrane tension $T_3$ and the $M$-theory fivebrane
tension $ T_6$ are related by
\be
T_6=\frac{1}{2\pi}T_3{}^2
\ee  

\item

Write down the dictionary that relates the fields of $M$-theory on 
$T^{2}е$ to $IIB$ on $S^{1}е$.

\end{enumerate}

\section{\bf THE M2-BRANE, D3-BRANE AND M5-BRANE}
\la{Branes}

\subsection{The M2-brane}
\la{M2}

The $8+8$ zero modes of the $M2$-brane are described by a 
supersymmetric 2+1 dimensional worldvolume action.  However, this 
obscures the underlying $D=11$ spacetime supersymmetry. It is 
possible to construct a covariant action with the scalars and spinors 
being given by the 11 bosonic coordinates $X^{M}е$ and the 32 
fermionic coordinates $\theta^{\alpha}е$ of a $D=11$ superspace 
\cite{BST1,BST2}. Indeed, in the case of the $M2$ brane this came first. 
The $8+8$ physical degrees of freedom, and the worldvolume 
supersymmetry then emerge by going to a physical 
gauge, as discussed in Section \ref{Bootstrap}.    
We begin with the bosonic sector of the $d=3$ worldvolume of the $D=11$
supermembrane which follows from Appendix \ref{M2rules}:
\begin{equation}
S_3=T_3\int d^3\xi\biggl[-{1\over2}\sqrt{-\gamma}\gamma^{ij}
\partial_i x^M\partial_j x^N g_{MN}(x) +{1\over2}\sqrt{-\gamma}
+{1\over3!}\epsilon^{ijk}\partial_i x^M\partial_j x^N\partial_k x^P
A_{MNP}(x)\biggr]\ ,
\la{membranebose}
\end{equation}
where $T_3$ is the membrane tension, $\xi^i$ ($i=0,1,2$) are the
worldvolume coordinates, $\gamma^{ij}$ is the worldvolume metric and
$x^M(\xi)$ are the spacetime coordinates $(M=0,1,\ldots,10)$.  Kappa
symmetry, discussed in Appendix \ref{M2rules}, demands that the
background metric $g_{MN}$ and background 3-form potential $A_{MNP}$
obey the classical field equations of $D=11$ supergravity 
(\ref{einstein}) and (\ref{tensor}).

Varying (\ref{membranebose}) with respect to $x^{M}$е yields
\be
\partial_{i}е(\sqrt{-\gamma}\gamma^{ij}е\partial_{j}еx^{N}еg_{MN}е)-
\frac{1}{2}\sqrt{-\gamma}\gamma^{ij}е\partial_{i}еx^{N}е\partial_{j}еx^{P}е
\partial_{M}е
g_{NP}е+
\frac{1}{3!}\epsilon^{ijk}е\partial_{i}еx^{N}е\partial_{j}еx^{P}е
\partial_{k}еx^{Q}е
F_{MNPQ}е=0
\la{membraneboseeq}
\ee
while varying with respect to $\gamma_{ij}е$ yields the embedding 
equation
\be
\gamma_{ij}е=\partial_{i}еx^{M}е\partial_{j}еx^{N}еg_{MN}е
\la{embedding}
\ee

An important issue is the existence of membrane configurations 
preserving some supersymmetry \cite{BDPS,BDPS2}. We usually  seek supersymmetric 
membrane vacuum states for which both the spacetime 
gravitino $\Psi_{M}е$ and the fermionic coordinates $\theta(\xi)$ are zero, so 
that the membrane action reduces to (\ref{membranebose}). In 
this case, the criterion for surviving supersymmetry is that the 
vacuum expectation values of $\Psi_{M}е$ and $\theta(\xi)$ remain zero 
under some appropriate combination of $\kappa$ symmetry and 
supersymmetry transformations. We shall work to linear 
order in fermions, which is as high as we need go to investigate the 
supersymmetry of a purely bosonic background. From Appendices \ref{D=11rules} and 
\ref{M2rules}, these transformations rules are
\be
\delta \Psi_{M}е={\tilde D}_{M}е\epsilon(x)
\la{susy} 
\ee
\be
\delta \theta=(1+\Gamma) \kappa(\xi) +\epsilon(x)
\la{kappa}
\ee
where $\epsilon$ and $\kappa$ are the supersymmetry and $\kappa$ 
symmetry parameters, respectively. Here ${\tilde D}_{M}$ is the 
$D=11$ supercovariant derivative (\ref{covariantderiv}) 
and $\Gamma$ is given by
\be
\Gamma=\frac{1}{3!}\sqrt{-\gamma}\epsilon^{ijk}\partial_{i}еx^{M}е
\partial_{j}еx^{N}е
\partial_{k}еx^{P}е\Gamma_{MNP}
\ee
As a consequence of the embedding equation (\ref{embedding}), $\Gamma$ 
satisfies $\Gamma^{2}е=1$. Since $\Gamma$ is tracefree, this implies 
that $(1\pm \Gamma)/2$ are projection operators with 16 zero eigenvalues. 
Irrespective of the background geometry the $\kappa$ symmetry may 
therefore be used to set 16 of the 32 components of $\theta$ to zero.
A convenient choice is
\be
(1+\Gamma)\theta=0
\la{kappagauge}
\ee
Acting with (\ref{kappa}), we see that around a purely bosonic 
background, (\ref{kappagauge}) is preserved if $(1+\Gamma)\delta 
\theta=0$, i.e., combined $\kappa$ and $\epsilon$ transformations for 
which 
\be
2(1+\Gamma)\kappa + (1+\Gamma)\epsilon=0
\ee
which implies that $\delta \theta=(1-\Gamma)\epsilon/2$. Thus a bosonic 
vacuum has residual supersymmetries corresponding to solutions of \cite{BDPS,BDPS2}   
\be
\tilde D_{M}е\epsilon=0,~~~~~~~~\Gamma\epsilon(x)=\epsilon(x)
\la{branesusy}
\ee
We shall make use of this in seeking the membrane at the end of the 
universe in Section \ref{Universe}.

\subsection{The M5-brane}
\la{M5}
The bosonic $M5$ worldvolume equations of motion follow from Appendix \ref{M5rules}. 
We work with equations of motion rather than an action because of the 
appearance of a self-dual worldvolume 3-form field strength $h_{abc}$.

The suitable pullbacks of the spacetime 3-form potential, and the induced metric 
are 
\[
A_{ijk}=\del_i x^{M}\del_j x^{N}\del_k x^{P}A_{PNM}
\]
\be
\gamma_{ij}=\del_i x^{M}E_{M}{}^{A}\del_j x^{N}E_{N}{}^{B}\eta_{AB}
\ee
$\xi^i$ ($i=0,\ldots,5$) are the
worldvolume coordinates, $\gamma^{ij}$ is the worldvolume metric and
$x^M(\xi)$ are the spacetime coordinates $(M=0,1,\ldots,10)$. 
We introduce the worldvolume $2$-form $A_{ij}е$ and corresponding 
worldvolume $3$-form:
\be
\cF_{ijk}\equiv 3\partial_{[i}еA_{jk]}-A_{ijk}.
\la{cf3}
\ee
The field equation for $A_{ij}е$ is
\be
G^{mn}\nabla_m
{\cF}_{npq}\se Q^{-1}\left[4Y-2(mY+Ym)+mYm\right]_{pq}\ ,
\ee
and the field equations for the $x^{M}$ are
\be
G^{mn}\nabla_m{\cE}_n{}^{C} \se {Q\over \sqrt{-g}} \e^{m_1\cdots m_6
}\left(\ft1{6!}F^{A}{}_{m_1\cdots m_6} + \ft1{(3!)^2}
F^{A}{}_{m_1m_2m_3}\,{\cF}_{m_4m_5m_6}\, \right)P_{A}{}^{C}\ .
\la{e1}
\ee
Several definitions are in order. To begin with,
\bea
m_a{}^b \equiv \d_a{}^b-2 k_a{}^b\ ,\qquad
k_a{}^b \equiv h_{acd} h^{bcd}\ ,\qquad Q \equiv (1-\ft23 \tr\,k^2)\ ,
\nn\w2
Y_{ab} \equiv \left[4*F-2(m*F+*F m)+m*F m\right]_{ab}\ ,
\nn\w2
 P_{A}{}^{C} \equiv
\d_{A}{}^{C}-{\cE}_{A}{}^m{\cE}_m{}^{C}\ ,
\quad\quad *F^{ab} \de \ft1{4!\sqrt{-g}}\e^{abcdef}F_{cdef}\ ,
\la{defs}
\eea
The fields ${\cF}_{3}$, $F_{4}$ and its Hodge dual
$F_{7}$ are given by
\be
{\cF}_3 \se dA_2-{A}_3\ , \qquad F_4 \se dA_3 \ , \qquad F_7 \se dA_6
+ \ft12\,A_3\wedge F_4\ .
\la{h7}
\ee
The target space indices on $F_4$ and $F_7$ have been converted to
worldvolume indices with factors of ${\cE}_m{}^{A}$ defined as
\be
{\cE}_m{}^{A}(x) \equiv \del_m x^{M} E_{M}{}^{A}
\ee
The metric
\be
\gamma_{mn}(x) \equiv {\cE}_m{}^{A}{\cE}_n{}^{B}\eta_{AB}\se
e_m{}^a e_n{}^b \eta_{ab}
\ee
is the standard induced metric with determinant $\gamma$, and $G^{mn}$
is another metric defined as
\be
G^{mn} \equiv (m^2)^{ab}e_a{}^m e_b{}^n\ .
 \la{gmn}
 \ee
Let us note that the connection in the covariant derivative $\nabla_m$
occurring in (\ref{e1}) is the Levi-Civita connection for the induced
metric $\gamma_{mn}$.

A key relation between $h_{abc}$ and ${\cF}_{abc}$ follows from the Bianchi 
identity $d{\cF}_3=-{F}_4$,and is given by
\be
h_{abc}\se \ft14\,m_a{}^d {\cF}_{bcd}\ .
\la{hf}
\ee
We conclude this Section by elucidating the consequences of the central
equation (\ref{hf}). To this end, we first note the useful identities
\[
h_{abe}h^{cde}=\delta^{[c}_{[a}k_{b]}^{\ \;d]}
\]
\[
k_{ac}k_b{}^c=\ft16 \eta_{ab}\tr~k^2
\]
\be
k_a{}^dh_{bcd}=k_{[a}{}^d h_{bc]d}
\la{hk1}
\ee
which are consequences of the linear self-duality of $h_{abc}$. Taking
the Hodge dual of (\ref{hf}) one finds $*\cF_{abc} = -\cF_{abc} + 2
Q^{-1} m_a{}^d \cF_{bcd}$. Using the identity $m^2=2m-Q$, we readily
find the nonlinear self-duality equation
\be
*\cF_{mnp}=Q^{-1}G_m{}^q \cF_{npq}
\la{sd}
\ee
This equation can be expressed solely in terms of $\cF_3$. To do this, we
first insert (\ref{hf}) into (\ref{hk1}), which yields the identities
\[
\cF_{abe}\cF^{cde} =2\delta_{[a}^{[c} X_{b]}^{\ \;d]} + \ft12
K^{-2} X_{[a}{}^{c} X_{b]}{}^{d} + 2(K^2-1) \delta_{[a}^{c}
\delta_{b]}^{d}
\]
\[
X_{ac}X_b{}^c = 4K^2(K^2-1)\eta_{ab}
\]
\be
X_a{}^d \cF_{bcd} = X_{[a}{}^d \cF_{bc]d}
\la{x}
\ee
where we have defined
\[
K \equiv  \sqrt{1+\ft1{24}\cF^{abc}\cF_{abc}}
\]
\be
X_{ab} \equiv \ft12 K *\cF_a{}^{cd}\cF_{bcd}\ .
\la{xx}
\ee
Next we derive the identities
\[
Q(K+1) \equiv 2
\]
\be
X_{ab} \equiv \ft12 \cF_{acd}\cF_{b}{}^{cd} -
\ft1{12} \eta_{ab} \cF_{cde}\cF^{cde} \se 4 K(1+K) k_{ab}
\ee
We can now express (\ref{sd}) entirely in terms of $\cF_3$
by deriving the identity
\be
Q^{-1}G_{mn}=K \eta_{mn} - \ft12 K^{-1} X_{mn}
\ee
Another way of writing (\ref{sd}) is
\be 
\cF^-_{abc}=\ft12 (1+K)^{-2} \cF^+_{ade}\cF^{+def}\cF^+_{fbc}\ ,
\ee
where $K$ is a root of the quartic equation
\be
(K+1)^3(K-1)= \ft1{24} \cF^{+ abc}\cF^{+}_{ade}\cF^{+def}\cF^+_{fbc}\ .
\ee

\subsection{The D3-brane}
\la{D3}

From Appendix \ref{D3rules}, the bosonic sector of the $D3$-brane  
coupled the background of Type $IIB$ supergravity is given by the action
\be
S_{4}= -T_{4}е\int d^{4}\xi\, 
e^{-\phi}\sqrt{-\det (\gamma_{ij} + {\cal F}_{ij})}+
T_{4}\left[ е\int A_{4}е +\int A_{2}е\wedge {\cal F} +\frac{1}{2!}\int 
A_{0}е\wedge {\cal F} \wedge {\cal F}\right]
\ee
where ${\cal F}_{ij}$ are the components of a modified $2$-form field strength 
\be
{\cal F} = F-B\ ,
\ee
where $F=dV$ is the usual field strength $2$-form of the Born-Infeld field $V$ 
and $B$ is the pullback to the worldvolume of the NS-NS Type $IIB$ $2$-form 
potential $B$, and where $A_{0}е$, $A_{2}е$ and $A_{4}е$ are the R-R 
forms. We use the same letter for superspace forms and their pullbacks to the
worldvolume.

\subsection{Problems 4}
\la{Problems4}

\begin{enumerate}

\item
Show that the equations for the bosonic sector of the Type $IIA$ superstring
in $D=10$ follow from those of the supermembrane in $D=11$ by assuming 
an $M_{10} \times S^{1}е$ topology, wrapping the membrane around 
the $S^{1}е$ and taking the small radius limit \cite{BDPS}. 

\end{enumerate}

\section{\bf ADS/CFT : THE MEMBRANE AT THE END OF THE UNIVERSE}
\la{Singleton}

\subsection{Singletons live on the boundary}
\la{Boundary}

As emphasized by Fronsdal et al. \cite{Fronsdal,Flato}, singletons are best 
thought of as 
living not in the $(d+1)$-dimensional bulk of the $AdS_{d+1}е$ spacetime 
but rather on the $d$-dimensional $S^{1} \times S^{d-1}$ boundary where the 
$AdS$ group $SO(d-1,2)$ plays the role of the {\it conformal} group. 
Remaining 
for the moment 
with our $4$-dimensional example, consider a scalar field 
$\Phi(t,r,\theta,\phi)$ on $AdS_{4}е$ with metric (\ref{metric}), 
described by 
the action
\be
S_{bulk}=\int_{AdS_{4}}d^{4}еx \sqrt{-g} 
\frac{1}{2}\Phi\left[ g^{\mu\nu}е\nabla_{\mu}е \nabla_{\nu}е 
-M^{2}е\right]\Phi
\la{bulk}
\ee
Note that this differs from the conventional Klein-Gordon action by a 
boundary term. Since the scalar Laplacian on $AdS_{4}е$ has eigenvalues 
$E_{0}(E_{0}-3)a^{2}$, the critical value of $M^{2}е$ for a singleton with 
$(E_{0},s)=(1/2,0)$ is
\be
M^{2}=\frac{5}{4}a^{2}е
\ee
In this case, one can show with some effort \cite{Fronsdal,Flato} that as 
$r 
\rightarrow \infty$,
\be
\Phi(t,r,\theta,\phi) \rightarrow 
r^{{-1/2}}\phi(t,\theta,\phi)
\ee
and hence that the radial dependence drops out:
\be
S_{boundary}е=\int_{S^{1} \times S^{2}}d^{3}е\xi 
\sqrt{-h}[-\frac{1}{2} h^{ij}\nabla_{i}\phi \nabla_{j}\phi 
-\frac{1}{8}a^{2}\phi^{2}]
\la{boundary}
\ee
Here we are integrating over a $3$-manifold with $S^{1} \times S^{2}$ 
topology and with metric
\be
h^{ij}еd\xi^{i}еd\xi^{j}е=-dt^{2} + 
\frac{1}{a^{2}е}(d\theta^{2}е+\sin^{2}е \theta d\phi^{2}е)
\la{boundarymetric}
\ee
This $3$-manifold is sometimes referred to as the {\it boundary} of 
$AdS_{4}$ 
but note that the metric $h_{ij}$ is not obtained by taking the 
$r \rightarrow \infty$ limit of $g_{\mu\nu}$ but rather the $r \rightarrow 
\infty$ limit of the conformally rescaled metric $\Omega^{2}еg_{\mu\nu}$ 
where $\Omega=1/ar$. The radius of the $S^{2}е$ is $a^{-1}е$, not 
infinity. Most particle physicists are familiar with the conformal 
group in flat Minkowski space. It is the group of coordinate 
transformations which leave invariant the Minkowski lightcone. In the 
case of three-dimensional Minkowski space, $M_{3}$, it is $SO(3,2)$. 
In the present context, however, the spacetime is curved with 
topology $S^{1} \times S^{2}е$, but still admits $SO(3,2)$ as its 
conformal group\footnote{One sometimes finds the statement in 
the physics literature that the only compact spaces admitting 
conformal Killing vectors are those isomorphic to spheres. By a 
theorem of Yano and Nagano \cite{Yano}, this is true for {\it 
Einstein spaces}, but $S^{1} \times S^{2}е$ is not Einstein.}, i.e. 
as the group which leaves invariant the 
three-dimensional lightcone $h_{ij}е 
d\xi^{i}еd\xi^{j}е=0$.  The 
failure to discriminate between these different kinds of conformal 
invariance is, we believe, a source of confusion in the singleton 
literature. In particular, the $\phi^{2}е$ ``mass'' term appearing in 
the action (\ref{boundary}) would be incompatible with conformal 
invariance if the action were on $M_{3}е$ but is essential for 
conformal invariance on $S^{1} \times S^{2}$. Moreover, the 
coefficient $-a^{2}е/8$ is uniquely fixed \cite{BD}. In Section 
\ref{Universe} we shall derive the singleton action (\ref{boundary}) 
including the scalar mass terms, starting from the membrane action 
(\ref{membranebose}).

So although singleton actions of the form (\ref{bulk}) and their 
superpartners appeared in the Kaluza-Klein harmonic expansions on $AdS_{4} 
\times S^7$ \cite{Sezgin,NS,GRW2}, they could be
gauged away everywhere except on the boundary where the above 
$OSp(4|8)$ 
corresponds to the superconformal group \cite{Nahm}. One 
finds an $(n=8,d=3)$ supermultiplet with $8$ scalars $\phi^{A}$ and $8$ 
spinors
$\chi^{\dot{A}}$, where the indices $A$ and $\dot{A}$ range over $1$ to 
$8$ and denote the 
$8_{s}$ and $8_{c}$ representations of $SO(8)$, respectively. The 
$OSp(4|8)$ 
action is a generalization of (\ref{boundary}) and is given by \cite{BD} 
\be
S_{singleton}=\int_{S^{1} \times S^{2}}d^{3}е\xi 
\sqrt{-h}[-\frac{1}{2}h^{ij}\nabla_{i}\phi^{A} \nabla_{j}\phi^{A} -\frac{1}{8}a^{2}
\phi^{A}\phi^{A}
+\frac{i}{2}{\bar{\chi}}^{\dot{A}}(1-\gamma)\gamma^{i}еD_{i}е\chi^{\dot{A}}]
\la{singletonaction}
\ee
where $\gamma=-\gamma_{0}е\gamma_{1}е\gamma_{2}е$ and where $D_{i}$ is the 
covariant derivative appropriate to the $S^{1}е \times S^{2}е$ background. 

In the
case of $AdS_{5} \times S^5$ one finds a $(n=4,d=4)$ supermultiplet with 
$1$ vector $A_{i}$, $(i=0,1,2,3)$, a complex spinor 
$\lambda^{a}{}_{+}$, $(a=1,2,3,4)$, obeying 
$\gamma_{5}\lambda^{a}{}_{+}=\lambda^{a}{}_{+}$ and $6$ real scalars 
$\phi^{ab}$, obeying $\phi^{ab}=-\phi^{ab}е$, 
$\phi^{ab}=\epsilon^{abcd}е \phi_{cd}е/2$.  The corresponding action for 
the 
doubletons of $SU(2,2|4)$ is 
\cite{NST}
\be  
S_{doubleton}= \int_{S^{1} \times S^{3}} 
[-\frac{1}{4}F_{ij}F^{ij} -\frac{1}{4}a^{2}е\phi_{ab}\phi^{ab} 
-\frac{1}{4}\partial_{i}\phi_{ab}\partial^{i}\phi^{ab}
+i{\bar{\lambda}}_{+a}\gamma^{i}D_{i}\lambda^{a}{}_{+}]
\la{doubletonaction} 
\ee
where $F_{ij}=2\partial_{[i}еA_{j]}е$. However, in contrast to the 
singletons, 
we know of no derivation of this doubleton action on the boundary 
starting from an action in the bulk analogous to (\ref{boundary}).

In the case of $AdS_{7} \times S^4$ one finds a
$((n_+,n_-)=(2,0),d=6)$ supermultiplet with a $2$-form $B_{ij}$, 
$(i=0,1,\ldots5)$, whose field strength is self-dual, $8$ spinors 
$\lambda^{A}{}_{+}$, $(A=1,2,3,4)$, obeying 
$\gamma^{7}\lambda^{A}{}_{+}=\lambda^{A}{}_{+}$ and $5$ scalars 
$\phi^{a}$, $(a=1,2,\ldots 5)$. The $OSp(6,2|4)$ tripleton covariant field 
equations on $S^{1} \times S^{5}$ are \cite{NST}:
\[
(\nabla^{i}\nabla_{i}е-4a^{2}е)\phi^{a}е=0
\]
\[
\gamma^{i}еD_{i}е\lambda^{A}{}_{+}=0
\]
\be
H^{ijk}=\frac{1}{3!}\sqrt{-h} \epsilon^{ijklmn}H_{lmn}
\la{tripletonequations}
\ee
where $H_{ijk}=3\partial_{[i}еB_{jk]}е$.  Once again, we know of no 
derivation 
of these tripleton field equations on the boundary starting from 
equations in 
the bulk.

\subsection{The membrane as a singleton: the membrane/supergravity 
bootstrap}
\la{Bootstrap}

Being defined over the boundary of $AdS_{4}е$, the $OSp(4|8)$ singleton 
action (\ref{singletonaction}) is a {\it three dimensional} theory 
with signature $(-,+,+)$ describing $8$ scalars and $8$ spinors.    
With the discovery of the eleven-dimensional supermembrane 
\cite{BST1,BST2}, it was noted that $8$ scalars and $8$ spinors on a 
three-dimensional worldvolume with signature $(-,+,+)$ is just what 
is obtained after gauge-fixing the supermembrane action! Moreover, 
kappa-symmetry of this supermembrane action forces the  
background fields to obey the field equations of $(N=1,D=11)$ 
supergravity. It was therefore 
suggested \cite{Fifteen} that on the $AdS_{4} \times S^{7}е$ 
supergravity background, the supermembrane whose worldvolume occupies 
the $S^{1}е\times S^{2}е$ boundary of the $AdS_{4}е$ could be regarded as the 
singleton of 
$OSp(4|8)$ .  Noting that these singletons also appear in the 
Kaluza-Klein harmonic expansion of this supergravity background, this 
further suggested a form of bootstrap \cite{Fifteen} in which the 
supergravity gives rise to the membrane on the boundary which in turn 
yields 
the 
supergravity in the bulk. This conjecture received further support with 
the subsequent discovery of the ``membrane at the end of the 
universe'' \cite{Fifteen,BDPS,BD,BDPS2,Sutton} to be discussed in 
Section \ref{Universe}, and 
the 
realisation \cite{dust} that the eleven-dimensional supermembrane 
emerges as a solution of the $D=11$ supergravity field equations. 

The possibility of a similar $3$-brane/supergravity bootstrap 
arising for the $SU(2,2|4)$ doubletons on $AdS_{5} \times 
S^{5}$ and a similar $5$-brane/supergravity bootstrap arising for the 
$OSp(6,2|4)$ tripletons on $AdS_{7} \times S^{4}е$ was also 
considered \cite{Fifteen}.
Ironically, however, it was (erroneously as we now know) rejected 
since the only supermembranes that were known at the time \cite{AETW} had 
worldvolume theories described by {\it scalar} supermultiplets, 
whereas the doubletons and tripletons required vector and tensor 
supermultiplets, respectively. See Section \ref{Revisit}. 

Nevertheless, since everything seemed to fit nicely for the $(d=3,D=11)$ 
slot 
on the 
brane-scan of supersymmetric extended objects with worldvolume 
dimension $d$, there
followed a good deal of activity relating other super $p$-branes in 
other dimensions to singletons and superconformal field theories
\cite{BDPS,BDPS2,BD,Classical,BD2,BSTan,BSS,NST,DPS}. In 
particular, it was pointed out \cite{BD,NST,BD2} that there was a 
one-to one-correspondence between the $12$ points on the brane-scan 
as it was then known 
\cite{AETW} and the $12$ superconformal groups in Nahm's 
classification \cite{Nahm} admitting singleton 
representations, as shown in Table \ref{singletons}. 
To understand these $12$ points on the brane-scan, we recall
the matching of physical bose and fermi degrees of freedom on the
worldvolume.  As the $p$-brane
moves through spacetime, its trajectory is described by the
functions $X^M (\xi)$ where $X^M$ are the spacetime coordinates ($M = 0, 1,
\ldots, D - 1$) and $\xi^i$ are the worldvolume coordinates ($i = 0, 1,
\ldots, d - 1$).  It is often convenient to make the so-called
 {\it static gauge choice} by making the $D = d + (D - d)$ split
\begin{equation}
X^M (\xi) = (X^{\mu} (\xi), Y^m (\xi)),
\end{equation}
where $\mu = 0, 1, \ldots, d - 1$~and~$m = d, \ldots, D - 1$,
and then setting
\begin{equation}
X^{\mu} (\xi) = \xi^{\mu}.
\end{equation}
Thus the only physical worldvolume degrees of freedom are given
 by the $(D - d)~Y^m (\xi)$.  So the number of on-shell bosonic degrees of
freedom is
\begin{equation}
{N_B = D - d.}
\end{equation}
To describe the super $p$-brane we augment the $D$ bosonic coordinates $X^M
(\xi)$ with anticommuting fermionic coordinates $\theta^{\alpha} (\xi)$.
Depending on $D$, this spinor could be Dirac, Weyl, Majorana or
Majorana-Weyl. The fermionic kappa symmetry means that half of the spinor
degrees of freedom are redundant and may be eliminated by a physical gauge
choice.  The net result is that the theory exhibits a {\it $d$-dimensional
worldvolume supersymmetry} where the number of fermionic
generators is exactly half of the generators in the original spacetime
supersymmetry.  This partial breaking of supersymmetry is a key idea.  Let
$M$ be the number of real components of the minimal spinor and $N$ the
number of supersymmetries in $D$ spacetime dimensions and let $m$~and~$n$
be the corresponding quantities in $d$ worldvolume dimensions.  Let us
first consider $d > 2$.  Since kappa symmetry always halves the number of
fermionic degrees of freedom and going on-shell halves it again, the
number of on-shell fermionic degrees of freedom is
\begin{equation}
{N_F = {1\over 2}mn = {1\over 4}MN.}
\end{equation}
Worldvolume supersymmetry demands $N_B = N_F$ and hence
\begin{equation}
{D - d = {1\over 2}mn = {1\over 4}MN.}
\la{bosefermi}
\end{equation}
A list of dimensions, number of real dimensions of the minimal spinor and
possible supersymmetries is given in Table \ref{minimal}, from which we
see that there are only $8$ solutions of (\ref{bosefermi}) all with 
$N = 1$, which exactly match the singletons 
shown in Table \ref{singletons}.  We note in particular that $D_{{\rm max}} =
11$ since $M \geq 64$ for $D \geq 12$ and hence (\ref{bosefermi}) cannot be
satisfied.  Similarly
 $d_{{\rm max}} = 6$ since $m \geq 16$ for $d \geq 7$.  The case $d = 2$ is
special because of the ability to treat left and right moving modes
independently.  If we require the sum of both left and right moving bosons
and fermions to be equal, then we again find the condition (\ref{bosefermi}). 
This provides a further $4$ solutions all with $N = 2$, corresponding to
Type $II$ superstrings in $D = 3, 4, 6$~and~$10$ (or 8 solutions
in all if we treat Type $IIA$ and Type $IIB$ separately).  Both the
gauge-fixed Type $IIA$ and Type $IIB$ superstrings will display $(8, 8)$
supersymmetry on the worldsheet. If we require only left (or right) matching,
then  (\ref{bosefermi}) is replaced by
\begin{equation}
{D - 2 = n = {1\over 2}MN,}
\end{equation}
which allows another $4$ solutions in $D = 3, 4, 6$~and~$10$,
all with $N = 1$. The gauge-fixed theory will display $(8,0)$ worldsheet
supersymmetry.  The heterotic string falls into this category.  

The number of dimensions transverse to the brane, $D-d$, equals the number of scalars 
in the singleton supermultiplet. (The two factors appearing in the $d=2$ 
case
is simply a reflection of the ability of strings to have right and left 
movers.
For brevity, we have written the Type $II$ assignments in Table 
\ref{singletons}, but 
more
generally we could have $OSp(2|p) \times OSp(2|q)$ where $p$ and $q$ are 
the
number of left and right supersymmetries \cite{GNST}.) Note that the $d=6$ 
upper limit on the worldvolume dimension is consistent with the 
requirement of renormalizability \cite{Classical}.     
\begin{table}
\begin{center}
\begin{tabular}{cccccccccc}
~&D$\uparrow$&&&&&&\\
~&11&.&~~~~~~~~~~~~~~~~~~&&${\bf OSp(4|8)}$&&&\\
~&10&.&~~~~~~~~&$[OSp(2|8)]^2$&&&&$OSp(6,2|2)$\\
~&~9&.&~~~~~~~~&&&&$F(4)$&\\
~&~8&.&~~~~~~~~&&&$SU(2,2|2)$&&\\
~&~7&.&~~~~~~~~&&$OSp(4|4)$&&&\\
~&~6&.&~~~~~~~~&$[OSp(2|4)]^2$&~&$SU(2,2|1)$&&\\
~&~5&.&~~~~~~~~&&$OSp(4|2)$&&&\\
~&~4&.&~~~~~~~~&$[OSp(2|2)]^2$&$OSp(4|1)$&&&\\
~&~3&.&~~~~~~~~&$[OSp(2|1)]^2$&~&&&\\
~&~2&.&~~~~~~~~&&&&&&\\
~&~1&.&~~~~~~~~&~&~&~&~&\\
~&~0&.&.&.&.&.&.&.\\
~&~~&0&1&2&3&4&5&6&d$\rightarrow$
\end{tabular}
\end{center}
\caption{The brane scan of superconformal groups admitting 
singletons.}
\la{singletons}
\end{table}
Note, however, that the $(d=3,D=11)$, $OSp(4|8)$ slot (written in 
boldface) occupies a privileged position in that the corresponding $D=11$ supergravity  
theory admits the $AdS_{4} \times S^{7}$ solution with $OSp(4|8)$ 
symmetry, whereas the other supergravities do not admit 
solutions with the superconformal group as a symmetry. For example, 
$D=10$ supergravity admits an $AdS_{3} \times S^{7}$ solution 
\cite{DTV,DGT}, but it does not have the full $[OSp(2|8)]^{2}е$ symmetry 
because 
the dilaton is non-trivial and acts as a conformal Killing vector on 
the $AdS_{3}$. This is slightly mysterious, since the bulk theory has 
less symmetry than the boundary theory. We shall return to this in 
Sections \ref{Aristocrat} and \ref{Maldacena}.

\subsection{Doubletons and tripletons revisited}
\la{Revisit}

These early works focussed on scalar supermultiplets because these
were the only $p$-branes known in 1988 \cite{AETW}. However, with
the discovery in 1990 of Type $II$ $p$-brane solitons
\cite{CHS1,CHS2,HS,DLgauge,Luscan}, vector
and tensor multiplets were also seen to play a role. In particular, the
worldvolume fields of the self-dual Type IIB superthreebrane were shown to
be described by an $(n=4,d=4)$ gauge theory \cite{DLgauge}, which on the
boundary of $AdS_5$ is just the doubleton supermultiplet of the
superconformal group $SU(2,2|4)$! Thus one can after all entertain a 
$3$-brane-doubleton-supergravity bootstrap similar to the
membrane-singleton-supergravity bootstrap of Section \ref{Bootstrap}, 
and we may now draw the doubleton
brane scan of Table \ref{doubletons}. Once again, the restriction to $d=4$ is 
consistent with renormalizability.  Note, however, that the $(d=4,D=10)$, 
$SU(2,2|4)$ 
slot
(written in boldface) occupies a privileged position in that the 
corresponding
$D=10$ Type $IIB$ supergravity admits the $AdS_5 \times S^5$ solution with 
$SU(2,2|4)$ symmetry, whereas the other supergravities do not admit 
solutions
with the superconformal group as a symmetry since, as discussed in Section
\ref{Aristocrat}, the dilaton is again non-trivial.

\begin{table}
\begin{center}
\begin{tabular}{cccccccccc}
~&D$\uparrow$&&&&&&\\
~&11&.&~~~~~~&~~~~~~~~~~~~&&~~~~~~~~~~~~&~~~~~~~~~~~~&\\
~&10&.&~~~~~~&&&${\bf SU(2,2|4)}$&~&\\
~&9&.&~~~~~~&&&&&\\
~&8&.&~~~~~~&&&$SU(2,2|2)$&&\\
~&7&.&~~~~~~&&&&&\\
~&6&.&~~~~~~&&&&~&~\\
~&5&.&~~~~~~&&&&&\\
~&4&.&~~~~~~&&&$SU(2,2|1)$~&&\\
~&3&.&~~~~~~&&~&&&\\
~&2&.&~~~~~~&&&&&&\\
~&1&.&~~~~~~&~&~&~&~&\\
~&0&.&~~~~~~.&.&.&.&.&.\\
~&~&0&~~~~~~1&2&3&4&5&6&d$\rightarrow$
\end{tabular}
\end{center}
\bigskip
\caption{The brane scan of superconformal groups 
admitting doubletons}
\la{doubletons}
\end{table}

Similarly, with the discovery of the $M$-theory fivebrane \cite{Gueven}, 
it was
realized \cite{GT} that the zero modes are described by an
$((n_+,n_-)=(2,0),d=6)$ multiplet with a chiral $2$-form, $8$ spinors and 
$5$
scalars, which on the boundary of $AdS_7$ is just the tripleton 
supermultiplet
of the superconformal group $OSp(6,2|4)$! (These zero modes are the 
same as
those of the Type $IIA$ fivebrane, found previously in \cite{CHS1,CHS2}). 
Thus one
can after all also entertain a $5$-brane-tripleton-supergravity bootstrap
similar to the membrane-singleton-supergravity bootstrap of Section
\ref{Bootstrap}. Thus we may now draw the tripleton
brane scan of Table \ref{tripletons}. Note once again, however, that the $(d=6,D=11)$, 
$OSp(6,2|4)$ slot
(written in boldface) occupies a privileged position in that the 
corresponding
$D=11$ supergravity admits the $AdS_7 \times S^4$ solution with 
$OSp(6,2|4)$ symmetry, whereas the other supergravities do not admit 
solutions
with the superconformal group as a symmetry since, as discussed in Section
\ref{Aristocrat}, the dilaton is again non-trivial.
\begin{table}
\begin{center}
\begin{tabular}{cccccccccc}
~&D$\uparrow$&&&&&&\\
~&11&.&~~~~~~&~~~~~~~~~~~~&~~~~~~~~~~~~&~~~~~~~~~~~~&~~~~~~~~~~~~&
${\bf OSp(6,2|4)}$\\
~&10&.&~~~~~~&&&&~&\\
~&9&.&~~~~~~&&&&&\\
~&8&.&~~~~~~&&&&&\\
~&7&.&~~~~~~&&&&&$OSp(6,2|2)$\\
~&6&.&~~~~~~&&&&~&~\\
~&5&.&~~~~~~&&&&&\\
~&4&.&~~~~~~&&&~&&\\
~&3&.&~~~~~~&&~&&&\\
~&2&.&~~~~~~&&&&&&\\
~&1&.&~~~~~~&~&~&~&~&\\
~&0&.&~~~~~~.&.&.&.&.&.\\
~&~&0&~~~~~~1&2&3&4&5&6&d$\rightarrow$
\end{tabular}
\end{center}
\bigskip
\caption{The brane scan of superconformal groups 
admitting tripletons}
\label{tripletons}
\end{table}
 
With the inclusion of branes with vector and tensor supermultiplets on 
their
worldvolume, another curiosity arises. Whereas the singleton brane
scan of Table \ref{singletons} exhausts all the scalar branes and the tripleton brane scan 
of
Table \ref{tripletons} exhausts all the tensor branes, the doubleton brane 
scan of Table \ref{doubletons}
is only a subset of all the vector branes \cite{super}. The Type 
$IIB$ 
$3$-brane is special because gauge theories are conformal only in $d=4$.  

\subsection{The membrane at the end of the universe}
\la{Universe}

As further evidence of the membrane/supergravity bootstrap idea, solutions 
of the $D=11$ supermembrane equations (\ref{membraneboseeq}) and (\ref{embedding}) 
were sought for which the spacetime is $AdS_4 \times M^7$ and for which the 
supermembrane occupies the $S^{1}е \times S^{2}е$ boundary of the 
$AdS_4$. As we shall now recall, the BPS condition was achieved only 
as $r\rightarrow \infty$, hence the name
{\it the Membrane at the End of the Universe} 
\cite{BDPS,BDPS2,CKKTV,Sutton}.

We substitute (\ref{adsmetric}) and (\ref{FRansatz2}) into 
(\ref{membraneboseeq}) and (\ref{embedding}) and look for 
solutions of the form
\be
t=\xi^{0}е,~~~~~~~\theta=\xi^{1}е,~~~~~~~\phi=\xi^{2}е
\la{braneansatz}
\ee
so that the membrane is embedded in the $AdS_{4}$ as
\be
ds_{4}^{2}=-\cosh^{2}е\rho 
(d\xi^{0})е^{2}е+\frac{1}{a^{2}}еd\rho^{2}е+\frac{1}{a^{2}е} 
\sinh^{2}е\rho 
((d\xi^{1}е)^{2}е+\sin^{2}е \xi^{1}е (d\xi^{2})е^{2}е)
\la{adsmetric1}
\ee
In order to show that this configuration does indeed satisfy the 
requirements of unbroken supersymmetry (\ref{branesusy}), we first exhibit a 
spinor $\epsilon$ satisfying ${\tilde D}_{M}е\epsilon=0$ everywhere 
and then show how it satisfies $\Gamma \epsilon=\epsilon$ as $r 
\rightarrow \infty$. It is not difficult to show that
\be
\Gamma=\gamma \otimes I
\la{Gamma}
\ee
where $\gamma \equiv \gamma_{012}$.  First we look for spinors of the form 
(\ref{spinordecomp}), then the general solution for $\epsilon(x)$ is
\be
\epsilon=\sqrt{2}(\sinh\rho/2+\gamma 
\cosh \rho/2)\zeta(t,\theta,\phi)
\la{epsilon}
\ee
where $\zeta$ satisfies
\be
(1+\gamma)(\nabla_{i}е+\frac{1}{2}a\gamma_{i}е\gamma_{3}е)\zeta =0
\la{zeta}
\ee
where $\nabla_{i}е$ is the covariant derivative on the $S^{1}е\times 
S^{2}е$ boundary of $AdS_{4}е$ with metric (\ref{boundarymetric}).
One can show \cite{BDPS} that this equation has four solutions, 
implying the well-known result that $AdS_{4}е$ has four Killing 
spinors.

Since $r=\infty$ corresponds to $\rho= \infty $, it follows from 
(\ref{epsilon}), (\ref{Gamma}) that $\epsilon$ becomes an eigenstate  
of $\gamma$ and hence $\Gamma\epsilon=\epsilon$ ``at the end of the 
universe''. Thus we have shown that the membrane at the end of the 
universe is supersymmetric whenever $M_{7}е$ admits Killing spinors. 
As we have seen in Section \ref{AdS4}, the number $N$ of such spinors depends 
on the Weyl holonomy of $M_{7}е$ and lies between $0$ and $8$. It remains to 
show that the action for such a membrane is indeed the $OSp(4|N)$ singleton 
action (\ref{singletonaction}). To simplify 
matters we shall now demonstrate this for the bosonic radial mode 
$r(\xi)$. Substituting the $AdS_{4} \times M_{7}$ solution and the 
brane ansatz (\ref{braneansatz}) into the membrane action 
(\ref{membranebose}) we find
\be
S=T_{3}е\int_{S^{1}е \times S^{2}е}е d^{3}е\xi 
\left[-\sqrt{-det(g_{ij}е+a^{-2}\partial_{i}е\rho\partial_{j}е\rho)}
+е\sinh^{3}е\rho \right]
\ee
where from (\ref{embedding}) and (\ref{adsmetric})
\be
g_{ij}е=
\left(
\begin{array}{ccc}
-\cosh^{2}е\rho&0&0\\
0&\frac{1}{a^{2}е}\sinh^{2}е\rho&0\\
0&0&\frac{1}{a^{2}е}\sinh^{2}е\rho \sin^{2}е\theta
\end{array}
\right)
\ee
Hence
\be
S=T_{3}е\int _{S^{1} \times S^{2}} d^{3}е\xi \sqrt{-h}\left[-
\frac{1}{2a^{2}е}\cosh \rho 
~h^{ij}е\partial_{i}е \rho \partial_{j}е\rho -\cosh \rho \sinh^{2}е \rho 
+\sinh^{3}е \rho \right]
\ee
Since we are interested in the $r \rightarrow \infty$ limit, we 
consider only large $\rho$, for which
\be
S=\frac{T_{3}е}{8}\int _{S^{1} \times S^{2}} d^{3}е\xi \sqrt{-h}\left[
-\frac{2}{a^{2}е}e^{\rho}е 
h^{ij}е\partial_{i}е \rho \partial_{j}е \rho -2e^{\rho}е \right]
\ee
So, bearing in mind that $T_{3}е \sim a^{3}е$ and making the change of variable
\be
e^{\rho}е \sim \frac{1}{a}\phi^{2}е
\ee
we find
\be
S=\int_{S^{1} \times S^{2}}d^{3}е\xi 
\sqrt{-h}\left[-\frac{1}{2}h^{ij}\nabla_{i}\phi \nabla_{j}\phi 
-\frac{1}{8}a^{2}\phi^{2}\right]
\la{boundary2}
\ee
which is just the singleton action (\ref{boundary}), including the 
scalar mass terms necessary for conformal invariance on $S^{1}е \times 
S^{2}е$. 

Following \cite{Seibergwitten}, this result may be generalized to 
arbitrary $(d-1)$-branes occupying the conformal boundary $M$ of an 
arbitrary Einstein $(d+1)$-dimensional manifold $W$.  We shall adapt the 
notation and the 
signature used in \cite{Seibergwitten} for our convenience.  The 
boundary $M$ 
has a natural conformal structure but not
a natural metric.  Let $h_{ij}е$ be an arbitrary metric on
the boundary in its conformal class.  Here the $\xi^i$, $i=0,\dots,d-1$
are an arbitrary set of local coordinates on the boundary.  
There is then a unique way \cite{graham} to extend the $\xi^i$
to coordinates on $W$ near the boundary, adding an
additional coordinate $\rho$ that tends to infinity on the boundary,  
such that the metric in a neighborhood of the boundary is
\be
ds^{2}е=\frac{1}{a^{2}е}\left(d\rho^{2}е+
\frac{1}{4}e^{2\rho}еh_{ij}еd\xi^{i}еd\xi^{j}е-
P_{ij}е d\xi^{i}еd\xi^{j}+O(e^{-2\rho}е)\right)
\ee
where
\be
{P_{ij}={2(D-1)R_{ij}-g_{ij}R\over 2(D-1)(D-2)},}
\ee
and $R_{ij}е$ is the Ricci tensor of $M$,
which implies
\be
{g^{ij}P_{ij}= {R\over 2(d-1)}.}
\ee
We find
\be
S=\frac{T_{d}е}{2^{d}}\int_{M}d^{d}е\xi \sqrt{-h}\left[
-\frac{2}{a^{2}е}e^{\rho}е 
h^{ij}е\partial_{i}е \rho \partial_{j}е \rho -
\frac{2}{(d-1)(d-2)}e^{\rho}еR \right]
\ee
we recognise the curvature term as that required for Weyl invariance 
of the action \cite{BD}.

\subsection{Near horizon geometry and p-brane aristocracy}
\la{Aristocrat}

More recently, $AdS$ has emerged as the near-horizon geometry of black
$p$-brane solutions \cite{GT,DGT,GHT,String} in $D$ dimensions. The dual
brane, with worldvolume dimension ${\tilde d}=D -d-2$, interpolates
between $D$-dimensional Minkowski space $M_{D}е$ and $AdS_{{\tilde d}+1}
\times
S^{d+1}$ (or $M_{{\tilde d}+1}\times S^{3}$ if $d=2$).  To see this, 
we recall that such branes arise generically as  
solitons of the following action \cite{lublack}:
\be
I=\frac{1}{2\kappa_{D}е^{2}}е \int d^{D}еx \sqrt{-g}\left[R-\frac{1}{2} 
(\partial 
\phi)^{2}е-\frac{1}{2(d+1)!}e^{-\alpha \phi} F_{d+1}{}^{2}\right]
\ee
where $F_{d+1}е$ is the field strength of a $d$-form potential 
$A_{d}е $ and $\alpha$ is the constant
\be
\alpha^{2}е=4-\frac{2d \tilde d}{d+\tilde d}
\la{alpha}
\ee
Written in terms of the 
$(d-1)$-brane sigma-model metric $e^{-{\alpha/d} \phi}еg_{MN}е$, the 
solutions are \cite{lublack,String}
\[
ds^{2} =H^{\frac{2-d}{d}}еdx^{\mu}еdx_{\mu}е+
H^{2/d}е(dy^{2}е+y^{2}еd\Omega_{d+1}е{}^{2})е
\]
\[
e^{2\phi }е=H^{\alpha}е
\]
\be
F_{d+1}е=dL^{d}е\epsilon_{d+1}е
\la{branesolution}
\ee
where 
\be
H=1+\frac{L^{d}}{y^{d}}
\ee
For a stack of $N$ singly charged branes $L^{d}е=Nb^{d}е$ and the near 
horizon, or large $N$, geometry corresponds to 
\be
ds^{2}е\sim{\frac{y}{L}}^{2-d}еdx^{\mu}еdx_{\mu}+\frac{L^{2}е}{y^{2}}еdy^{2}е
+L^{2}еd\Omega_{d+1}^{2}ее
\ee
Or, defining the new coordinate
\be
y=Le^{\zeta/L}
\ee
we get
\[
ds^{2}\sim e^{\frac{2-d}{L}\zeta}dx^{\mu}еdx_{\mu} +d\zeta^{2}е
+L^{2}еd\Omega_{d+1}{}^{2}
\]
\[
\phi \sim   \frac{d\alpha}{2L}\zeta
\]
\be
F_{d+1}е \sim dL^{d}е\epsilon_{d+1}е
\la{nearhorizon}
\ee
Thus for $d\neq 2$ the near-horizon geometry is $AdS_{\tilde d +1}е \times 
S^{d+1}е$. Note, however, that the gradient of the dilaton is 
generically non-zero and plays the role of a conformal Killing vector on 
$AdS_{\tilde d +1}$. Consequently, there is no enhancement of 
symmetry in the near-horizon limit. The unbroken supersymmetry remains 
one-half and the bosonic symmetry remains $P_{\tilde d}е\times 
SO(d+2)$.
(If $d=2$, then (\ref{nearhorizon}) reduces to
\[
ds^{2} \sim dx^{\mu}еdx_{\mu} + d\zeta^{2}+L^{2}еd\Omega_{3}{}^{2}  
\]
\[
\phi \sim \frac{\alpha}{L}\zeta
\]
\be
F_{3}е \sim 2L^{2}е\epsilon_{3}е
\ee
which is $M_{\tilde d+1}е\times S^{3}е$, with a linear dilaton 
vacuum. The bosonic symmetry remains $P_{\tilde d}е\times 
SO(4)$.)

Of particular interest are the ($\alpha=0$) subset of solitons for which 
the 
dilaton is zero or constant: the {\it non-dilatonic $p$-branes}. From 
(\ref{alpha}) we see that for branes with one kind of charge there are only 3 
cases:
\[
D=11: d=6, \tilde d=3
\]
\[
D=10: d=4, \tilde d=4
\] 
\[
D=11: d=3, \tilde d=6
\]
which are precisely the three cases that occupied privileged positions on 
the
singleton, doubleton and tripleton brane-scans of Tables 
\ref{singletons}, \ref{doubletons} and \ref{tripletons}. 
Then the near-horizon geometry coincides with the $AdS_{\tilde d +1}е 
\times 
S^{d+1}е$ non-dilatonic maximally symmetric compactifications of the 
corresponding 
supergravities. The supersymmetry doubles and the bosonic symmetry is 
also enhanced to $SO(\tilde d,2) \times SO(d+2)$. Thus the total 
symmetry is given by the conformal supergroups $OSp(4|8)$, 
$SU(2,2|4)$ and $OSp(6,2|4)$, respectively. 

For bound states of branes with $M$ kinds of charge, the constant $\alpha$ gets 
replaced by \cite{DR2,DR,KKLP}
\be
\alpha^{2}е=\frac{4}{M}-\frac{2d \tilde d}{d+\tilde d}
\la{alpha2}
\ee
A non-dilatonic solution ($\alpha$=0) occurs for $M=2$: 
\[
D=6: d=2, \tilde d=2 
\]
which is just the dyonic string \cite{Rahmfeld}, of which the self-dual 
string 
\cite{lublack} is a special case, whose near-horizon geometry is 
$AdS_{3}е\times S^{3}$. For $M=3$ we have
\[
D=5: d=2, \tilde d=1
\]
which is the 3-charge black hole \cite{T}, whose near-horizon geometry is 
$AdS_{2} \times S^{3}$, and
\[
D=5: d=1, \tilde d=2
\]
which is the 3-charge string \cite{T} whose near-horizon geometry is 
$AdS_{3} \times S^{2}$. For $M=4$ we have
\[
D=4: d=1, \tilde d=1
\]
which is the 4-charge black hole \cite{CT1,CT2}, of which the 
Reissner-Nordstr\"{o}m  
solution is a special case \cite{DR2}, and whose near-horizon geometry 
is $AdS_{2} \times S^{2}е$ \cite{FG}.

Thus we see that not all branes are created equal. A {\it $p$-brane 
aristocracy} obtains whose members are those branes whose near-horizon
geometries have as their symmetry the conformal supergroups. As an example
of a plebian brane we can consider the ten-dimensional  superstring:
\[
D=10: d=6, \tilde d=2
\]
whose near-horizon geometry is $AdS_{3} \times S^{7}$ but with a 
non-trivial dilaton of Section \ref{Bootstrap} which does not have 
the conformal group $[OSp(2|8)]^{2}$ as its symmetry, even though 
this group appears in the $(D=10, \tilde d=2)$ slot on the singleton 
brane-scan of Table \ref{singletons}. In which case, of course, one may ask what role do 
these singletons play. We shall return to this in Section 
\ref{Maldacena}.  

The original membrane at the end of the universe was a spherical 
brane embedded in the 
$AdS_{4}$ geometry as given in \ref{Universe}.  Alternatively, one could take as 
the membrane at the end of 
the
universe to be the flat near-horizon membrane, which is embedded as
\be
ds^{2}=e^{4\zeta/L} (-(d\xi^{0})е^{2}+(d\xi^{1})е^{2}+(d\xi^{2}е)^{2}) 
+ d\zeta^{2}
\la{near}
\ee
and has $M_{3}$ topology. It is still possible to associate an 
$OSp(4|8)$ action but this time it is defined over $M_{3}$ and has no 
scalar mass terms \cite{CKKTV,DFFFTT}. One can continue to call these 
fields 
``singletons'', of course, if by singleton one simply means anything 
transforming according to the $D(1/2,0)$ and $D(1,1/2)$ representations 
of
$SO(3,2)$.  A comparison of these two approaches is discussed in some 
detail in 
\cite{CKKTV}. 

\subsection{Supermembranes with fewer supersymmetries. Skew-whiffing.}
\la{fewer}

So far we have focussed attention on compactifications to $AdS_{\tilde 
d+1}е$ on 
round spheres $S^{d+1}$ which have maximal supersymmetry, but the 
supergravity 
equations admit infinitely many other compactifications on Einstein 
spaces $X^{d+1}$ which have fewer supersymmetries \cite{DNP}. Indeed 
generic 
$X^{d+1}$ have no supersymmetries at all\footnote{Thus in the early 
eighties, 
the most 
highly prized solutions were those with many supersymmetries. Nowadays, 
bragging rights seem to go those which have none!}. We note in this 
connection the {\it skew-whiffing theorem} \cite{DNP}, which states that 
for 
every $AdS_{\tilde 
d+1}е$ compactification preserving supersymmetry, there exists one 
with no supersymmetry simply obtained by reversing the orientation of 
$X^{d+1}$ (or, equivalently, reversing the sign of $F_{d+1}$). The 
only exceptions are when $X^{d+1}$ are round spheres which preserve 
the maximum supersymmetry for either orientation. A 
corollary is that other {\it symmetric spaces}, which necessarily admit an 
orientation-reversing isometry, can have no supersymmeties. Examples 
are provided by products of round spheres.
   
The question naturally arises as to whether these compactifications 
with fewer supersymmetries also arise as near-horizon geometries of 
$p$-brane solitons.  The answer is yes and the soliton 
solutions are easy to construct \cite{DLPS,ccdfft}. One simply makes 
the replacement
\be
d\Omega_{d+1}{}^{2}е \rightarrow d{\hat \Omega}_{d+1}{}^{2}е
\ee
in (\ref{branesolution}), where $d{\hat \Omega}_{d+1}{}^{2}е$ is the 
metric 
on an arbitrary Einstein space $X^{d+1}$ with the same scalar curvature 
as the round $S^{d+1}$. The space need only be Einstein; it need not 
be homogeneous \cite{DLPS}. (There also exist brane solutions on Ricci 
flat 
$X^{d+1}$ \cite{DLPS} but we shall not discuss them here).  Note, 
however, that these non-round-spherical solutions do not tend to 
$(D-d)$-dimensional Minkowski space as $r\rightarrow \infty$. Instead 
the metric on the $(D-\tilde d)$-dimensional space transverse to the 
brane is asymptotic to a generalized cone
\be
ds_{D-\tilde d}е{}^{2}е=dr^{2}е+r^{2}еd{\hat \Omega}_{d+1}{}^{2}е
\ee
and $(D-d)$-dimensional translational invariance is absent except when 
$X^{d+1}$ is a round sphere. The number of supersymmetries preserved 
by these $p$-branes is determined by the number of Killing spinors 
on $X^{d+1}$.

To illustrate these ideas let us focus on the eleven-dimensional 
supermembrane. The usual supermembrane interpolates between $M_{11}$   
and $AdS_{4}е \times$ round $S^{7}е$, has symmetry $P_{3}е \times SO(8)$ 
and 
preserves $1/2$ of the spacetime supersymmetries for either orientation 
of the round $S^{7}е$. Replacing the round $S^{7}е$ by  generic 
Einstein spaces $X^{7}$ leads to membranes with symmetry $P_{3} 
\times G$, where G is the isometry group of $X^{7}$. For example 
$G=SO(5) \times SO(3)$ for the squashed $S^{7}$ \cite{ADP,DNP2}. For 
one orientation of $X^{7}$, they preserve $N/16$ spacetime 
supersymmetries where $1 \leq N \leq 8$ is the number of Killing 
spinors on $X^{7}$; for the opposite orientation they preserve no 
supersymmetries since then $X^{7}$ has no Killing spinors. For 
example, $N=1$ for the left-squashed $S^7$ owing to its $G_{2}е$ 
holonomy \cite{ADP,DNP,DNP2}, whereas $N=0$ for the right-squashed 
$S^{7}$. However, all these solutions satisfy the same Bogomol'nyi
bound between the mass and charge as the usual supermembrane 
\cite{DLPS}.  Of course, skew-whiffing is not the only way to obtain vacua 
with less than maximal supersymmetry. A summary of known $X^{7}$, 
their supersymmetries and stability properties is given in \cite{DNP}. 
Note, however, that skew-whiffed vacua are automatically stable at the 
classical level since skew-whiffing affects only the spin $3/2$, $1/2$ 
and $0^{-}е$ towers in the Kaluza-Klein spectrum, whereas the 
criterion for classical stability involves only the $0^{+}е$ tower 
\cite{DNP}.

\subsection{The Maldacena conjecture}
\la{Maldacena}

The year 1998 marks a revolution in anti-de Sitter space 
brought about by Maldacena's conjectured duality between physics in the 
bulk of $AdS$ and a conformal field theory on the boundary 
\cite{Maldacena1}.
In particular, $M$-theory on $AdS_{4}\times S^{7}$ is dual to a 
non-abelian 
$(n=8,d=3)$ superconformal theory, Type $IIB$ string theory on 
$AdS_{5}\times S^{5}$ is dual to a $d=4$ $SU(N)$ super Yang-Mills theory 
theory and 
$M$-theory on $AdS_{7}е\times S^{4}$ is dual to a non-abelian 
$((n_+,n_-)=(2,0),d=6)$ 
conformal theory. In particular, as has been spelled out most clearly in 
the $d=4$ 
$SU(N)$ Yang-Mills case, there is seen to be a correspondence between the
Kaluza-Klein mass spectrum in the bulk and the conformal dimension of
operators on the boundary \cite{Gubserklebanovpolyakov,Wittenads}. 

One immediately recognises that the dimensions and supersymmetries of 
these 
three conformal theories are exactly the 
same as the singleton, doubleton and tripleton supermultiplets of 
Section \ref{singletons}.  Moreover,  both the old and new $AdS/CFT$ 
correspondences are {\it holographic} in the sense of 
\cite{thooft,Susskind1}. 
Following Maldacena's conjecture \cite{Maldacena1}, therefore, 
a number of papers appeared reviving the old singleton-$AdS$-membrane-
superconformal field theory connections
\cite{Ferrara2,KKR,Boonstra,CKKTV,IMSY,Gunaydin,Gubserklebanovpolyakov,%
Horowitz,Wittenads,Wittenads2,Kachru,Berkooz,Ferrara,Rey,Maldacena2,lnv} and applying
them to this new duality context. What are the differences? 

The first difference is that the branes were spherical as opposed to 
flat, since the ``boundary'' in question for the 
singletons and $p$-branes at the end of the universe was the $S^{1}е 
\times S^{p}е$ with finite radius $1/a$ for the $S^{p}е$. As such, 
superconformal invariance requires mass terms for the scalars, as discussed in 
Section \ref{Boundary}. The branes in the Maldacena conjecture, on the other hand, are 
the flat Minkowski space branes embedded in AdS in the way dictated by the 
near-horizon geometry.
  
Another curious difference is that, with the exception of the three 
aristocratic branes, all the slots on the three brane-scans of superconformal 
field theories corresponded to bulk supergravities whose brane solutions  
are dilatonic, and hence have a symmetry smaller than the boundary 
theory. It seems that the branes at the end of the universe do not care
about the dilaton because the 
$\rho=$constant surfaces in (\ref{adsmetric1}) (or the $\zeta=$ constant surfaces in 
(\ref {near})) posess the full superconformal symmetry even 
though the bulk $AdS$ solution does not. In other words, 
they admit the maximal set of conformal Killing vectors even though   
the bulk admits less than the maximal set of Killing vectors. 
This contrasts 
with the new $AdS/CFT$ conjecture where a non-conformal supergravity 
solution 
in the bulk \cite{DGT} is deemed to be dual to non-conformal field theory 
on 
the boundary \cite{IMSY}. It is not obvious at the moment whether 
this difference is real or apparent and it would be interesting to 
pursue the matter further. 

Thirdly, although the {\it membrane/supergravity bootstrap} idea, that physics 
in the bulk of $AdS$ should be dictated by the membrane on the boundary and 
vice-versa, might nowadays be called ``holographic'', it was not 
motivated by considerations of entropy. Rather, in analogy with string 
theory where there are self-consistency requirements between the 
spacetime and worldsheet dynamics, it was expected that a similar 
phenomenon should hold for membranes and the membrane on the boundary 
of $AdS$ was simply a special case.  

Fourthly, attention was focussed on {\it free} superconformal theories on 
the boundary as opposed to the {\it interacting} theories currently 
under consideration.  For example, although the worldvolume 
fields of the Type $IIB$ $3$-brane were known to be described by an 
$(n=4,d=4)$ gauge theory \cite{DLgauge}, we now know that this brane 
admits the interpretation of a Dirichlet brane \cite{Polchinski} and 
that the superposition of $N$ such branes yields a {\it non-abelian} 
$SU(N)$ gauge theory
\cite{Wittenbound}. These observations are crucial to the new duality conjecture 
\cite{Maldacena1}. For earlier related work on coincident threebranes 
and $n=4$ super Yang Mills, see \cite{Gubser1,Klebanov2,Gubser2,Gubser3}.

Let us consider the solution for $N$ coincident $3$-branes 
corresponding to $N$ units of $5$-form flux \cite{HS,DLgauge}. As we 
have already seen in section \ref{AdS5}, $L=N^{1/4}еb$ and the near
horizon limit 
may equally well be regarded as the large $N$ limit.  We find $AdS_{5}е \times 
S^{5}е$, but with an $AdS$ radius proportional to $N^{1/4}$.
The philosophy is that Type $IIB$ supergravity is a good approximation for 
large $N$ and that Type $IIB$ stringy 
excitations correspond to operators whose dimensions diverge for 
$N \rightarrow \infty $. This makes contact with the whole industry of 
large $N$ QCD. These large $N$, non-abelian features were absent in the 
considerations of a $3$-brane/supergravity bootstrap discussed 
in Section \ref{Revisit}, as was the precise correspondence between the
Kaluza-Klein mass spectrum in the bulk and the conformal dimension of
operators on the boundary \cite{Gubserklebanovpolyakov,Wittenads}. Nevertheless, 
as we hope these 
lectures show, there are sufficently many similarities 
between the current bulk/boundary duality and the old Membrane at 
the End of the Universe idea to merit further comparisons. 

\subsection{Problems 5}
\la{Problems5}
\begin{enumerate}
\item

Repeat the analysis of Section \ref{Aristocrat} for {\it rotating} 
branes. 

\end{enumerate}

\section{\bf ANTI-DE SITTER BLACK HOLES}
\la{Blackhole}

\subsection{Introduction}
\la{Introduction}

Anti-de Sitter black hole solutions of gauged extended supergravities 
\cite{romans} are currently attracting a good deal of attention
\cite{Behrndt1,birm,Caldarelliklemm1,Klemm,Behrndt2,%
Duffliu,Chamblin,Cveticgubser1,%
Cveticgubser2,Sabra,Caldarelliklemm2} due, in large
part, to the $AdS/CFT$ correspondence.  As we have seen, these gauged
extended supergravities arise as the massless modes of various
Kaluza-Klein compactifications of both $D=11$ and $D=10$
supergravities. The three examples studied in these lectures are gauged
$D=4$, $N=8$ $SO(8)$ supergravity \cite{deWit1,deWit2} arising from
$D=11$ supergravity on $S^{7}$ \cite{duffpope3,DNP} whose black hole
solutions are discussed in \cite{Duffliu}; gauged $D=5$, $N=8$
$SO(6)$ supergravity \cite{PPV,GRW} arising from Type IIB supergravity
on $S^{5}$ \cite{Schwarzcompact,GM,KRV} whose black hole solutions are
discussed in \cite{Behrndt1,Behrndt2}; and gauged $D=7$, $N=4$ $SO(5)$
supergravity \cite{PPV,TV} arising from $D=11$ supergravity on $S^{4}$
\cite{PTV} whose black hole solutions are given in Section \ref{S4}
and in \cite{Cveticgubser1,lm}.\footnote{BPS black holes arising in
the $SU(2)\times SU(2)$ version of gauged $N=4$ supergravity in $D=4$,
which is the massless sector of the $S^3\times S^3$ compactification
of $N=1$ supergravity in $D=10$, were discussed in
\cite{Klemm}.  These solutions are not asymptotically $AdS$.}
In the absence of the black holes,  these three $AdS$
compactifications are singled out as arising from the near-horizon
geometry of the extremal non-rotating $M2$, $D3$ and $M5$ branes
\cite{GT,DGT,GHT,Duffads}. One of our goals will be to embed these
known lower-dimensional black hole solutions into ten or eleven
dimensions, thus allowing a higher dimensional interpretation in terms
of {\it rotating} $M2$, $D3$ and $M5$-branes.  

Since these gauged supergravity theories may be obtained by
consistently truncating the massive modes of the full Kaluza-Klein
theories, it follows that all solutions of the lower-dimensional
theories will also be solutions of the higher-dimensional ones
\cite{Duffpope,Pope}. In principle, therefore, once we know the
Kaluza-Klein ansatz for the massless sector, it ought to be
straightforward to read off the higher dimensional solutions.  It
practice, however, this is a formidable task. The correct massless
ansatz for the $S^{7}$ compactification took many years to finalize
\cite{deWitnicolaiwarner,deWitnicolai}, and is still highly implicit,
while for the $S^{5}$ compactifications the complete
massless ans\"atze are still unknown. The $S^{4}$ case has recently 
been given in its entirety in \cite{Nastase1,Nastase2}.  For our present purposes, it
suffices to consider truncations of the gauged supergravities to
include only gauge fields in the Cartan subalgebras of the full gauge
groups, namely $U(1)^{4}$, $U(1)^{3}$ and $U(1)^{2}$ for the $S^{7}$,
$S^{5}$ and $S^{4}$ compactifications, respectively. These truncated
theories will admit respectively the 4-charge $AdS_{4}$, 3-charge
$AdS_{5}$ and 2-charge $AdS_{7}$ black hole solutions.

The simplest of the three is perhaps the $D=5$, $N=8$ maximal gauged
supergravity, for which there is a consistent $N=2$ (\ie minimal)
truncation to supergravity coupled to two abelian vector multiplets.
This has the bosonic field content of a graviton, three $U(1)$ gauge
fields and two scalars.  In these lectures we obtain the complete
non-linear Kaluza-Klein ansatz for the compactification of $D=10$
Type IIB supergravity on $S^5$, truncated to the $U(1)^3$ Cartan
subgroup of $SO(6)$.

In four dimensions there is a consistent truncation of gauged $N=8$
maximal supergravity to gauged $N=2$ supergravity coupled to three
vector multiplets. The bosonic sector consists of a graviton, four
vectors and three complex scalars, whose real and imaginary parts
correspond to three ``axions'' and three ``dilatons.''
\footnote{Interestingly enough, the ungauged version of this theory,
obtained by switching off the gauge coupling and performing some
dualisations, appears in the $T^{2}$ compactification of $D=6$, $N=1$
string theory. The four vectors are the two Kaluza-Klein and two
winding gauge fields, while the three complex scalars $S$, $T$ and $U$
correspond to the axion-dilaton, the K\"{a}hler form and complex structure
of the torus. This $STU$ system plays a crucial role in
four-dimensional string/string/string triality
\cite{Duffliurahmfeld1}.  The black hole solutions of this theory
\cite{cvyod,Duffliurahmfeld1}, and
their embedding in ungauged $N=8$ supergravity \cite{lpsol,khor} 
arising from the
$T^{7}$ compactification of $M$-theory as intersections
\cite{cveticts,tsey} are also well known.}.
The inclusion of the axions is necessary for providing a consistent
truncation; the full bosonic Lagrangian in this case is obtained in
Appendix \ref{D=4}.  This truncation corresponds to the $U(1)^4$ Cartan
subgroup of the non-abelian $SO(8)$, for which there exist $AdS$ black
hole solutions with four electric charges \cite{Duffliu}.  While one
would ideally wish to obtain a complete Kaluza-Klein ansatz for the
$N=2$ truncation, in practice the complexity arising from the
inclusion of the axions is considerable.  Thus in these lectures we
omit the axions in the Kaluza-Klein reduction.  This is of course
sufficient for the embedding of the electric black hole solutions in
$D=11$ as they do not involve the axions.

Finally, in seven dimensions, maximal $N=4$ gauged $SO(5)$
supergravity admits a consistent truncation to $N=2$ supergravity,
comprising the metric, a 2-form potential, three vectors and a
dilaton, coupled to a vector multiplet comprising a vector and three
scalars.  We obtain the Kaluza-Klein ansatz for an $S^4$ reduction of
$D=11$ supergravity, including two $U(1)^2$ gauge fields and two
dilatonic scalars.  This is sufficient for the consideration of the
embedding of the $D=7$ black holes in $D=11$.

  Having obtained the explicit Kaluza-Klein reduction ans\"atze, this
allows an investigation of the embedding of the various $AdS$ black
holes of $D=4$, $D=5$ and $D=7$ in the respective higher-dimensional
supergravities.  An important point here is that one must know the
exact Kaluza-Klein reduction ansatz for the reduction of the
supergravity theory itself, and not just for a specific solution, in
order to show that the metric, gauge fields and scalar fields of the
lower-dimensional solution are indeed precisely embeddable in the
higher-dimensional theory.  It is worth remarking, in this regard,
that it is the scalar fields that present most of the subtleties and
complexities in the implementation of the reduction procedure.

Having embedded these black holes in ten or eleven dimensions, an
interesting question then arises as to their higher-dimensional
interpretation.  It was noted some time ago \cite{BDPS2}, in the
context of a ``test'' membrane moving in a fixed $AdS_{4} \times
S^{7}$, that a 4-dimensional BPS state (whose $AdS$ energy is equal to
its electric charge) admits the eleven-dimensional interpretation of
an $M2$-brane \cite{BST1,BST2,dust} that is rotating in the extra
dimensions. Moreover, the electric charge is equal to the spin.

Recently there has been an upsurge of activities on the study of
rotating $p$-branes
\cite{Cveticyoum3,Cveticyoum2,Gubserthermo,Csaki,Kraus,Cai,%
Chamblin,Cveticgubser1,Cveticgubser2,klt,rosf,Csaki2,sfetsos}.  In
particular, in \cite{Chamblin} $AdS$ Reissner-Nordstr\"{o}m black holes
({\it i.e.}~the charged black holes without scalars) of $AdS$
supergavity in $D=4$ and $D=5$ were studied, and shown to be related
to the rotating solutions of M-/string theory.  In
\cite{Cveticgubser1} the near-extreme spinning $D3$-brane with one
angular momentum was shown to reproduce the metric and the gauge
fields of the $k=0^+$ limit of $D=5$ gauged supergravity black
holes \cite{Behrndt2}, with the anticipation that the result would
generalize to multiple angular momenta.  However, the identification
of the scalar fields was not given. In addition, in
\cite{Cveticgubser1,Cveticgubser2}, the equivalence of the
thermodynamics of the near-extreme spinning branes and the
corresponding large black holes of $D=4,\ 5,\ 7$ gauged supergravity
was given.  While incomplete, these works provided some initial
stages in the investigation of the sphere compactifications of
M-/string theory.

Unlike black holes that are asymptotically Minkowskian, for which the
horizons are always spherical, it is known that $AdS$ black holes can
also admit horizons of more general topology.  Following the embedding
procedure described above, we demonstrate that the $AdS_{4}$ black 
holes
with toroidal horizon can indeed be interpreted as the near-horizon
structures of an $M2$ brane rotating in the extra dimensions. The four
charges corresponding to the $U(1)^{4}$ Cartan subgroup are just the
four angular momenta. Similarly, the 5-dimensional charged black hole
with toroidal horizon corresponds to a rotating $D3$-brane and the
7-dimensional charged black hole with toroidal horizon to a rotating
$M5$-brane.  In each case, the event horizon coincides with the
worldvolume of the brane.\footnote{This is a concrete realisation of
the ``Membrane Paradigm'' \cite{thorne}.}
Additionally, one may use the Kaluza-Klein 
ansatz to obtain the higher-dimensional interpretation of $AdS$ black holes
with horizons of other topologies.  We conjecture that these
correspond to the near-horizon limits of rotating $p$-branes whose
world-volumes have these topologies.  (In fact the rotating
``test'' membrane in \cite{BDPS2} had $S^2$ topology.)

In these lectures we also obtain the general rotating
$p$-brane solutions in arbitrary dimensions, supported by a single
$(p+2)$-form charge, and discuss their sphere
reductions.  These rotating $p$-branes are easily constructed, merely
by performing standard diagonal dimensional oxidations of the general
rotating black holes that were constructed in \cite{Cveticyoum3}.

\subsection{S$^5$ reduction of Type IIB supergravity}
\la{S5}
        The $S^5$ reduction \cite{Schwarzcompact,GM,KRV} of Type $IIB$
supergravity gives rise to $N=8$, $D=5$ gauged supergravity, with
$SO(6)$ Yang-Mills gauge group \cite{PPV,GRW}.  The complete
details of this reduction, as with any sphere reduction, would be of
great complexity, and in fact no example has ever been fully worked
out.  For our present purposes, however, it suffices to consider the
truncation of the five-dimensional theory to $N=2$ supersymmetry.  In
this truncation, which is of course a consistent one, the gauge group
is reduced down to the $U(1)\times U(1)\times U(1)$ Cartan subgroup of
$SO(6)$.  The bosonic sector of the theory comprises these three gauge
bosons, the metric, and two scalar fields.  (The consistency of the
truncation to this field content can be seen by considering the $S^1$
reduction of ungauged minimal non-chiral supergravity in $D=6$, whose
bosonic fields $(g_{\mu\nu}, A_\2,\phi)$ reduce to give precisely the
field content we are considering here in $D=5$.  After gauging, one
would obtain the $U(1)^3$ gauged theory.)

Even to construct the $S^5$ reduction ansatz for this truncated
$N=2$ theory is somewhat non-trivial, owing to the presence of the
scalar fields.  It is most conveniently expressed in terms of the
parameterisation of sphere metrics given in \cite{mype}.

We find that the ansatz for the reduction of the
ten-dimensional metric is
\be
ds_{10}^2 = \sqrt{\wtd\Delta}\, ds_5^2 + \fft1{g^2\,
\sqrt{\wtd\Delta}}\,  \sum_{i=1}^3 X_i^{-1}\,
\Big(d\mu_i^2 + \mu_i^2\, (d\phi_i +g\, A^i)^2\Big)\ ,\label{2bs5met}
\ee
where the two scalars are parameterised in terms of the three
quantities $X_i$, which are subject to the constraint $X_1\, X_2\,
X_3=1$.  They can be parameterised in terms of two dilatons $\varphi_1$
and $\varphi_2$ as
\be
X_i=e^{-\ft12 \vec a_i\cdot \vec \varphi}\ ,
\ee
where $\vec a_i$ satisfy the dot products
\be
M_{ij}\equiv\vec a_i\cdot \vec a_j=4\delta_{ij} -\ft43\ .
\ee
A convenient choice is
\be
\vec a_1 = (\ft{2}{\sqrt6}, \sqrt2)\ , \qquad
\vec a_2 = (\ft{2}{\sqrt6}, -\sqrt2)\ ,\qquad
\vec a_3 = (-\ft{4}{\sqrt6}, 0)\ .
\ee
The three quantities $\mu_i$ are subject to the constraint $\sum_i
\mu_i^2=1$, and the metric on the unit round 5-sphere can be written
in terms of these as
\be
d\Omega_5^2 = \sum_i (d\mu_i^2 + \mu_i^2\, d\phi_i^2)\ .
\ee
The $\mu_i$ can be parameterised in terms of angles on a 2-sphere, for
example as
\be
\mu_1 = \sin\theta\ ,\qquad \mu_2 = \cos\theta\, \sin\psi\ ,\qquad
\mu_3 = \cos\theta\, \cos\psi\ .
\ee
Note that $\wtd\Delta$ is given by
\be
\wtd \Delta = \sum_{i=1}^3 X_i\, \mu_i^2\ ,
\ee
and is therefore expressed purely in terms of the scalar fields, and the
coordinates on the compactifying 5-sphere.  The constant $g$ in
(\ref{2bs5met}) is the inverse of the radius of the compactifying
5-sphere, and is equal to the gauge coupling constant.  We find that
the ansatz for
the reduction of the 5-form field strength is
$F_\5=G_\5+ {*G_\5}$, where
\bea
G_\5 &=& 2g\, \sum_i\Big(X_i^2\, \mu_i^2 -\wtd\Delta\, X_i\Big)\,
\ep_\5 - \fft1{2g}\, \sum_i X_i^{-1}\, {{\bar *} dX_i}\wedge
d(\mu_i^2) \nn\\
&&+ \fft1{2 g^2} \,
\sum_i X_i^{-2}\, d(\mu_i^2)\wedge (d\phi_i +g\, A_\1^i)
\wedge {{\bar *} F_\2^i}\ .\label{5fs5red}
\eea
Here, $F_\2^i=dA_\1^i$, $\ep_\5$ is the volume form of the 5-dimensional
metric $ds_5^2$, and ${\bar *}$ denotes the Hodge dual with respect to
the five-dimensional metric $ds_5^2$.

   Substituting these ans\"atze into the equations for motion for the
Type $IIB$ theory, we obtain five-dimensional
equations of motion that can be derived from the
Lagrangian\footnote{We shall make some more detailed comments on
certain general features of these spherical Kaluza-Klein reductions in
Section \ref{S7}, where we consider the $S^7$ reduction of $D=11$ supergravity.}
\be
e^{-1}\, {\cal L}_5 = R - \ft12(\del\varphi_1)^2 -\ft12(\del\varphi_2)^2
+ 4g^2\, \sum_i X_i^{-1}- \ft14 \sum_i X_i^{-2}\, (F_\2^i)^2  +\ft14
\ep^{\mu\nu\rho\sigma\lambda}\, F^1_{\mu\nu}\, F^2_{\rho\sigma}\,
A^3_\lambda\ .
\label{d5gauged}
\ee
(The other bosonic fields of the type IIB theory are set to zero in
this $U(1)^3$ truncated reduction.)  Note that the ten-dimensional
Bianchi identity $dF_\5=0$ gives rise to the equations of motion for
the scalars and gauge fields in five dimensions.

   Thus we have established that the reduction ans\"atze
(\ref{2bs5met}) and (\ref{5fs5red}) describe the exact embedding of
the five-dimensional $N=2$ gauged $U(1)^3$ supergravity into Type 
$IIB$
supergravity.

        The bosonic Lagrangian (\ref{d5gauged}) can be further
truncated down to smaller sectors.  For example, we can consistently
set $\varphi_2=0$, implying that $X_1=X_2=X_3^{-1/2}$, provided that
$F_\2^1=F_\2^2=F_\2/\sqrt2$.  The Lagrangian then becomes
\bea
e^{-1} {\cal L}_5 &=& R-\ft12 (\del\varphi_1)^2 +
4g^2\, (2e^{\ft1{\sqrt6}\varphi_1} + e^{-\ft2{\sqrt6}\varphi_1}) -
\ft14e^{\ft2{\sqrt6}\varphi_1}\, (F_\2)^2-
\ft14e^{-\ft4{\sqrt6}\varphi_1}\, (F_\2^3)^2\nn\\
&& + \ft18 \ep^{\mu\nu\rho\sigma\lambda}\, F_{\mu\nu}\, F_{\rho\sigma}\,
A^3_\lambda\ .
\eea
It is also possible to set both scalars to zero, implying that
$X_i=1$, provided that $F_\2^i
=F_\2/\sqrt3$.  The Lagrangian is then given by
\be
e^{-1}{\cal L}_5 = R + 12 g^2 - \ft14 F_\2^2 +
\ft1{12\sqrt3} \ep^{\mu\nu\rho\sigma\lambda}\,
F_{\mu\nu}\, F_{\rho\sigma}\, A_\lambda\ .\label{d5trunc}
\ee
The embedding of the truncated Lagrangian (\ref{d5trunc}) in $D=10$
dimensions was discussed in \cite{Chamblin}.

\subsection{D=5 AdS black holes}
\la{AdS5black}

       The Lagrangian (\ref{d5gauged}) admits a three-charge $AdS$ black
hole solution, given by \cite{Behrndt2}
\bea
ds_5^2 &=& -(H_1H_2H_3)^{-2/3}\, f\, dt^2 +
(H_1H_2H_3)^{1/3}\, (f^{-1}\, dr^2 + r^2 d\Omega_{3,k}^2)\ ,\nn\\
X_i&=& H_i^{-1}\, (H_1H_2H_3)^{1/3}\ ,\qquad
A^i_\1 = \sqrt{k}\, (1-H_i^{-1})\, \coth\beta_i\, dt\ ,
\label{d5adsbh}
\eea
and
\be
f=k-\fft{\mu}{r^2} + g^2\, r^2\, (H_1H_2H_3)\ ,\qquad
H_i = 1 + \fft{\mu\, \sinh^2\beta_i}{k\, r^2}\ .
\ee
Here $k$ can be 1, 0 or $-1$, corresponding to the foliating surfaces
of the transverse space being $S^3$, $T^3$ or $H^3$, with unit metric
$d\Omega_{3,k}^2$, where $H^3$ denotes the hyperbolic 3-space of
constant negative curvature.  In the case of $k=0$, one first needs to
make the rescaling \cite{Cveticgubser1} $\sinh^2\beta_i
\longrightarrow k\, \sinh^2\beta_i$, followed by sending $k$ to zero.
The gauge potential for the $k=0$ case is then given by
\be
A^i_\1 =\fft{1-H_i^{-1}}{\sinh\beta_i}\, dt\ .
\ee

\subsection{Rotating D3-brane}
\la{RotatingD3}

         In this section, we show that the $k=0$ three-charge $AdS$
black hole of the $N=2$ gauged supergravity in $D=5$ given in
(\ref{d5adsbh}) can be embedded in $D=10$ as a solution that is
precisely the decoupling limit of the rotating $D3$-brane.
The higher-dimensional solutions corresponding to five-dimensional 
$AdS$
black holes with $k=1$ and $k=-1$ can also be easily obtained, by
substituting the five-dimensional solutions into the $S^5$ reduction
ans\"atze.

          There can be three angular momenta, $\ell_i$, $i=1,2,3$, in
the rotating $D3$-brane.  The generic single-charge rotating $p$-branes
can be obtained by dimensional oxidation of the generic
single-charge rotating black holes constructed in \cite{Cveticyoum3}. 
See problems of Section \ref{Problems6}.  We find that the metric of the rotating
$D3$-brane is given by%
\footnote{This metric agrees with previously obtained results
\cite{klt,rosf}, after correcting some typographical errors.}
%
\bea
ds_{10}^2 &=& H^{-\ft12}\Big(-(1 -\fft{2m}{r^4\Delta})
\, dt^2 + dx_1^2 + dx_2^2 + dx_3^2 \Big)
+ H^{\ft12}\Big[\fft{\Delta\, dr^2}{H_1H_2H_3 -2m\, r^{-4}}\nn\\
&&+ r^2\, \sum_{i=1}^3 H_i\,(d\mu_i^2 + \mu_i^2\, d\phi_i^2)
-\fft{4m\, \cosh\a}{r^4\, H\, \Delta}\, dt\, (\sum_{i=1}^3
\ell_i\, \mu_i^2\, d\phi_i) \nn\\
&& + \fft{2m}{r^4\, H\, \Delta}\, (\sum_{i=1}^3
\ell_i\, \mu_i^2\, d\phi_i)^2\Big]\ ,
\label{d3rotate}
\eea
where the functions $\Delta$, $H$, and $ H_i$ are given by
\bea
\Delta &=& H_1H_2H_3\, \sum_{i=1}^3 \fft{\mu_i^2}{H_i}\ , \qquad
H = 1 + \fft{2m\, \sinh^2\a}{r^4 \Delta}\ ,\nn\\
H_i &=& 1 + \fft{\ell_i^2}{r^2}\ ,\qquad i=1,2,3\ .
\label{d3rotatefun}
\eea
The rotating $D3$-brane is supported by the self-dual 5-form field
strength $F_\5$ of the type IIB theory.  It is given by $F_\5 = G_\5 +
*G_\5$, where $G_\5=dB_\4$ and
\be
B_\4 = \fft{1-H^{-1}}{\sinh\a}\Big(-\cosh\a\, dt + \sum_{i=1}^3 \ell_i\,
\mu_i^2\, d\phi_i\Big)\wedge d^3x\ .
\ee

       As is well known, the non-rotating $D3$-brane has a ``decoupling
limit'' where the spacetime of the $D3$-brane becomes a product space
$M_5\times S^5$.  If the $D3$-brane is extremal, $M_5$ is a
five-dimensional anti-de Sitter spacetime.  More generally, when the
$D3$-brane is non-extremal, $M_5$ is the Carter-Novotny-Horsky metric
\cite{clp1}, which can thus be viewed as a ``non-extremal''
generalisation of $AdS_5$.  A similar limit also exists for the
rotating $D3$-brane, and can be achieved by making the rescalings
\bea
&&m\longrightarrow \ep^4 \, m\ ,\qquad
\sinh\a \longrightarrow \ep^{-2}\, \sinh\a\ ,\nn\\
&&r\longrightarrow \ep\, r\ ,
\qquad x^\mu\longrightarrow \ep^{-1}\, x^\mu\ ,
\qquad \ell_i \rightarrow \ep\, \ell_i\ ,
\eea
and then sending $\ep\longrightarrow 0$.  (Note that when this limit
is taken, we also have $\cosh\a\longrightarrow \ep^{-2}\, \sinh\a$.)
This has the effect that the last term in (\ref{d3rotate}) is set to
zero and that
\be
H = 1 + \fft{2m\, \sinh^2\a}{r^4 \Delta} \longrightarrow
\fft{2m\, \sinh^2\a}{r^4 \Delta}\ .
\ee
In this limit, the metric (\ref{d3rotate}) becomes
\bea
ds_{10}^2 &=& \sqrt{\wtd\Delta}\, \Big[-(H_1 H_2 H_3)^{-2/3}\,
f\, dt^2 + (H_1 H_2 H_3)^{1/3}(f^{-1}\, dr^2 + r^2\,
d\vec y\cdot d\vec y)\Big]\nn\\
&&+\fft1{g^2\,\sqrt{\wtd \Delta}}\, \sum_{i=1}^3 X_i^{-1}\,
\Big(d\mu_i^2 + \mu_i^2\, (d\phi_i +g\, A^i)^2\Big)
\ ,\label{d3rotatehor}
\eea
where
\be
\vec y = g\, \vec x\ ,\qquad
g^2 =\fft{1}{\sqrt{2m}\,\sinh\a}\ ,\qquad
\mu = 2m\,g^2\ .
\ee
The metric (\ref{d3rotatehor}) precisely matches the dimensional
reduction ansatz (\ref{2bs5met}), with the lower dimensional fields
given by
\bea
ds_5^2 &=& -(H_1 H_2 H_3)^{-2/3}\,
f\, dt^2 + (H_1\, H_2\, H_3)^{1/3}(f^{-1}\, dr^2 + r^2\,
d\vec y\cdot d\vec y)\ ,\nn\\
X_i &=& H_i^{-1}\, (H_1 H_2 H_3)^{1/3}\ , \qquad
A^i_\1 = \fft{1 - H_i^{-1}}{g\,\ell_i\,\sinh\a}\, dt\ ,
\label{d5adsbhk0}
\eea
where
\be
f=-\fft{\mu}{r^2} + g^2\, r^2\, H_1 H_2 H_3\ ,\qquad
g^2=\fft{1}{\sqrt{2m}\sinh\a}\ ,\qquad \mu = 2m\, g^2\ .
\ee
         To complete the story, we note that the 5-form field strength
in the decoupling limit is given by $F_\5=G_\5+{*G_\5}$, where
$G_\5=dB_\4$ and
\be
B_\4 = -g^4\, r^4\, \Delta\, dt\wedge d^3x + \fft{1}{\sinh\a}\,
(\sum_{i=1}^3 \ell_i\, \mu_i^2\, d\phi_i)\wedge d^3x\ .
\ee
This gives precisely the field strength in the
dimensional reduction ansatz (\ref{5fs5red}).

   Thus we see that the solution (\ref{d5adsbhk0}) is precisely the
$k=0$ three-charge $AdS$ black hole given in the previous subsection,
after reparameterising the angular momenta $\ell_i^2 = \mu
\sinh^2\beta_i$.  This shows that the embedding of the three-charge
$AdS$ $k=0$ black hole in gauged $N=2$ supergravity in five dimensions
gives a ten-dimensional solution that is precisely the decoupling
limit of the rotating $D3$-brane.  Single-charge $AdS$ black holes coming
from the reduction of the metric of a rotating $D3$-brane with one
angular momentum were obtained in \cite{Cveticgubser1}, however
without the explicit embedding of the scalar fields.  The connection
between the thermodynamics of $AdS$ black holes and rotating $p$-branes
was discussed in \cite{Cveticgubser1,Cveticgubser2}.

         It is also straightforward to oxidise the $k=1$ and $k=-1$
$AdS$ black holes back to $D=10$ type IIB.  The metric is the same form
as (\ref{d3rotatehor}) with $d\vec{y}\cdot d\vec y$ replaced by the
unit metric for $S^3$ or $H^3$ respectively.  The 5-form field
strength follows by substituting the five-dimensional fields into
(\ref{5fs5red}).

\subsection{S$^7$ reduction of D=11 supergravity}
\la{S7}

The $S^7$ reduction of eleven-dimensional supergravity gives rise
to $SO(8)$ gauged $N=8$ supergravity in four dimensions.  One may
again consider a consistent truncation to $N=2$, for which the bosonic
sector comprises the metric, four commuting $U(1)$ gauge potentials,
three dilatons and three axions.  (That this is a consistent
truncation can be seen by reducing minimal non-chiral six-dimensional
supergravity on $T^2$, for which the reduction of $(g_{\mu\nu}, A_\2,
\phi)$ will give precisely the field content we are considering.
After gauging, this would give the $U(1)^4$ gauged theory.  See
Appendix \ref{D=4} for an extended discussion of this.) We have not yet
determined the complete reduction ansatz for the entire truncated
theory where the axions are included, but we can give the exact ansatz
in the case where one sets the axions to zero.  This will not, of
course, be a consistent truncation, since the $U(1)$ gauge fields will
provide source terms of the form $\ep^{\mu\nu\rho\sigma}\,
F_{\mu\nu}\, F_{\rho\sigma}$ for the axions.  Nevertheless, one can
use the axion-free ansatz for discussing the exact embedding of
four-dimensional solutions for which the axions are zero.  The full
$N=2$ four-dimensional theory, including the axions, is obtained in
Appendix \ref{D=4}.

   The reduction ansatz for the eleven-dimensional metric is
\be
ds_{11}^2 = \wtd\Delta^{2/3}\, ds_4^2 +g^{-2}\,
\wtd\Delta^{-1/3}\, \sum_i X_i^{-1}\, \Big( d\mu_i^2 + \mu_i^2\,
 (d\phi_i + g\,
A^i_\1)^2 \Big)\ .\label{s7metred}
\ee
where $\wtd \Delta = \sum_{i=1}^4 X_i\, \mu_i^2$.  The four quantities
$\mu_i$ satisfy $\sum_i \mu_i^2 =1$.  They can be parameterised in
terms of angles on the 3-sphere as
\be
\mu_1 =\sin\theta\ ,\quad \mu_2=\cos\theta\, \sin\varphi\ ,\quad
\mu_3=\cos\theta\, \cos\varphi\, \sin\psi\ ,\quad
\mu_4=\cos\theta\, \cos\varphi\, \cos\psi\ .
\ee
The four $X_i$, which satisfy $X_1X_2X_3X_4=1$, can be parameterised
in terms of three dilatonic scalars $\vec\varphi =(\varphi_1,
\varphi_2, \varphi_3)$:
\be
X_i=e^{-\ft12 \vec a_i \cdot \vec \varphi}\ ,
\ee
where the $\vec a_i$ satisfy the dot products
\be
M_{ij} \equiv \vec a_i\cdot \vec a_j = 4\delta_{ij} -1\ .
\ee
A convenient choice, corresponding to the combinations of
(\ref{eq:lambdas}), is
\be
\vec a_1 =(1,1,1)\ ,\quad \vec a_2=(1,-1,-1)\ ,\quad
\vec a_3=(-1,1,-1)\ ,\quad \vec a_4=(-1,-1,1)\ .
\ee

   The reduction ansatz for the 4-form field strength is
\bea
F_\4 &=& 2g\,\sum_i \Big(X_i^2\, \mu_i^2 - \wtd\Delta\, X_i \Big)\,
\ep_\4 +\fft1{2g}\, \sum_i X_i^{-1}\, {{\bar *}dX_i}\wedge d(\mu_i^2)
\nn\\
&&-\fft1{2g^2}\, \sum_i X_i^{-2}\, d(\mu_i^2)\wedge (d\phi_i + g\,
A^i_\1) \wedge
{{\bar *} F_\2^i}\ .\label{s7f4red}
\eea
Here, ${\bar *}$ denotes the Hodge dual with respect to the
four-dimensional metric $ds_4^2$, and $\ep_\4$ denotes its
 volume form.\footnote{If the ans\"atze (\ref{s7metred}) and
(\ref{s7f4red}) are linearised around an $AdS_4\times S^7$ background,
they can be seen to be in agreement with previous results that were
derived at the linear level \cite{duffpope3}.  The full non-linear
metric ansatz (\ref{s7metred}) should be in agreement with the
appropriate specialisation of the ansatz
given in \cite{nilsson}.}

         It is of interest to note that the eleven-dimensional Bianchi
identity $dF_\4=0$ already gives rise to the four-dimensional
equations of motion for the scalars and gauge potentials, namely
\bea
d{{\bar *}d\log(X_i)} &=& \ft14 \sum_{j} M_{ij}\, X_j^{-2}\,
{{\bar *}F_\2^j}\wedge F_\2^j +g^2\sum_{j,k} M_{ij}\, X_j\, X_k
-g^2\sum_j M_{ij} X_j^2\ ,\nn\\
d(X_i^{-2}\, {{\bar *}F_\2^i}) &=&0\ .
\eea
It is straightforward to see that these equations of motion can be
obtained from the four-dimensional Lagrangian
\be
e^{-1}{\cal L}_4 = R -
\ft12 (\del\vec\varphi)^2 +
8g^2 (\cosh\varphi_1 +\cosh\varphi_2+\cosh\varphi_3)
-\ft14 \sum_{i=1}^4 e^{\vec a_i\cdot\vec \varphi}\,
(F_\2^i)^2\ .
\label{d4lagxx}
\ee

    One might think that it would be possible to obtain the
four-dimensional Lagrangian by substituting the ans\"atze
(\ref{s7metred}) and (\ref{s7f4red}) into the eleven-dimensional
Lagrangian.  In fact this is not the case, and one must work at the
level of the eleven-dimensional equations of motion.  One way of
understanding this is from the fact that the ansatz for $F_\4$ does
not {\sl identically} satisfy the Bianchi identity.  Rather, as we
have seen, it satisfies it modulo the use of the four-dimensional
equations of motion for the scalars and gauge fields.  In other words,
the ansatz is made on the eleven-dimensional 4-form $F_\4$ rather than
on the fundamental potential $A_\3$ itself.  Consequently, it would
not be correct to insert the ansatz for $F_\4$ into the Lagrangian.

     We may further illustrate this point by showing, as an example,
how the scalar potential arises in the four-dimensional Einstein
equation.  This comes from considering the eleven-dimensional Einstein
equation,
\be
\hat R_{AB} - \ft12 \hat R\, \eta_{AB} = \ft1{12} \Big( F^2_{AB}
- \ft18 F^2\, \eta_{AB}\Big)\ .\label{d11einst}
\ee
with vielbein indices $A,B$ ranging just over the four-dimensional spacetime
directions $\a,\b$.  From the ansatz (\ref{s7metred}), the relevant terms in
the eleven-dimensional Ricci tensor and Ricci scalar are given by
\bea
\hat R_{\a\b}&=&
\fft{4g^2}{3\wtd\Delta^{8/3}}\, \Big[ -\Big( \sum_i X_i^2\,
\mu_i^2\Big)^2 +\wtd \Delta\, \sum_i X_i^2\, \mu_i^2\, \sum_j X_j +
\wtd\Delta\, \sum_i X_i^3\, \mu_i^2 \nn\\
&&\qquad  - \wtd\Delta^2\, \sum_i
X_i^2 \Big]\, \eta_{\a\b} + \wtd \Delta^{-2/3}\, \, R_{\a\b} +
\cdots\ ,\label{ricci2}\\
\hat R &=& \fft{2g^2}{3\wtd\Delta^{8/3}}\, \Big[ -\Big( \sum_i X_i^2\,
\mu_i^2\Big)^2 - 2\wtd \Delta\, \sum_i X_i^2\, \mu_i^2\, \sum_j X_j +
4 \wtd\Delta\, \sum_i X_i^3\, \mu_i^2 \nn\\
&&\qquad + 6\wtd\Delta^2\, \Big(\sum_i X_i\Big)^2 - 7\wtd\Delta^2\, \sum_i
X_i^2 \Big] + \wtd\Delta^{-2/3}\, R +\cdots\ ,\label{ricci1}
\eea
where $R_{\a\b}$ and $R$ are the four-dimensional Ricci tensor and
scalar, and the ellipsis indicate that terms not involving purely the
undifferentiated scalars have been omitted for the purposes of the
present illustrative discussion.  From the ansatz
(\ref{s7f4red}) for the 4-form, the eleven-dimensional energy-momentum
tensor vielbein components in the four-dimensional spacetime
directions are given by
\be
\ft1{12}(F^2_{\a\b} - \ft18 F^2\, \eta_{a\b}) =
-g^2\, \wtd\Delta^{-8/3}\, \Big(\sum_i(X_i^2\, \mu_i^2 -\wtd\Delta\,
X_i)\Big)^2 \, \eta_{\a\b}\ .\label{enmom}
\ee

   Substituting (\ref{ricci2}), (\ref{ricci1}) and (\ref{enmom}) into
(\ref{d11einst}), we find that all the angular dependence coming from
the $\mu_i$ variables cancels, and that the scalar
potential terms in the four-dimensional Einstein equation are given by
\be
R_{\a\b} -\ft12 R\, \eta_{\a\b} = -\ft12 g^2\, V\, \eta_{\a\b}\ ,
\label{einpot}
\ee
with $V$ given by
\be
V= - 4 \sum_{i<j} X_i\, X_j = -8 (\cosh\varphi_1 + \cosh\varphi_2 +
\cosh\varphi_3)\ .
\ee
Since (\ref{einpot}) derives from the Lagrangian $R -g^2\, V$, we see
that we have precisely produced the hoped-for potential terms of the
gauged supergravity Lagrangian (\ref{d4lagxx}).  This sample
calculation also serves to illustrate that the angular dependence
coming from the $\mu_i$ variables would not have cancelled if we had
merely substituted the ans\"atze (\ref{s7metred}) and (\ref{s7f4red})
into the eleven-dimensional Lagrangian.  It also shows that the
cancellation of the $\mu_i$ dependence in the higher-dimensional
equations of motion depends crucially on ``conspiracies'' between the
contributions from the metric and the 4-form field strength.  This is
quite different from the situation in toroidal reductions, where each
term in the higher-dimensional theory reduces consistently by itself,
without the need for any such conspiracies.  Note, furthermore, that
the required conspiracies needed for the success of the spherical
reduction depend on the 4-form field strength occurring with precisely
the correct coefficient relative to the Einstein-Hilbert term.
This normalisation is not a free parameter, but is governed by the
strength of the $FFA$ term in the eleven-dimensional theory.  Thus
ultimately the consistency of the spherical reduction ansatz can be
traced back to the supersymmetry of the eleven-dimensional theory.

    Note that the Lagrangian (\ref{d4lagxx}) can be further truncated,
to pure Einstein-Maxwell with a cosmological constant, by setting all
the field strengths equal, $F_\2^i = \ft12 F_\2$, and setting all the
scalars to zero:
\be
e^{-1}{\cal L}_{4} = R -\ft14 (F_\2)^2 + 24 g^2\ .
\ee
The embedding of this theory into $D=11$ supergravity was obtained in
\cite{Pope}. The ansatz for the metric and field strength for the
embedding in \cite{Pope} was given in terms of a decomposition of the
7-sphere as a $U(1)$ bundle over $CP^3$.  This is identical, after a
transformation of coordinates, to the Einstein-Maxwell embedding given
in \cite{Chamblin}.  In the same spirit, the $S^5$ reduction to
five-dimensional Einstein-Maxwell can be described using the method
presented in \cite{Pope}, with $S^5$ viewed as a $U(1)$ bundle over
$CP^2$.  (An analogous consistent embedding of four-dimensional
Einstein-Yang-Mills with an $SU(2)$ gauge group, and a cosmological
constant, in $D=11$ supergravity was obtained in \cite{Pope2}. This
involves a decomposition of $S^7$ as an $SU(2)$ bundle over $S^4$.)

\subsection{D=4 AdS black holes}
\la{AdS4black}

    The $D=4$, $N=2$ gauged supergravity coupled to three vector
multiplets  admits 4-charge $AdS$ black hole solutions, given by
\cite{Duffliu,Sabra}
\bea
ds_4^2 &=& -(H_1H_2H_3H_4)^{-1/2}\, f\, dt^2 +
(H_1H_2H_3H_4)^{1/2}\, (f^{-1}\, dr^2 + r^2 d\Omega_{2,k}^2)\ ,\nn\\
X_i&=& H_i^{-1}\, (H_1H_2H_3H_4)^{1/4}\ ,\qquad
A^i_\1 = \sqrt{k}\, (1-H_i^{-1})\, \coth\beta_i\, dt\ ,
\label{d4adsbh}
\eea
and
\be
f=k-\fft{\mu}{r} + 4g^2\, r^2\, (H_1H_2H_3H_4)\ ,\qquad
H_i = 1 + \fft{\mu\, \sinh^2\beta_i}{k\, r}\ .\label{d4adsbh2}
\ee
Here, $k$ can be 1, 0 or $-1$, corresponding to the cases where the
foliations in the transverse space have the metric $d\Omega_{2,k}^2$
on the unit $S^2$, $T^2$ or $H^2$, where $H^2$ is the unit hyperbolic
2-space of constant negative curvature.
In the case of $k=0$, one must first make the
rescaling $\sinh^2\beta_i \longrightarrow k\, \sinh^2\beta_i$
before sending $k$ to zero.  The gauge potential for the $k=0$ case is
then given by
\be
A^i_\1 =\fft{1-H_i^{-1}}{\sinh\beta_i}\, dt\ .
\ee

\subsection{Rotating M2-brane}
\la{RotatingM2}

      There are four angular momenta, $\ell_i$, $i=1,2,3,4$, in the
rotating $M2$-brane.  The solution can be obtained by oxidising the
$D=9$ rotating black hole \cite{Cveticyoum3}.  After the oxidation, we
find that the metric of the rotating $M2$-brane is given by
\bea
ds_{11}^2&=& H^{-\ft23}\, \Big(-(1 -\fft{2m}{r^6\, \Delta})\, dt^2 +
dx_1^2 + dx_2^2 \Big) + H^{\ft13}\Big[
\fft{\Delta\, dr^2}{H_1\, H_2\, H_3\, H_4 -\fft{2m}{r^6}}\nn\\
&&+ r^2 \sum_{i=1}^4 H_i\, (d\mu_i^2 + \mu_i^2\, d\phi_i^2) -
\fft{4m\, \cosh\a}{r^6\, H\, \Delta}\, dt
(\sum_{i=1}^4 \ell_i\, \mu_i^2\, d\phi_i)\nn\\
&&+\fft{2m}{r^4\, H\, \Delta}\, (\sum_{i=1}^4 \ell_i\,
\mu_i^2\, d\phi_i)^2 \Big]\ ,\label{m2rotate}
\eea
where the functions $\Delta$, $H$ and $H_i$ are given by
\bea
\Delta&=& H_1 H_2 H_3 H_4\, \sum_{i=1}^4 \fft{\mu_i^2}{H_i}
\ ,\qquad H=1 +\fft{2m\, \sinh^2\a}{r^6\, \Delta}\ ,\nn\\
H_i&=& 1 + \fft{\ell_i^2}{r^2}\ ,\qquad i=1,2,3,4\ .\label{funct2}
\eea
The 3-form gauge potential is given by
\be
A_\3 = \fft{1-H^{-1}}{\sinh\a} (-\cosh\a\, dt +
\ell_i\, \mu_i^2\, d\phi_i)\wedge d^2x\ .\label{a3form}
\ee

        Following the previous $D3$-brane example, we consider the
decoupling limit, which is obtained by making the rescaling
\bea
&&m\longrightarrow \ep^6 m\ ,\qquad
\sinh\a \longrightarrow \ep^{-3}\, \sinh\a\ ,\nn\\
&&r\longrightarrow \ep\, r\ ,
\qquad x^\mu\longrightarrow \ep^{-2}\, x^\mu\ ,
\qquad \ell_i \rightarrow \ep\, \ell_i
\eea
and then sending $\ep\longrightarrow 0$.  This has the effect that the
last term in (\ref{m2rotate}) is set to zero and that the 1 in
function $H$ (\ref{funct2}) is removed.   In this limit,
the rotating $M2$-brane (\ref{m2rotate}) becomes
\bea
ds_{11}^2 &=& \wtd\Delta^{2/3}\, \Big[ -(H_1 H_2 H_3 H_4)^{-1/2}\, f\,
dt^2 + (H_1 H_2H_3H_4)^{1/2}\, (f^{-1}\, d\rho^2 + \rho^2 d\vec y
\cdot d\vec y)\Big] \nn\\
&&+ g^{-2}\, \wtd\Delta^{-1/3}\, \sum_{i=1}^4 X_i^{-1}\Big(d\mu_i^2 +
\mu_i^2\, (d\phi_i + g\, A^i)^2\Big)\ ,\label{m2limmet}
\eea
where
\bea
&&\rho= \ft12 g\, r^2\ , \qquad \vec y= 2g\, \vec x\ ,\qquad
f=-\fft{\mu}{\rho} + 4g^2\, \rho^2\, H_1 H_2 H_3 H_4\ ,\nn\\
&&g^2=(2m\,\sinh^2\a)^{-1/3}\ ,\qquad \mu = m\, g^5\ ,\qquad
\wtd\Delta = \sum_i\, X_i\, \mu_i^2 \ .
\eea
This is precisely of the form of the metric ansatz in the dimensional
reduction given by (\ref{s7metred}).  The lower dimensional fields are
given by
\bea
&&ds^2_4 = -(H_1 H_2 H_3 H_4)^{-1/2}\, f\,
dt^2 + (H_1 H_2H_3H_4)^{1/2}\, (f^{-1}\, d\rho^2 + \rho^2 d\vec {\td
x}\cdot d\vec {\td x})\nn\\
&&X_i = H_i^{-1}\, (H_1 H_2 H_3 H_4)^{1/4}\ , \qquad
A^i = \fft{1 - H_i^{-1}}{g\,\ell_i\,\sinh\a}\, dt\ .
\eea

   In the decoupling limit, the gauge potential $A_\3$ given in
(\ref{a3form}) for the rotating $M2$-brane becomes, after a gauge
transformation,
\be
A_\3 = -g^6\, r^6\, \Delta\, dt\wedge d^2x + \fft1{\sinh\a}\,
\sum_i \ell_i\, \mu_i^2\, d\phi_i\wedge d^2x\ .
\ee
We find that its field strength $F_\4=dA_\3$ is also of the form given
in (\ref{s7f4red}) for the dimensional reduction ansatz.  Thus we have
established an exact embedding of four-dimensional non-extremal
4-charge $AdS$ black holes into eleven-dimensional supergravity, and
furthermore, that they become precisely the decoupling limit of the
rotating $M2$-branes.  It should, of course, be emphasised that the
four-dimensional $AdS$ black holes that we are considering at this point
have $T^2$ rather than $S^2$ horizons, corresponding to $k=0$ in
(\ref{d4adsbh}) and (\ref{d4adsbh2}).

        It is also straightforward to oxidise the $k=1$ and $k=-1$ $AdS$
black hole solutions back to $D=11$, by substituting the
four-dimensional fields into the ans\"atze (\ref{s7metred}) and
(\ref{s7f4red}).

\subsection{S$^4$ reduction of D=11 supergravity}
\la{S4}

The Kaluza-Klein reduction of eleven-dimensional supergravity on
$S^4$ gives rise to $N=4$ gauged $SO(5)$ supergravity in seven
dimensions.  In a similar manner to the $S^5$ and $S^7$ reductions
that we discussed previously, we may consider an $N=2$ truncation of
this seven-dimensional theory.  As described in the introduction, the
truncated theory comprises $N=2$ supergravity coupled to a vector
multiplet, comprising the metric, 2-form potential, four vector
potentials and four scalars in total.  For our present purposes, we 
shall focus on a further truncation where only the metric, two gauge
potentials (which are associated with the $U(1)\times U(1)$ Cartan
subgroup of $SO(5)$) and two scalars are retained.  This is not in general a
consistent truncation, but, as in the case of the neglect of the
axions in the $S^7$ reduction, it is consistent for a subset of
solutions where the truncated fields are not excited by the ones that
are retained.   In particular, solutions of the $N=2$ theory for which
$F_\2^1\wedge F_\2^2=0$, such as the $AdS$ black holes, will also be
solutions of this truncated theory.

    We find that we can obtain this truncated theory by making the
following Kaluza-Klein $S^4$-reduction ansatz:
\bea
ds_{11}^2 &=& \wtd\Delta^{1/3}\, ds_7^2 + g^{-2}\, \wtd\Delta^{-2/3}\,
\Big(X_0^{-1}\, d\mu_0^2 + \sum_{i=1}^2 X_i^{-1}\, (d\mu_i^2 + \mu_i^2\,
(d\phi_i + g\, A_\1^i)^2) \Big)\ ,\label{s4metred}\\
{*F_\4} &=&
2g\,\sum_{\a=0}^2 \Big(X_\a^2\, \mu_\a^2 - \wtd\Delta\, X_\a \Big)\,
\ep_\7 + g\, \wtd\Delta\, X_0\, \ep_\7
+\fft1{2g}\, \sum_{\a=0}^2 X_\a^{-1}\, {{\bar *}dX_\a}
\wedge d(\mu_\a^2) \nn\\
&&+\fft1{2g^2}\, \sum_{i=1}^2 X_i^{-2}\, d(\mu_i^2)\wedge
(d\phi_i + g\, A^i_\1) \wedge
{{\bar *} F_\2^i}\ ,\label{s4f4red}
\eea
where we have defined the auxiliary variable $X_0\equiv (X_1 X_2)^{-2}$.
Here, ${\bar *}$ denotes the Hodge dual
with respect to the seven-dimensional metric $ds_7^2$, $\ep_\7$
denotes its volume form, and $*$ denotes the Hodge dualisation in the
eleven-dimensional metric.  The quantity $\wtd\Delta$ is given by
\be
\wtd\Delta = \sum_{\a=0}^2 X_\a\, \mu_\a^2\ ,
\ee
where $\mu_0$, $\mu_1$ and $\mu_2$ satisfy
$\mu_0^2+\mu_1^2+\mu_2^2=1$.   The two scalar fields $X_i$ can be
parameterised in terms of two canonically-normalised dilatons
$\vec\varphi=(\varphi_1,\varphi_2)$ by writing
\be
X_i = e^{-\fft12\vec a_i\cdot\vec\varphi}\ ,
\ee
where the dilaton vectors satisfy the relations $\vec a_i\cdot\vec a_j
= 4\delta_{ij} -\ft85$.  A convenient parameterisation is given by
\be
\vec a_1 = (\sqrt2, \sqrt{\ft25})\ ,\qquad \vec a_2 = (-\sqrt2,
\sqrt{\ft25}) \ .\label{d7adef}
\ee
Note that the two $X_i$ are independent here, unlike in the cases of
the three $X_i$ in $D=5$ or the four $X_i$ in $D=4$, which satisfied
$\prod_i X_i =1$.  The auxiliary variable $X_0$ that we have
introduced in order to make the expressions more symmetrical can be
written as $X_0=e^{-\fft12\vec a_0\cdot\vec\varphi}$, where $\vec a_0=
-2(\vec a_1+\vec a_2)= (0,-4\sqrt{2/5})$.

    After substituting into the eleven-dimensional equations of
motion, one obtains seven-dimensional equations that can be derived
from the Lagrangian
\be
e^{-1}{\cal L}_7 = R -\ft12 (\del\vec\varphi)^2 - g^2\, V -
\ft14 \sum_{i=1}^2 e^{\vec a_i\cdot\vec\varphi}\, (F_\2^i)^2\ ,
\label{d7lag}
\ee
where the potential $V$ is given by
\be
V= -4 X_1 X_2 - 2X_1^{-1}\, X_2^{-2} - 2 X_2^{-1}\, X_1^{-2} +\ft12
(X_1 X_2)^{-4}\ .
\ee
This potential has a more complicated structure than those in the
$D=5$ and $D=4$ gauged theories, and in particular it has not only a
maximum at $X_1=X_2=1$, but also a saddle point at $X_1=X_2=
2^{-1/5}$ \cite{ppv}.  Note that by making use of the auxiliary variable
$X_0=(X_1 X_2)^{-2}$, the potential can be re-expressed as
\be
V = -4 X_1 X_2 -2 X_0 X_1 -2 X_0 X_2 + \ft12 X_0^2\ .
\ee

   It is interesting to note that the Lagrangian (\ref{d7lag}) can be
further consistently truncated, by setting $X_1=X_2=X$, and $F_\2^1 =
F_\2^2=F_\2/\sqrt2$.  This implies that the dilatonic scalar
$\varphi_1$ is set to
zero, in terms of the parameterisation defined by (\ref{d7adef}).
This gives
\be
e^{-1}{\cal L}_7 = R - \ft12(\del\varphi_2)^2 + g^2\, ( 4X^2 +
4X^{-3} -\ft12 X^{-8}) -\ft14 X^{-2}\, (F_\2)^2\ ,
\ee
where
\be
X = e^{-\fft1{\sqrt{10}}\varphi_2}\ .
\ee
This scalar potential was used in \cite{lpss} to construct
supersymmetric domain
wall solutions.

\subsection{D=7 AdS black holes}
\la{AdS7black}

    This Lagrangian (\ref{d7lag}) admits 2-charge $AdS$ black-hole
solutions, given by
\bea
ds_7^2 &=& -(H_1H_2)^{-4/5}\, f\, dt^2 +
(H_1H_2)^{1/5}\, (f^{-1}\, d r^2 + r^2\, d\Omega_{5,k}^2)\ ,\nn\\
f&=& k -\fft{\mu}{r^4} + \ft14 g^2\, r^2\, H_1 H_2\ ,\qquad
X_i= (H_1H_2)^{2/5}\, H_i^{-1}\ ,\nn\\
A_\1^i &=&\sqrt k \, \coth\beta_i\, (1-H_i^{-1})\, dt\ ,\qquad
H_i = 1+ \fft{\mu\, \sinh^2\beta_i}{r^4}\ ,
\eea
where $d\Omega_{5,k}^2$ is the metric on a unit $S^5$, $T^5$ or $H^5$
according to whether $k=1,0$ or $-1$.  As in the previous cases we
discussed, the $k=0$ solution is obtained by first rescaling
$\sinh^2\beta_i \longrightarrow k\, \sinh^2\beta_i$ before setting
$k=0$.  The metric of the $D=7$ $AdS$ black hole was
obtained in \cite{Cveticgubser1}, by isolating the spacetime
direction of the rotating $M5$-brane metric.

\subsection{Rotating M5-brane}
\la{RotatingM5}

      There are two angular momenta, $\ell_1$ and $\ell_2$, in the
rotating $M5$-brane \cite{Cveticyoum2,Csaki2}.  Its metric is given by
\bea
ds_{11}^2 &=& H^{-1/3}\Big(-(1-\fft{2m}{r^3\Delta})\, dt^2 +
dx_1^2 + \cdots + dx_5^2\Big) +
H^{2/3}\Big[\fft{\Delta\, dr^2}{H_1H_2 - \fft{2m}{r^3}} \nn\\
&&+r^2\Big(d\mu_0^2 +\sum_{i=1}^2 H_i(d\mu_i^2 + \mu_i^2\,
d\phi_i^2)\Big) - \fft{4m\, \cosh\a}{r^3\, H\, \Delta}\, dt\,
(\sum_{i=1}^2 \ell_i\, \mu_i^2\, d\phi_i)^2\nn\\
&&+\fft{2m}{r^3\, H\, \Delta}(\sum_{i=1}^2\ell_i\, \mu_i^2\,
d\phi_i)^2\Big]\ ,
\eea
where $\Delta$, $H$ and $H_i$ are given by
\bea
&&\Delta = H_1 H_2(\mu_0^2 + \fft{\mu_1^2}{H_1} + \fft{\mu_2^2}{H_2})\ ,
\qquad H= 1 + \fft{2m\, \sinh^2\a}{r^3\, \Delta}\ ,\nn\\
&&H_1=1 + \fft{\ell_1^2}{r^2}\ ,\qquad
H_2=1 + \fft{\ell_2^2}{r^2}\ .
\eea
The three quantities $\mu_0$, $\mu_1$ and $\mu_2$ satisfy
$\mu_0^2+\mu_1^2 +\mu_2^2=1$.
The 4-form field strength is given by $F_4={*dA_6}$, where
\be
A_6 = \fft{1-H^{-1}}{\sinh\a}(\cosh\a\, dt +
\ell_1\, \mu_1^2\, d\phi_1+ \ell_2\, \mu_2^2\, d\phi_2)\wedge
d^5x\ .
\ee

   The decoupling limit is defined by
\bea
&&m\longrightarrow \ep^3\, m\ ,\qquad
\sinh\a \longrightarrow \ep^{-3/2}\, \sinh\a\ ,\nn\\
&&r\longrightarrow \ep\, r\ ,\qquad
x^{\mu} \longrightarrow \ep^{-1/2}\, x^\mu\ ,\qquad
\ell_i \longrightarrow \ep\, \ell_i\ ,
\eea
with $\ep \longrightarrow 0$.  In this limit, the metric becomes
\bea
ds_{11}^2&=& \wtd \Delta^{1/3}\Big[-(H_1H_2)^{-4/5}\, f\, dt^2 +
(H_1H_2)^{1/5}\, (f^{-1}\, d\rho^2 + \rho^2\, d\vec y\cdot d\vec y)
\Big]\nn\\
&&+g^{-2}\, \wtd \Delta^{-2/3}\,
\Big((X_1X_2)^{2}\, d\mu_0^2 +
\sum_{i=1}^2 X_i^{-1}\, (d\mu_i^2 + \mu_i^2(d\phi_i +
      g\, A_\1^i)^2 )\Big)
\ .\label{m5rotatehor}
\eea
where
\bea
&&\rho^2 =4r\, g^{-1}\ ,\qquad \vec y= \ft12g\, \vec x\ ,\qquad
\wtd \Delta=(X_1 X_2)^{-2}\, \mu_0^2 +
X_1\, \mu_1^2 + X_2\, \mu_2^2 \ ,\nn\\
&&g^2 =(2m\,\sinh^2\a)^{-2/3}\ ,\qquad \mu = 32 m\, g^{-1}\ .
\eea
The metric (\ref{m5rotatehor}) fits precisely the dimensional
reduction ansatz given in (\ref{s4metred}). The lower dimensional
fields are given by
\bea
ds_7^2&=&-(H_1H_2)^{-4/5}\, f\, dt^2 + (H_1H_2)^{1/5}\,
(f^{-1}\, d\rho^2 + \rho^2\, d\vec y\cdot d\vec y)\ ,\nn\\
X_i&=& (H_1 H_2)^{2/5}\, H_i^{-1}\ ,\qquad
 f=-\fft{\mu}{\rho^4} + \ft14 g^2\rho^2\, H_1H_2\ ,\nn\\
A_\1^i &=& \fft{1-H_i^{-1}}{g\, \ell_i\, \sinh\a}\, dt\ .
\eea
This is precisely the $k=0$ $AdS_7$ black hole obtained in the previous
section, with the angular momenta reparameterised as $\ell_i=\mu\,
g^2\, \sinh^2\beta_i/16$.  This establishes that the 2-charge $k=0$
$AdS$ black hole in $D=7$ can be reinterpreted as the decoupling limit
of the rotating $M2$-brane.  (Of course in this example, one can {\it
only} discuss the embedding when the scalar fields are included, since
there is no choice of charge parameters for which the scalar fields
vanish in the seven-dimensional black holes. This contrasts with the
cases of the rotating $D3$-branes and $M5$-branes, where the special
choice of setting all the charges equal allows the discussion of a
simplified ansatz where the scalars are omitted.)

\subsection{Charge as angular momentum}

To those students of Kaluza-Klein theory who are used to the idea 
that electric charge is {\it momentum} in an extra dimension, it may 
come as a bit of a shock to learn that the electric charges of the 
$AdS$ black holes correspond to {\it angular momenta} in extra 
dimensions.  My understanding of this is as follows and it relies 
crucially on the interpretation of $AdS_{\tilde d +1}е\times S^{d+1}е$ 
vacua as the near-horizon limit of $(\tilde d -1)$-brane geometries.

Momentum and angular momentum correspond to generators of the 
Poincar\'{e} 
group or $AdS$ group, and hence to spacetimes that are asymptotically 
Minkowski or $AdS$. If the spacetime is 
asymptotically $M^{D}$, for example, it makes sense to talk about 
angular momentum in $D$ dimensions. On the other hand, if the spacetime is 
asymptotically $AdS_{\tilde d+1} \times S^{d+1}$, it makes sense to talk about 
angular momentum in $(\tilde d +1)$ dimensions but not in $D=(\tilde 
d +d +2)$ dimensions. Of course, there 
is an $SO(d+2)$ symmetry but this is just the isometry group of 
$S^{d+1}$; I see no reason (yet) to call this angular momentum \footnote{Somewhat 
confusingly, some authors refer to the black holes as 
``spinning in the $S^{d+1}е$ directions'' but this is, in my opinion, 
an abuse of language.}.

The ability to call $SO(d+2)$ angular momentum depends crucially on 
the the fact that the $AdS_{\tilde d+1}е$ vacua are near horizon 
geometries of $(\tilde d-1)$-branes and that these brane geometries tend 
asymptotically to $M^{\tilde d +d +2}е$. We can now legitimately refer to 
$S^{d+1}$ as the rotation group in the $d$ dimensions transverse to the 
branes, and if we allow these branes to spin it will describe their 
angular momentum.

(Having said this, one should probably not refer to the electric charge 
of ordinary five-dimensional Kaluza-Klein as ``momentum in the fifth 
direction'' either. After all, the symmetry group of $M^{4}е \times 
S^{1}е$ is $P_{4}е\times U(1)$ and not $P_{5}е$.)

\subsection{Magnetic black holes}
\la{Magnetic}

We have seen that the $N=8$ gauged supergravity naturally admits a
four-electric-charge black hole solution.  In fact it turns out that
this solution is easily generalized to give magnetically charged black
holes; although the full theory involves non-abelian $SO(8)$ gauge
  fields, the $U(1)^4$ truncation of (\ref{d4lagxx}) gives rise to bosonic
equations of motion that are symmetric under the electric-magnetic duality
\begin{equation}
F^{i}\to X_{i}^{-2}ее*F^{i},
\qquad
X_{i}е\to X_{i}е^{-1}е
\end{equation}
The resulting four magnetic charge solution in the case $k=1$ has the form
\begin{eqnarray}
&&ds^2=-(H_1H_2H_3H_4)^{-1/2}fdt^2+(H_1H_2H_3H_4)^{1/2}
({f^{-1}еdr^2}+r^2d\Omega^2),\nonumber\\
&&X_{i}е=H_{i}е(H_{1}еH_{2}еH_{3}еH_{4}е)^{-1/4}е\nonumber\\
&&H_i=1+{\mu\sinh^2\beta_i\over r},
\qquad f=1-{\mu\over r}+4g^2r^2(H_1H_2H_3H_4),\nonumber\\
&&F_{\theta\phi}^{i}=
\mu\cosh\beta_i\sinh\beta_\i\sin\theta.
\la{magnetic}
\end{eqnarray}

While the extremal limit is once again reached by taking $\mu\to0$ and
$\beta_{i}е\to\infty$ with $P_\alpha\equiv \mu\sinh^2\beta_i$ fixed, the
resulting extremal black hole is in fact not supersymmetric whenever
$g\ne0$! In the case of the magnetic Reissner-Nordstr\"{o}m black hole,
this phenomenon was previously found in \cite{Romans}.  (Note,
however, that
it is possible to obtain magnetic black holes that do preserve some
supersymmetry if one allows for event horizons with non-spherical
topologies \cite{Caldarelli})
To see that (\ref{magnetic}) admits no Killing spinors,
we note that while the scalar potential is symmetric under
$X_{i}е\to X_{i}е^{-1}е$, the scalar related
terms in the supersymmetry variations are not.  In particular, focusing 
on
$\delta\chi$, we find for example
\begin{eqnarray}
\delta(2\chi^{(3)i_{(1)}})&=&{1\over\sqrt{2}}\epsilon^{ij}\gamma^r
\Biggl\{\partial_r\log{H_1H_3\over H_2H_4} [\delta^{jk}-i\eta f^{-1/2}
\gamma_{\overline{0}}\gamma^5\epsilon_{jk}]\nonumber\\
&&\qquad\qquad+\partial_r((H_1+H_3)-(H_2+H_4))
[\sqrt{2}grf^{-1/2} \gamma_{\overline{r}}\delta^{jk}] \Biggr\}
\epsilon_{k_{(1)}}
\label{eq:magsusy}
\end{eqnarray}
(where $\eta\equiv\eta_1=\eta_2=\eta_3=\eta_4$),
indicating explicitly that the $g$-dependent term on the last line has a
different structure than the others.  This is in contrast to the 
electric solution.  Additionally, note that the matrices
$[i\gamma_{\overline{0}}\gamma^5\epsilon_{ij}]$ and
$[\gamma_{\overline{r}}\delta^{ij}]$ now commute, while previously, for the
electric black hole, they had anticommuted in the absence of $\gamma^5$.

For both of the above reasons, we see that whenever $g\ne0$ none of the
supersymmetry variations vanish, and hence the magnetic solution is
non-BPS (regardless of the choice of signs of the magnetic charges).  In the
$g\to0$ limit, on the other hand, the last line of (\ref{eq:magsusy}) drops
out, and we are left with
\begin{equation}
\delta(2\chi^{3i_{(1)}})=\sqrt{2}\epsilon^{ij}\gamma^r\partial_r
\log{H_1H_3\over H_2H_4}\tilde P_\eta\epsilon_{j_{(1)}},
\end{equation}
where $\tilde P_\eta^{ij}={1\over2}[\delta^{ij}-i\eta\gamma_{\overline{0}}
\gamma^5\epsilon_{ij}]$ is the projection appropriate to a magnetically
charged solution.  

Just as for the electrically charged solutions, we may embed the 
magnetically charged solutions back in eleven dimensions. Although the 
electrically charged black holes then have the interpretation as the 
near-horizon limit of a rotating $M2$-brane, we do not have any 
such   simple interpretation in the magnetic case.

\subsection{Kaluza-Klein states as black holes}
\la{KKstates}

For {\it ungauged} $N=8$ supergravity, the supersymmetry algebra
admits 4 central charges $Z_{1},Z_{2},Z_{3},Z_{4}$. States fall
into 5 categories according as they are annihilated by $4 \geq q \geq
0$ supersymmetry generators. $q$ also counts the number of $Z$'s that
obey the bound $M=Z_{max}$. Non-rotating black holes (in the
sense of vanishing bosonic Kerr angular momentum $L$) belong to
superspin $L=0$ supermultiplets \cite{Rahmfeld3,Duffliurahmfeld1}. Starting
with a spin $J=0$ member, the rest of the black hole multiplet may
then be filled out using the fermionic zero-modes
\cite{Aichelburg,Duffliurahmfeld2}. The spin will run from $J=0$ up
to $J=(8-q)/2$. For {\it gauged} supergravity, the algebra is different
with no central charges but the same multiplet shortening phenomenon
still occurs \cite{Freedmannicolai,Kaluza}. So we can be
confident that the above black holes preserving $4,2,1,0$ supersymmetries will
belong to supermultiplets with maximum spins $2,3,7/2,4$
\footnote{The maximum spin $5/2$ solutions are (mysteriously?) absent
just as for ungauged supergravity.}. Unfortunately, as far as we know,
the analogue of the $M=Z_{max}$ condition has never been been spelled out
in the literature. It is presumably some relation between the $AdS$ quantum
numbers $(E_{0},s)$ and the $SO(8)$ Casimirs.

It seems entirely consistent, therefore, to identify a subset of the maximum
spin 2 black hole supermultiplets with the $S^{7}$ Kaluza-Klein
spectrum, in analogy with the black hole Kaluza-Klein correspondence
of ungauged supergravity \cite{Rahmfeld1,Kaluza}. The subset in
question will correspond to electric black holes whose mass is quantized
in units of the inverse $S^7$ radius. However, this raises the
puzzle of how the black holes carrying
only $U(1)$ charges can be identified with the Kaluza-Klein particles
carrying non-trivial $SO(8)$ representations. Although we have not
demonstrated this explicitly, it seems reasonable to suppose that it
is the fermion zero modes that provide the non-trivial $SO(8)$ quantum
numbers just as they provide the non-trivial spin. The fact that
these nonabelian charges arise from
{\it fermionic hair} also
nicely circumvents the usual no-hair theorems of classical
relativity. In this connection, it would be interesting to repeat the
gyromagnetic ratio calculations of \cite{Duffliurahmfeld2} and verify
that the fermionic hair again yields a gyromagnetic ratio equal to 1, as
demanded by Kaluza-Klein reasoning.

It is furthermore tempting, in analogy with the ungauged case, to
identify the 2, 3 and 4 charge solutions as 2, 3 and 4-particle
bound states of the singly charged solution
\cite{Duffliurahmfeld1,Rahmfeld2}. However, although the quantum number
assignments are consistent with this, we do not have multi-center
solutions in the $AdS$ case. Such a bound state interpretation would,
of course, lead to states of arbitrarily high spin.

Another difference between the $S^{7}$ and the $T^{7}$
compactifications is that the $g\rightarrow 0$ limit of the gauged
supergravity does not directly coincide with the massless sector of
the $T^{7}$ compactification. They differ by various dualizations. Thus
it was possible, for example, to find 4-charge solutions with all
charges electric as opposed to the 2-electric and 2-magnetic charges
of the ungauged theory. Moreover, whereas 2 charges are Kaluza-Klein
modes and 2 are winding modes in the Type $IIA$ string theory
context, there is no T or U-duality associated with the $S^7$
compactification.

One might also generalize the purely electric and purely
magnetic solutions discussed here to dyonic black hole solutions of
gauged $N=8$ supergravity.  Neither magnetic nor dyonic black holes
have any Kaluza-Klein interpretation and do not appear in the spectrum
of the $S^{7}$ compactification of $D=11$ supergravity. It would be
interesting to provide their $M$-theory interpretation and to determine their
role in the $AdS/CFT$ correspondence.

\subsection{Recent Developments}
\la{recent}

Since these lectures were delivered, several papers have appeared on 
related topics:

* Explicit Kaluza-Klein ans\"{a}tze for deriving various gauged 
supergravites have recently been given: $N=4$, $D=7$ gauged supergravity 
from $D=11$ supergravity \cite{Nastase1,Nastase2}; $N=1$, $D=7$ gauged 
supergravity 
from $D=11$ supergravity \cite{Lupope1}; $N=4$, $SU(2) \times U(1)$, 
$D=5$ gauged supergravity from $D=10$ Type $IIB$ supergravity 
\cite{Lupope2}; $N=4$, $SO(4)$, $D=4$ gauged supergravity from $D=11$ 
supergravity \cite{Lupope3}. The complete ansatz for $N=8$, $SO(6)$, 
$D=5$ gauged supergravity from $D=10$ Type $IIB$ supergravity 
remains unsolved.

* The CFT duals of $AdS$ black holes rotating in spacetime and carrying 
electric charge as a consequence of the rotation of the corresponding 
branes in the transverse dimensions, as discussed in Section 
\ref{Blackhole}, are treated in \cite{Hawkingreall} and 
\cite{Landsteiner}. These papers also exploit the appearance of the 
$N=8$ superconformal singleton action \cite{BD} as the membrane on the 
boundary, discussed in Section \ref{Singleton}.  

* The thermodynamics of spinning branes and their field theory duals 
are discussed in \cite{Harmark1}, while the phase structure of 
non-commutative field theories and spinning brane bound states are 
discussed in \cite{Harmark2}.

* Unitary supermultiplets of $Osp(6,2|4)$, discussed in Section 
\ref{Singleton},  and the $AdS_{7}/CFT_{6}$ 
correspondence are treated in \cite{Gunaydintakemae}.

* An application of the correspondence between $AdS_{4}$ and a $d=3$ 
conformal field theory involving the squashed $S^{7}е$ \cite{ADP,DNP} 
discussed in Section \ref{AdS4} is given in \cite{Ahn}.

* The {\it Membrane at the End of the Universe} finds application in 
\cite{Yau}, which resolves some puzzles in the $AdS/CFT$ correspondence.

* A review of $M$-theory cosmology may be found in \cite{Banks}.

\subsection{Problems 6}
\la{Problems6}

\begin{enumerate}

\item
Calculate the Ricci tensor for the general Kaluza-Klein ansatz for odd 
sphere $S^{2k-1}$ reductions of the $D$-dimensional metric.

\item

Compute the near-horizon limit of the rotating $p$-brane carrying a 
single charge given in \ref{Solutions5}.

\end{enumerate}

\section{\bf SOLUTIONS TO PROBLEMS}
\la{Solutions}

\subsection{Solutions 1}
\la{Solutions1}

\begin{enumerate}

\item In fact we will solve a slightly more general problem by starting 
with the $D$-dimensional Einstein action
\be
I_{D}е=\frac{1}{2\kappa_{D}{}^{2}}\int d^{D}еx \sqrt{ -{\hat g}}{\hat R}
\ee
which after integration by parts may be written
\be
I_{D}е=\frac{1}{2\kappa_{D}{}^{2}}\int d^{D}еx \sqrt{ -{\hat g}}
(\Omega_{ABC}е\Omega^{ABC}е-2\Omega_{ABC}е\Omega^{CAB}е
-4\Omega_{CA}е{}^{A}е\Omega^{C}е{}_{B}е{}^{B}е)
\ee
where $\Omega_{MNP}е$ are the anholonomy coefficients \ref{anholo}.
We write the vielbein as
\be
e_{M}е{}^{A}е=\left( \begin{array}{cc}
e_{\mu}е{}^{\alpha}е&e_{\mu}е{}^{a}е\\
e_{m}е{}^{\alpha}е&e_{m}е{}^{a}е
\end{array} \right)
\ee
so that the metric takes the form
\begin{equation}
\hat{g}_{\hat{\mu}\hat{\nu}}=
\pmatrix{g_{\mu\nu}+g_{mn}еA_{\mu}{}^{m}еA_{\nu}{}^{n}е&
g_{mn}еA_{\mu}{}^{m}е\cr
g_{mn}еA_{\nu}{}^{n}е&g_{mn}е}
\end{equation}
where
\be
A_{\mu}е{}^{m}е=e^{m}е{}_{a}еe_{\mu}е{}^{a}е
\ee
Note that
\be
det~g_{MN}е=det~g_{\mu \nu}еdet~g_{mn}е\equiv det~g_{\mu \nu}е\Delta
\ee
If we choose a gauge $e_{m}е{}^{\alpha}=0$, the only non-vanishing 
components are $\Omega_{\alpha \beta \gamma}$,
\[
\Omega_{\beta \alpha c}= 
e_{mc}e^{\mu}е{}_{\alpha}еe^{\nu}е_{\beta}F_{\mu \nu}{}^{m}ее
\]
\be
\Omega_{\alpha bc}е=-\Omega_{b\alpha 
c}е=-e^{m}е{}_{b}еe^{\mu}е_{\alpha}е\partial_{\mu}еe_{mc}е
\ee
where 
\be
Fе_{\mu\nu}е{}^{m}е=\partial_{\mu}еA_{\nu}е{}^{m}е
-\partial_{\nu}еA_{\mu}е{}^{m}
\ee
Here we have assumed that all fields are independent of the extra 
coordinates $y^{m}е$. 
This yields the four-dimensional action
\be
I_{4}е=\frac{1}{2\kappa_{4}е{}^{2}е}\int d^{4}еx \sqrt{-g} \sqrt{\Delta}
[R-
\frac{1}{4}g_{ij}еF_{\mu \nu}е{}^{i}ееF_{\rho \sigma}е{}^{j}е
g^{\mu \rho}еg^{\nu \sigma}е
+\frac{1}{4}g^{\rho \sigma}е(g^{ik}еg^{jl}е-g^{il}еg^{jk}е)
\partial_{\rho}еg_{ik}е\partial_{\sigma}еg_{jl}е]
\ee
We can eliminate $\sqrt{\Delta}$ by a Weyl rescaling
\be
e_{\mu}е{}^{a}е \rightarrow \Delta^{1/4}еe_{\mu}е{}^{a}е
\ee
to obtain
\be
I_{4}е=\frac{1}{2\kappa_{4}е{}^{2}е}\int d^{4}еx \sqrt{-g}
[R-
\frac{1}{4}\sqrt{\Delta}g_{ij}еF_{\mu \nu}е{}^{i}ееF_{\rho \sigma}е{}^{j}е
g^{\mu \rho}еg^{\nu \sigma}е
+\frac{1}{4}g^{\mu\nu}е
\partial_{\mu}еg_{ij}е\partial_{\nu}еg^{ij}
-\frac{1}{8}g^{\mu\nu}е\partial_{\mu}еln\Delta \partial_{\nu}еln\Delta]
\ee
In the special case of one extra dimension, 
$g_{11}=e^{-\sqrt{3} \phi}$, $\Delta=e^{-\phi/\sqrt{3}}$, we 
obtain the required result.

Note that, of the original $D$-dimensional diffeomorphism group with 
parameters $\xi^{M}е(x,y)$, the surviving symmetry is the 
$4$-dimensional diffeomorphism group with parameters $\xi^{\mu}е(x)$, 
an abelian $U(1)^{k}е$ gauge invariance with parameter $\xi^{m}е(x)$ 
and a global $SL(n,R)$ invariance with constant parameter 
$a^{m}е {}_{n}е$ with 
$a^{m}е {}_{m}е=0$:
\[
\xi^{\mu}е=\xi^{\mu}е(x)
\]
\be
\xi^{m}е=\xi^{m}е(x)+a^{m}е {}_{n}еy^{n}е
\ee
The choice of gauge $e_{m}е{}^{\alpha}е=0$ does not restrict these 
invariances, since
\be
\delta e_{m}е{}^{\alpha}е=\xi^{\mu}е\partial_{\mu}е e_{m}е{}^{\alpha}е
+a_{m}{}^{n}еe_{n}е{}^{\alpha}е=0
\ee

\item 

We make a $4+7$ split of the world indices:
\be
x^{M}е=(x^{\mu}е,y^{m}е)
\la{split2}
\ee
where $\mu=0,1,2,3$ and $m=4,5,6,7,8,9,10$ and write the $SO(1,10)$ 
spinor index as
\be
\alpha=(\alpha',\alpha'')
\ee
where $\alpha'=1,\ldots 4$ is a spinor index of $SO(1,3)$ and 
      $\alpha''=1,\ldots 8$ is a spinor index of $SO(7)$.
Then the fields split according to the $T^{7}е$ column of Table \ref{N=8}. 
\begin{table}
\[
\begin{array}{llllll}
D=11&D=4&\hbox{field}&\hbox{spin}&T^{7}&S^{7}\\
&&&&&\\
g_{MN}е&g_{\mu \nu}е&\hbox{tensor}&2&1&1\\
       &g_{\mu n}е&\hbox{vectors}&1&7&28\\
       &g_{mn}е&\hbox{scalars}&0&28&35\\
&&&\\
\Psi_{M}е{}^{\alpha}е&\Psi_{\mu}е{}^{\alpha'\alpha''}е&\hbox{vector-spinors}&3/2&8&8\\
&\Psi_{m}е{}^{\alpha'\alpha''}е &\hbox{spinors}&1/2&56&56\\
&&&\\
A_{MNP}е&A_{\mu \nu \rho}е&\hbox{3-form}&-&1&0\\
&A_{\mu \nu p}е&\hbox{2-forms}&0&7&0\\ 
&A_{\mu n p}е&\hbox{1-forms}&1&21&0\\ 
&A_{m n p}е&\hbox{0-forms}&0&35&35 
\end{array}
\]         
\caption{The fields of $(D=4,N=8)$ supergravity coming from 
$D=11,N=1$ on $T^{7}е$ and $S^{7}е$}
\la{N=8}
\end{table}

The spin assigments are obvious except for the $3$-form and $2$-form gauge 
fields. A $3$-form is dual to a non-propagating auxilary field $c$:
\be
\sqrt{-g}g^{\mu\alpha}еg^{\nu\beta}еg^{\rho\gamma}еg^{\sigma\delta}е
\partial_{\alpha}еA_{\beta\gamma\delta}е=\epsilon^{\mu\nu\rho\sigma}еc
\ee
and $c=constant$ as a result of the Bianchi identity $ddA \equiv 0$. 
Similarly, a $2$-form gauge field is dual to a scalar field $\phi$:
\be
\sqrt{-g}g^{\mu\alpha}еg^{\nu\beta}еg^{\rho\gamma}е
\partial_{\alpha}еA_{\beta\gamma}е=\epsilon^{\mu\nu\rho\sigma}е
\partial_{\sigma}е\phi
\ee
and $\del^{2}е\phi=0$ as a result of the Bianchi identity. 

To summarize: the particle content of $N=8,D=4$ supergravity is (1 
spin 2, 8 spin 3/2, 28 spin 1, 56 spin 1/2, 70 spin zero).
\end{enumerate}

\subsection{Solutions 2}
\la{Solutions 2}

\begin{enumerate}
\item
The Einstein equation follows from varying with respect to $g_{MN}е$ 
and the $3$-form equation follows by varying with respect to $A_{MNP}е$.
We shall need the identities:
\be
\delta (\sqrt{-g}еg^{MN}еR_{MN}е)=(\delta \sqrt{-g})еg^{MN}еR_{MN}е+
                                    \sqrt{-g}е(\delta g^{MN})еR_{MN}е
                                    \sqrt{-g}еg^{MN}е(\delta R_{MN}е)
\la{vary}
\ee
and
\be
\delta \sqrt{-g}=\frac{1}{2}\sqrt{-g}g^{MN}е\delta g_{MN}е
\ee
\be
\delta g^{MN}е=-g^{MP}еg^{NQ}е\delta g_{PQ}е
\ee
\be
\sqrt{-g}g^{MN}е\delta R_{MN}е=
\partial_{N}е(\sqrt{-g}g^{MN}е\delta \Gamma_{ML}е{}^{L}е)-
\partial_{L}е(\sqrt{-g}g^{MN}е\delta \Gamma_{MN}е{}^{L}е)
\ee
Since the last expression is a total divergence we can drop the final 
term in (\ref{vary}). Similarly
\be
\delta (\sqrt{-g}еg^{MR}g^{NS}еg^{PT}еg^{QU}еF_{MNPQ}еF_{RSTU})=
\sqrt{-g}(4F_{M}е{}^{PQR}еF_{NPQR}е-\frac{1}{2}g_{MN}еF^{PQRS}еF_{PQRS}е)
\ee
Bearing in mind that the $AFF$ term is independent of the metric, we 
obtain the Einstein equation (\ref{einstein}).

A straightforward variation of the action with respect to $A_{MNP}е$ 
then yields the $3$-form equation (\ref{3form}).

\item 

We begin by making an ansatz for the $D=11$ gauge fields 
$g_{MN}е=e_{M}е {}^{A}еe_{N}е{}^{B}е\eta_{AB}е$ and $A_{MNP}е$ 
corresponding to the most general six-five split invariant under 
$P_{6}е\times SO(5)$, where $P_{6}е$ is the $D=6$ Poincar\'{e} group. We 
split the $D=11$ coordinates
\be
x^{M}е=(x^{\mu}е,y^{m}е)
\la{split3}
\ee
where $\mu=0,1,2,3,4,5$ and $m=6,7,8,9,10$, and write the line element 
as 
\be
ds^{2}е=e^{2A}е\eta_{\mu\nu}еdx^{\mu}еdx^{\nu}е
+e^{2B}е\delta_{mn}еdx^{m}еdx^{n}е
\la{metricnew}
\ee
and the $4$-form gauge field strength as
\be
F_{mnpq}=\epsilon_{mnpqr}\partial^{r}еe^{C}е
\la{4form}
\ee
All other components of $F_{MNPQ}$ and all components of the 
gravitino $\Psi_{M}е$ are set equal to zero. $P_{6}е$ invariance 
requires that the arbitrary functions $A,B$ and $C$ depend only on 
$y^{m}е$; $SO(5)$ invariance then requires that this dependence be 
only through $y=\sqrt{\delta_{mn}еy^{m}еy^{n}е}$.

As we shall show, the three arbitrary functions $A$, $B$, and $C$ are 
reduced to one by the requirement that the field configuration 
(\ref{metricnew}) and (\ref{4form}) preserve some unbroken supersymmetry. 
In other words, there must exist Killing spinors $\epsilon$ satisfying
\be
{\tilde D}_{M}е \epsilon =0
\la{covariantnew}
\ee
where ${\tilde D}_{M}$ is the bosonic part of the supercovariant derivative 
appearing in 
the supersymmetry transformation rule of the gravitino 
(\ref{susytransform}).

We make the six-five split
\be
\Gamma^{A}е=(\gamma_{\alpha}е\otimes {\bf 1}, \gamma_{7}е\otimes \Sigma_{a}е)
\ee
where $\gamma_{\alpha}$ and $\Sigma_{a}е$ are the $D=6$ and $D=5$ 
Dirac matrices respectively and where
\be
\gamma_{7}е=\gamma_{0} \ldots \gamma_{5}е
\ee
so that $\gamma_{7}е{}^{2}е=1$.  We also decompose the spinor field as
\be
\epsilon(x,y)=\zeta(x) \otimes \eta(y)
\ee
where $\zeta$ is a constant spinor of $SO(1,5)$ which may further be 
decomposed into chiral eigenstates 
via the projection operators $(1\pm \gamma_{7}е)/2$ and $\eta$ is an 
$SO(5)$ spinor .

In our background (\ref{metricnew}) and (\ref{4form}), the supercovariant 
derivative becomes:
\[
{\tilde D}_{\mu}е=\partial_{\mu}е
-\frac{1}{2}\gamma^{\mu}\gamma_{7}е\Sigma^{m}е\partial_{m}еe^{2A}
- \frac{1}{12} \gamma_{\mu} \Sigma^{m}е\partial_{m}еe^{C}е,
\]
\be
{\tilde D}_{m}е=\partial_{m}е
+\frac{1}{2}\Sigma_{m}е\Sigma^{n}е\partial_{n}еe^{B}е
- 
\frac{1}{6}\partial_{m}еe^{C}\gamma_{7} 
- \frac{1}{6}(\Sigma_{m}е\Sigma^{n}е-\Sigma_{n}е\Sigma^{m}е)\partial_{m}еe^{C}
\ee
Hence we find that (\ref{covariant}) admits  non-trivial solutions
\be
(1 + \gamma_{7}е)\eta =0
\la{chiralnew}
\ee
where
\be
\eta=e^{-C/6}е\eta_{0}е
\ee
where $\eta_{0}е$ is a constant spinor and
\[
A=\frac{1}{3}C
\]
\be
B=-\frac{2}{3}C+constant
\la{ABnew}
\ee
In each case, (\ref{chiralnew}) means that one half of the maximal possible 
supersymmetry survives.

With the substitutions (\ref{metricnew}), (\ref{4form}) and (\ref{ABnew}), 
the Einstein equation and the $4$-form equation reduce to the single 
equation for one unknown:
\be
\delta^{mn}е \partial_{m}е\partial_{n}еe^{-C}е=0
\ee
and hence, imposing the boundary condition that the metric be 
asymptotically Minkowskian, we find
\be
e^{-C}е=1+\frac{b^{3}е}{y^{3}е}
\ee
where $k_{6}е$ is a constant, at this stage arbitrary. Thus the 
metric is given by
\be
ds^2=(1+b^{3}е/y^3)^{-1/3}dx^{\mu}dx_{\mu}+
(1+b^{3}е/y^3)^{2/3}(dy^2+y^2d\Omega_4{}^2)
\ee
and the four-form field strength by
\be
F_4= 3b^{3}е\epsilon_4
\ee
Here $\epsilon_4$ is the volume form on $S^4$ and $\Omega_4$ is the volume. 
\end{enumerate}

\subsection{Solutions 3}
\la{Solutions3}

\begin{enumerate}
\item

We use the duality between $M$-theory compactified on a $2$-torus $T^{2}е$ 
of area $A$ (measured with the $M$ theory metric $g^{M}е$) and the Type $IIB$ 
theory compactified on a circle $S^{1}$ of radius $R$ (measured with 
the $IIB$ metric $g^{B}е=\beta^{2}еg^{M}е$) \cite{Schwarzpower}. 
Thus $p$-brane tensions of 
dimension $d=p+1$ will be related to each other by 
$T_{d}^{M}=\beta^{d}еT_{d}е^{B}е$.

The $D=9$ particle spectrum will involve Kaluza-Klein $M0$-branes of 
mass $T_{I}^{M}(KK)=2\pi A^{-1/2}е$ which are identified with winding states 
of mass $T_{1}^{B}(W)=2\pi RT_{2}^{B}$, coming from wrapping the $IIB$ 
string with tension $T_{2}^{B}$ around $S^{1}$, so
\be
2\pi A^{-1/2}е=\beta 2\pi RT_{2}^{B}
\ee
There will also be winding $M0$-branes of mass $AT_{3}^{M}$ coming from wrapping 
the $M2$-brane with tension $T_{3}^{M}$ееaround $T^{2}е$, which are identified 
with the $IIB$ Kaluza-Klein 
states of mass $R^{-1}$, so
\be
AT_{3}е^{M}е=\beta R^{-1}е
\ee
Hence
\be
A=\beta^{4}е\left(\frac {T_{2}е^{B}е}{T_{3}е^{M}е}\right)^{2}е 
\ee
and
\be
R=\beta^{-3}е\frac {T_{3}е^{M}е}{(T_{2}е^{B}е)^{2}е}  
\ee

Next we identify the $M2$-brane in $D=9$ with the 
$D3$-brane with tension $T_{4}е^{B}$ wrapped around $S^{1}е$:
\be
T_{3}^{M}е=\beta^{3}е2\pi R T_{4}е^{B}е
\ee
and the $D3$-brane in $D=9$ with the $M5$-brane with tension 
$T_{6}^{M}$ wrapped around $T^{2}$:
\be
AT_{6}е^{M}е=\beta^{4}еT_{4}е^{B}е
\ee
Eliminating $R$ we find
\be
T_{4}е^{B}е=\frac{1}{2\pi} (T_{2}е^{B}е)^{2}е 
\ee
and eliminating $A$ we find
\be
\frac{T_{6}е^{M}}{(T_{3}е^{M}е)^{2}е}=\frac{T_{4}е^{B}}{(T_{2}е^{B}е)^{2}е}
\ee
and hence that
\be
T_{6}е^{M}е=\frac{1}{2\pi} (T_{3}е^{M}е)^{2}е 
\ee
in agreement with the purely $D=11$ result (\ref{tension}).

\item

The Lagrangian of $D=9$ $N=2$ supergravity as low-energy limit
of Type $IIA$ string compactified on a circle can be obtained from
dimensional reduction of Type $IIA$ supergravity in $D=10$, which itself
can be obtained from dimensional reduction of $D=11$
supergravity.  Using the notation adopted in \cite{lpsol}, the bosonic
sector of the theory contains the vielbeins, a dilaton $\phi$ together
with a second dilatonic scalar $\varphi$ (which measures the size of
the compactifying circle), one 4-form field strength $\tilde F_4 =dA_3$,
two 3-forms $\tilde F_3^{(i)}= dA_2^{(i)}$, three 2-forms $\tilde F_2^{(12)}
= dA_1^{(12)}$ and $\tilde {\cal F}_2^{(i)}= d {\cal A}_1^{(i)}$ and one
1-form $\tilde {\cal F}_1^{(12)} = d{\cal A}_{0}^{(12)}$.  The full
bosonic Lagrangian is given by 
\bea
e^{-1} {\cal L}_{\rm IIA} &=& R - \ft12 (\del \phi)^2 -
\ft12(\del\varphi)^2 - \ft12 ({\cal F}_1^{(12)})^2
e^{-\ft32\phi -\ft{\sqrt7}{2} \varphi} \nonumber\\
&& -\ft1{48} (F_4)^2 e^{-\ft12\phi -\ft3{2\sqrt7} \varphi} 
-\ft1{12} (F_3^{(1)})^2 e^{\phi -\ft1{\sqrt7} \varphi} 
- \ft1{12} (F_3^{(2)})^2 e^{-\ft12\phi + \ft{5}{2\sqrt7} \varphi}
\label{d92alag}\\
&& -\ft14 (F_2^{(12)})^2 e^{\phi + \ft3{\sqrt7}\varphi} 
-\ft14 ({\cal F}_2^{(1)})^2 e^{-\ft32 \phi -\ft1{2\sqrt7}\varphi} 
- \ft14 ({\cal F}_2^{(2)})^2 e^{-\ft4{\sqrt7} \varphi}\nonumber\\
&&-\ft12 \tilde F_4 \wedge \tilde F_4 \wedge A_1^{(12)} -
\tilde F_3^{(1)} \wedge \tilde F_3^{(2)} \wedge A_3\ .\nonumber
\eea
Here we are using the notation that field strengths without tildes
include the various Chern-Simon modifications, whilst field strengths
written with tildes do not include the mofications.  Thus we have
\bea 
F_4&=&\tilde F_4 - \tilde F_3^{(1)}\wedge {\cal A}_1^{(1)} -
\tilde F_3^{(2)}\wedge {\cal A}_1^{(2)} - \ft12 \tilde F_2^{(12)}
\wedge {\cal A}_{1}^{(1)}
\wedge {\cal A}_1^{(2)}\ ,\nonumber\\ F^{(1)}_3 &=& \tilde F^{(1)}_3 - \tilde
F_2^{(12)} \wedge {\cal A}_1^{(2)}\ ,
\nonumber\\ 
F_3^{(2)} &=& \tilde F_3^{(2)} + F_2^{(12)}\wedge {\cal
A}_1^{(1)} -
{\cal A}_0^{(12)} (\tilde F^{(1)}_3 -F_2^{(12)}\wedge {\cal A}_1^{(2)})
\ ,\label{cs9d}\label{csterms}\\ 
F_2^{(12)} &=& \tilde F^{(12)}_2\ ,\qquad {\cal
F}_2^{(1)} = {\cal F}_2^{(1)} +{\cal A}_0^{(12)} {\cal F}_1^{(2)}
\ ,\quad {\cal F}_2^{(2)} = \tilde {\cal F}_2^{(2)}\ , 
\quad {\cal F}_1^{(12)} = \tilde {\cal F}_1^{(12)}\ .\nonumber
\eea

            The lagrangian of the $D=9$ supergravity as
low-energy limit of Type $IIB$ string compactified on a circle can be
obtained from dimensional reduction of Type $IIB$ supergravity in
$D=10$. It is given by
\bea
e^{-1} {\cal L}_{\rm IIB} &=& R-\ft12 (\del\phi)^2 -\ft12 (\del
\varphi)^2 - \ft12 e^{-2\phi} (\del \chi)^2 \nonumber\\
&&-\ft1{48} e^{\ft2{\sqrt7}
\varphi} F_4^2 -\ft1{12} e^{\phi-\ft1{\sqrt7}\varphi} (F_3^{({\rm
NS})})^2 -\ft12 e^{-\phi -\ft1{\sqrt7} \varphi} (F_3^{({\rm R})})^2
\label{d92blag}\\
&&-\ft14 e^{-\ft4{\sqrt7} \varphi} ({\cal F}_2)^2 -
\ft14 e^{-\phi + \ft3{\sqrt7}\varphi} (F_2^{({\rm R})})^2 -
\ft14 e^{\phi + \ft3{\sqrt7} \varphi} (F_2^{({\rm NS})})^2\nonumber\\
&&-\ft12 \tilde F_4 \wedge \tilde F_4 \wedge {\cal A}_1 -
\tilde F_3^{({\rm NS})} \wedge \tilde F_3^{({\rm R})} \wedge A_3\ .\nonumber 
\eea
Note that in $D=10$, there are two 2-form antisymmetric tensors, one
of which is the NS-NS field $A_2^{({\rm NS})}$, and the other is the 
R-R
field $A_2^{({\rm R})}$.  The dimensional reduction of these two
2-form guage potentials also gives rise to two vector fields in $D=9$,
denoted by $A_1^{({\rm NS})}$ and $A_1^{({\rm R})}$ respectively.

        The $D=10$ $IIA$ string and $IIB$ string are related by by a
perturbative T-duality, in that Type $IIA$ string compactified
on a circle with radius $R$ is equivalent to Type $IIB$ string
compactified on a circle with radius $1/R$.  At the level of their
low-energy effective actions, it implies that there is only one $D=9$ $N=2$
supergravity.   The Lagrangians (\ref{d92alag}) and (\ref{d92blag})
are related to each other \cite{DLP1} by local field redefinitions.  
The relations between the gauge potentials of these two
nine-dimensional theories (including the axions) are summarized in 
Table \ref{gaugeII}.

\begin{table}
\bigskip\bigskip
\begin{center}
\begin{tabular}{|c|c|c|c|c|c|}\hline
    &\multicolumn{2}{|c|}{IIA} &
    &\multicolumn{2}{c|}{IIB} \\ \cline{2-6}
    & $D=10$ & $D=9$ &T-duality & $D=9$ & $D=10$ \\ \hline\hline
    & $A_3$ & $A_3$ & $\longleftrightarrow$ &
                   $A_3$ & $B_4$ \\ \cline{3-6}
R-R & &  $A_2^{(1)}$& $\longleftrightarrow$
                           & $A_2^{\rm R}$ & $A_2^{\rm R}$
                                               \\ \cline{2-5}
fields& ${\cal A}_1^{(1)}$ & ${\cal A}_1^{(1)}$ &
                $\longleftrightarrow$ &
        $A_1^{\rm R}$ & \\ \cline{3-6}
   & & ${\cal A}_0^{(12)}$ & $\longleftrightarrow$
                            & $\chi$ &$\chi$
                                 \\ \hline\hline
NS-NS & $G_{\mu\nu}$ & ${\cal A}_1^{(2)}$
                        & $\longleftrightarrow$ &
        $A_1^{\rm NS}$ & $A_2^{\rm NS}$ \\ \cline{2-5}
fields& $A_2^{(1)}$ & $A_2^{(1)}$ &
               $\longleftrightarrow$ & $A_2^{\rm NS}$ &
                                       \\ \cline{3-6}
      & & $A_1^{(12)}$ & $\longleftrightarrow$ &
                              ${\cal A}_1$ & $G_{\mu\nu}$
                                       \\ \hline
\end{tabular}
\end{center}
\caption{Gauge potentials of Type $II$ theories in $D=10$
and $D=9$}
\la{gaugeII}
\end{table}
\bigskip\bigskip
The relation between the dilatonic scalars of the two
nine-dimensional theories is given by
\be
\pmatrix{\phi \cr \varphi}_{IIA} =\pmatrix{\ft34 & -\ft{\sqrt7}{4} \cr
                                           -\ft{\sqrt7}{4} & -\ft34}
\pmatrix{\phi \cr \varphi}_{IIB} \ .\label{dils}
\ee
The dimensional reduction of the ten-dimensional string metric to
$D=9$ is given by
\bea
ds_{\rm str}^2 &=& e^{-\ft12\phi}\, ds_{10}^2 \nn\\
&=& e^{-\ft12\phi}\, (e^{\varphi/(2\sqrt7)}\, ds_9^2 +
e^{-\sqrt7\varphi/2} \, (dz_2 + {\cal A})^2 ) \ ,
\eea
where $ds_{10}^2$ and $ds_9^2$ are the Einstein-frame metrics in
$D=10$ and $D=9$.  The radius of the compactifying circle, measured using
the ten-dimensional string metric, is therefore given by $R=e^{-\ft14
\phi -\sqrt7\varphi /4}$.  It follows from (\ref{dils}) that the radii
$R_{IIA}$ and $R_{IIB}$ of the compactifying circles, measured using
their respective ten-dimensional string metrics, are related by
$R_{IIA}=1/R_{IIB}$.
\end{enumerate}

\subsection{Solutions 4}
\la{Solutions4}

\begin{enumerate}
\item

We begin with the bosonic sector of the $d=3$ worldvolume of the $D=11$
supermembrane \ref{membranebose}.  To see how a double worldvolume/spacetime 
compactification of the
$D=11$ supermembrane theory on $S^1$ leads to the Type $IIA$ string in
$D=10$, let us denote all $(d=3,D=11)$ quantities by a hat
and all $(d=2,D=10)$ quantities without.  We then make a ten-one split
of the spacetime coordinates
\be
{\hat x}^{\hat M}=(x^M,y)\qquad M=0,1,\ldots,9
\la{split4}
\ee
and a two-one split of the worldvolume coordinates
\begin{equation}
{\hat \xi}^{\hat i}= (\xi^i,\xi^{2}е)\qquad i=0,1
\la{split5}
\end{equation}
in order to make the partial gauge choice
\be
\xi^{2}е=y,
\la{choice}
\ee
which identifies the eleventh dimension of spacetime with the third
dimension of the worldvolume. The dimensional reduction is then
effected by taking the background fields ${\hat g}_{{\hat
M}{\hat N}}$ and ${\hat A}_{{\hat M}{\hat N}{\hat P}}$ to be independent of
$y$.  The string backgrounds of dilaton $\phi$, string $\sigma$-model metric
$g_{MN}$, $1$-form $A_M$, $2$-form $B_{MN}$ and $3$-form $A_{MNP}$ are given
by
\begin{eqnarray}
{\hat g}_{MN}&=& e^{-2\phi/3}\left(
\begin{array}{cc}
g_{MN}+e^\Phi A_MA_N&e^{2\phi}A_M\\
e^{2\phi}A_N&e^{\Phi}
\end{array}
\right)\nonumber\\
{\hat A}_{MNP}&=&A_{MNP}\nonumber\\
{\hat A}_{MNY}&=&A_{MN}\ .
\la{ansatz2}
\end{eqnarray}

The choice of dilaton prefactor, $e^{-2\phi/3}$, is dictated by the
requirement that $g_{MN}$ be the $D=10$ string $\sigma$-model 
metric. ( To
obtain the $D=10$ fivebrane $\sigma$-model metric, the prefactor is unity
because the reduction is then spacetime only and not simultaneous
worldvolume/spacetime.  This explains the remarkable ``coincidence''
between $\hat g_{MN}$ and the $D=10$ fivebrane $\sigma$-model
metric.)

Varying the supermembrane action (\ref{membranebose}) with respect to the metric 
${\hat \gamma}_{{\hat i}{\hat j}}е$ yields the embedding equation
\be
{\hat \gamma}_{{\hat i}{\hat j}}=\partial_{{\hat i}}{\hat x}^{{\hat M}}е
                                 \partial_{{\hat j}}{\hat x}^{{\hat N}}е
{\hat g}_{{\hat M}е{\hat N}}({\hat x})
\la{gammaeq}
\ee
while varying with respect to ${\hat x}^{{\hat M}}е$ yields the 
equation of motion
\be
\partial_{{\hat i}}е(\sqrt{-{\hat \gamma}}{\hat \gamma}^{{\hat 
i}{\hat j}}е\partial_{{\hat j}}е{\hat x}^{{\hat M}}е)+
\sqrt{-{\hat \gamma}} {\hat \gamma}^{{\hat i}{\hat j}}е
\partial_{{\hat i}}е{\hat x}^{{\hat N}}е\partial_{{\hat j}}е{\hat 
x}^{{\hat P}}е
{\hat \Gamma}_{{\hat N}{\hat P}}е{}^{{\hat M}}е=
\frac{1}{6}\epsilon^{{\hat i}{\hat j}{\hat k}}
е\partial_{{\hat i}}е{\hat x}^{{\hat N}}
е\partial_{{\hat j}}е{\hat x}^{{\hat P}}
е\partial_{{\hat k}}е{\hat x}^{{\hat Q}}е
{\hat F}^{{\hat M}}е_{{\hat N}{\hat P}{\hat Q}}е
\la{hat}
\ee
where ${\hat F}е_{{\hat M}{\hat N}{\hat P}{\hat Q}}е$ is the field strength 
of ${\hat A}_{{\hat M}{\hat N}{\hat P}}$,
\be
{\hat F}е_{{\hat M}{\hat N}{\hat P}{\hat Q}}е
=4\partial_{[{\hat M}}е{\hat A}_{{\hat N}{\hat P}{\hat Q}]}
\ee
Having made the two-one split of the worldvolume coordinates 
and the ten-one split of the spacetime coordinates, and 
having made the partial gauge choice, the double dimensional 
reduction is then effected by demanding that
\be
\frac{\partialе x^{M}е}{\partial \xi^{2}е}=0
\ee
and
\be
\frac{\partial {\hat g}_{{\hat M}{\hat N}е}}{\partial y}=0=
\frac{\partial {\hat A}_{{\hat M}{\hat N}{\hat P}}е}{\partial y}=0
\ee
From (\ref{ansatz2}), the induced metric on the worldvolume is now given by
\begin{eqnarray}
{\hat \gamma}_{{\hat i}{\hat j}}&=& e^{-2\phi/3}\left(
\begin{array}{cc}
\gamma_{ij}+e^2\phi A_iA_j&e^{2\phi}A_i\\
e^{2\phi}A_j&e^{2\phi}
\end{array}
\right)
\end{eqnarray}
where
\[
\gamma_{ij}е=\partial_{i}еx^{M}е\partial_{j}еx^{N}еg_{MN}е
\]
\be
A_{i}е=\partial_{i}еx^{M}еA_{M}е
\ee
Note that
\be
\sqrt{-{\hat \gamma}}=\sqrt{-\gamma}
\ee
Substituting these expressions into the field equations (\ref{hat}) 
yields in the case $\hat M=M$
\be
\partial_{{ i}}е(\sqrt{-{ \gamma}}{ \gamma}^{{ 
i}{ j}}е\partial_{{ j}}е{ x}^{{M}}е)+
\sqrt{-{ \gamma}} { \gamma}^{{ i}{ j}}е
\partial_{{ i}}е{ x}^{{N}}е\partial_{{ j}}е{ 
x}^{{ P}}е
{ \Gamma}_{{ N}{ P}}е{}^{{ M}}е=
\frac{1}{2}\epsilon^{{ i}{ j}}
е\partial_{{ i}}е{ x}^{{ N}}
е\partial_{{ j}}е{ x}^{{ P}}
{ F}^{{ M}}е_{{ N}{ P}}е
\la{unhat}
\ee
where ${F}е_{{ M}{ N}{ P}}е$ is the field strength 
of ${A}_{{ M}{ N}{ P}}$,
\be
{ F}е_{{ M}{ N}{ P}}е
=3\partial_{[ {M}}е{ A}_{{ N}{ P}]}
\ee
In the case $\hat M=y$, (\ref{hat}) is an identity as it must be for 
consistency.
 
But (\ref{unhat}) is just the ten-dimensional string 
equation of motion derivable from the action 
\begin{eqnarray}
S_2=T_2\int d^2\xi\biggl[&-&{1\over2}\sqrt{-\gamma}\gamma^{ij}
\partial_i x^M\partial_j x^N g_{MN}(x) 
\nonumber \\
&-& {1\over2!}\epsilon^{ij}\partial_i x^M\partial_j x^N
A_{MN}(x) \biggr]
\la{stringbose} 
\end{eqnarray}
One may repeat the procedure in superspace to obtain
\begin{eqnarray}
S_2=T_2\int
d^2\xi\biggl[-{1\over2}\sqrt{-\gamma}\gamma^{ij}{E_i{}^aE_j{}^b\eta_{ab}} 
+{1\over2!}\epsilon^{ij}\partial_i X^M\partial_j X^N
A_{MN}(Z)\biggr] 
\la{stringfermi}
\end{eqnarray}
which is just the action of the Type $IIA$ superstring.  Note that the  
ten-dimensional bosonic R-R fields $A_{MNP}е,A_{M}е$ 
and have decoupled in (\ref{stringbose}). They have not disappeared from the 
theory, however, since their coupling still survives in the R-R 
sector of (\ref{stringfermi}).

\end{enumerate}

\subsection{Solutions 5}
\la{Solutions5}

\begin{enumerate}
\item

Rotating $p$-branes in arbitrary dimensions, supported by
a single $(p+2)$-form charge, are all straightforwardly obtained
by diagonally oxidising the rotating black holes constructed in
\cite{Cveticyoum3}.  There are
two cases arising, depending on whether $\td d$ is even or odd.

In this case $\td d+2=2N$, there are $N$ angular momenta $\ell_i$, with
$i=1,2,\ldots, N$.  We find that the metric of the rotating
$(n-2)$-brane solution to the equations is
\bea
ds_{D}^2&=& H^{-\ft{\td d}{D-2}}\Big(-(1 - \fft{2m}{r^{\td d}\Delta})\,
dt^2 + d\vec x\cdot d\vec x\Big) +
H^{\ft{d}{D-2}}\Big[\fft{\Delta\, dr^2}{H_1\cdots H_N -2m\,
r^{-\td d} }\nn\\
&&+r^2 \sum_{i=1}^N H_i(d\mu_i^2 + \mu_i^2\, d\phi_i^2) -
\fft{4m\cosh\b}{r^{\td d}\, H\, \Delta}\, dt\,
(\sum_{i=1}^N \ell_i\, \mu_i^2\, d\phi_i)\nn\\
&&+\fft{2m}{r^{\td d}\, H\, \Delta}\,
(\sum_{i=1}^N \ell_i\, \mu_i^2\, d\phi_i)^2 \Big]\ ,
\label{tddeven}
\eea
where the functions $\Delta$, $H$ and $H_i$ are given by
\bea
&&\Delta = H_1\cdots H_N\, \sum_{i=1}^N \fft{\mu_i^2}{H_i}
\ ,\qquad H= 1 + \fft{2m\, \sinh^2\b}{r^{\td d}\, \Delta}\ ,
\nn\\
&&H_i = 1 + \fft{\ell_i^2}{r^2}\ ,\qquad i=1,2,\ldots, N\ .
\eea
The dilaton $\phi$ and gauge potential $A_{\sst{(n-1)}}$ are given by
\be
e^{2\phi/a} = H\ ,\qquad
A_{\sst{(n-1)}} = \fft{1-H^{-1}}{\sinh\b}\Big(\cosh\b \, dt +
\sum_{i=1}^N \ell_i\, \mu_i^2\, d\phi_i\Big )\wedge d^{n-2}x\ .
\ee
The $N$ quantities $\mu_i$, as usual, are subject to the constraint
$\sum_i \mu_i^2=1$.  One can parameterise the $\mu_i$ in terms of
$(N-1)$ unconstrained angles.  A common choice is
\bea
\mu_i &=& \sin\psi_i\, \prod_{j=1}^{i-1}\cos\psi_j\ ,
\qquad i\le N-1\ ,\nn\\
\mu_N &=& \prod_{j=1}^{N-1}\cos\psi_j\ .
\eea
Note that $\prod_{j=1}^n \cos\psi_j$ is defined to be equal to 1 if
$n\le 0$.

In the case $\td d +2 = 2N+1$, the solution has the same form as above, but 
with the
range of the index $i$ extended to include 0.  However, there is no angular
momentum parameter or azimuthal coordinate associated with the extra
index value, and so $\ell_0=0$ and $\phi_0=0$.  Otherwise, all the
formulae given above are generalised simply by extending the summation
to span the range $0\le i\le N$.  Of course $H_0=1$ as a consequence
of $\ell_0=0$.

\end{enumerate}

\subsection{Solutions 6}
\la{Solutions6}

\begin{enumerate}

\item

The general Kaluza-Klein ansatz for odd sphere $S^{2k-1}$ reductions
of the $D$-dimensional metric may be expressed in the form
\begin{equation}
ds_{D}^2 = \wtd\Delta^{a}\, ds_d^2 +
\wtd\Delta^{-b}\, \sum_{i=1}^k X_i^{-1}\, \Big( d\mu_i^2 + \mu_i^2\,
(d\phi_i + A^i_\1)^2 \Big)\ ,
\label{eq:bigmet}
\end{equation}
where we have set the radius of $S^{2k-1}$ to unity.  There are $k-1$
scalar degrees of freedom parameterised by the $k$ quantities $X_i$
satisfying the constraint $\prod_{i=1}^kX_i=1$.  This form of the line
element encompasses both the $S^5$ reduction of Type $IIB$ supergravity and
the $S^7$ reduction of eleven dimensional supergravity.  As defined
previously, $\wtd\Delta = \sum_{i=1}^k X_i\mu_i^2$ and $\sum_{i=1}^k
\mu_i^2=1$.

In the absence of the gauge fields, this metric has a block diagonal form,
with the blocks corresponding to the $d$-dimensional spacetime, the $k-1$
direction cosines $\mu_i$ and the $k$ azimuthal rotation angles $\phi_i$.
The main difficulty in computing the curvature of (\ref{eq:bigmet})
lies in the fact that the $\mu_i$'s are constrained.  Nevertheless, we may
perform an asymmetric choice of using the first $k-1$ of them as actual
coordinates, while expressing $\mu_k$ as the constrained quantity
$\mu_k = (1-\sum_{i=1}^{k-1}\mu_i^2)^{1/2}$.

Since numerous terms are involved in the computation, it is imperative to
clarify our notation.  We denote the lower-dimensional spacetime
indices by $\mu,\nu,\ldots=0,1,\ldots,d-1$, the direction cosine
indices by $\alpha,\beta,\gamma,\ldots=1,2,\ldots,k-1$ and the azimuthal
indices by $i,j,\ldots=1,2,\ldots,k$.  Note that for instance
implicit sums over $\alpha$ always run over $k-1$ values, while sums
over $i$ always run over the full $k$ values.

Thus (with vanishing gauge fields) the $D$-dimensional metric may be
expressed in the form
\begin{equation}
G_{MN} = {\rm diag}[\wtd\Delta^a g_{\mu\nu},\wtd\Delta^{-b}
\hat g_{ij},\wtd\Delta^{-b}\wtd g_{\alpha\beta}]\ ,
\end{equation}
where $\hat g_{ij} = X_i^{-1}\mu_i^2\delta_{ij}$ is diagonal and
\begin{eqnarray}
\wtd g_{\alpha\beta}&=& X_\alpha^{-1}\delta_{\alpha\beta} + X_k^{-1}
\hat\mu_\alpha\hat\mu_\beta\ ,\nonumber\\
\wtd g^{\alpha\beta}&=&X_\alpha\delta_{\alpha\beta}-\wtd\Delta^{-1}
X_\alpha X_\beta\mu_\alpha\mu_\beta\ ,
\end{eqnarray}
with $\hat\mu_\alpha\equiv\mu_\alpha/\mu_k$.  Note that ${\rm
det}\,\wtd g_{\alpha\beta}=\wtd\Delta/\mu_k^2$.  As the $\mu_\alpha$
themselves are coordinates, this allows the use of expressions such as
$\partial_\alpha\mu_k = - \hat\mu_\alpha$ and
$\partial_\alpha\hat\mu_\beta=\mu_k^{-1}(\delta_{\alpha\beta}+\hat\mu_\alpha
\hat\mu_\beta)$.  In addition, all $\alpha,\beta,\ldots$ indices are raised
and lowered with the metric $\wtd g_{\alpha\beta}$.

Using this specific form of the metric $\wtd g_{\alpha\beta}$ and the fact
that ${\rm det}\,\hat g_{ij}=\prod_{i=1}^k\mu_i^2$, we find ${\rm det}\,
G_{MN}={\rm det}\, g_{\mu\nu}\, 
\wtd\Delta^{\kappa+2a}\prod_{i=1}^{k-1}\mu_i^2$ where the product
provides the measure over the internal $S^{2k-1}$.  Here
$\kappa=a(d-2)-b(2k-1)+1$ so that $\sqrt{-G}R\sim \sqrt{-g}\, 
\wtd\Delta^{\kappa/2}$.
Hence one expects $\kappa=0$ in order to prevent any $\wtd\Delta$ dependence
from appearing in front of the lower-dimensional Einstein term.  Indeed we
see that $\kappa$ vanishes for both the $S^5$ and the $S^7$
reductions considered in the text.  

We have only computed selected components of the full $D$-dimensional
Ricci tensor which are of interest in the Kaluza-Klein reduction.  While we
have used an asymmetric parameterisation of the direction cosines, the
final results are symmetric in all $k$ of the $\mu_i$'s.  For the
lower-dimensional components of $R_{MN}$ we find
\begin{eqnarray}
R_{\mu\nu}&=& R^{(d)}_{\mu\nu} 
-\half\kappa\wtd\Delta^{-1}\partial_\mu\partial_\nu\wtd\Delta
+\ft14((a+2)\kappa-(a+b)b(2k-1)+a+2b)\wtd\Delta^{-2}\partial_\mu\wtd
\Delta\partial_\nu\wtd\Delta\nonumber\\
&&+\ft14 (\partial_\mu\wtd g^{\alpha\beta}\partial_\nu \wtd g_{\alpha\beta}
-X_i^{-2}\partial_\mu X_i\partial_\nu X_i)\\
&&+g_{\mu\nu}[-\ft{a}{2}\wtd\Delta^{-1}\nabla^2\wtd\Delta
-\ft{a}{4}(\kappa-2)\wtd\Delta^{-2}\partial^\rho\wtd\Delta\partial_\rho
\wtd\Delta]\nonumber\\
&&+g_{\mu\nu}\wtd\Delta^{a+b}[-\ft{a}{4}(\kappa+2a+2b-3)\wtd\Delta^{-2}
\partial^\alpha\wtd\Delta\partial_\alpha\wtd\Delta
-\ft{a}{2}(\wtd\Delta^{-1}\wtd\nabla^2\wtd\Delta+\wtd\Delta^{-1}
\mu_i^{-1}\partial^\alpha\mu_i\partial_\alpha\wtd\Delta)]\ ,
\kern-3pt
\nonumber
\end{eqnarray}
where $R^{(d)}_{\mu\nu}$ denotes the Ricci tensor of the
$d$-dimensional spacetime metric $g_{\mu\nu}$.  
We have also determined the internal
Ricci components $R_{ij}$ and $R_{\alpha\beta}$ necessary for computing the
$D$-dimensional scalar curvature.  For the former we find
\begin{eqnarray}
R_{ij}&=&\hat g_{ij}\wtd\Delta^{-a-b}[\ft14\kappa\wtd\Delta^{-1}X_i^{-1}
\partial^\rho
\wtd\Delta\partial_\rho X_i+\ft{b}{2}\wtd\Delta^{-1}\nabla^2
\wtd\Delta+\ft{b}{4}(\kappa-2)\wtd\Delta^{-2}\partial^\rho\wtd\Delta
\partial_\rho\wtd\Delta]\\
&&+\hat g_{ij}[-\half(\kappa+2a+2b-1)\wtd\Delta^{-1}\mu_i^{-1}\partial^\alpha
\mu_i\partial_\alpha\wtd\Delta-\mu_i^{-1}\wtd\nabla^2\mu_i+\wtd\Delta^{-1}
(X_i\sum X-X_i^2)\nonumber\\
&&\qquad+\ft{b}{4}(\kappa+2a+2b-3)\wtd\Delta^{-2}\partial^\alpha
\wtd\Delta\partial_\alpha\wtd\Delta+\ft{b}{2}\wtd\Delta^{-1}\wtd\nabla^2
\wtd\Delta +\ft{b}{2}\wtd\Delta^{-1}\mu_l^{-1}\partial^\alpha\mu_l
\partial_\alpha\wtd\Delta]\nonumber
\end{eqnarray}
(no sum on $i$), while for the latter we have
\begin{eqnarray}
R_{\alpha\beta}&=&
\wtd\Delta^{-a-b}[\half \wtd g^{\gamma\delta}\partial^\rho
\wtd g_{\alpha\gamma}
\partial_\rho \wtd g_{\beta\delta}-\ft14\kappa\wtd\Delta^{-1}\partial^\rho
\wtd g_{\alpha\beta}\partial_\rho\wtd\Delta-\half\nabla^2\wtd g_{\alpha\beta}]
\nonumber\\
&&+\wtd g_{\alpha\beta}\wtd\Delta^{-a-b}[\ft{b}{4}(\kappa-2)\wtd\Delta^{-2}
\partial^\rho\wtd\Delta\partial_\rho\wtd\Delta+\ft{b}{2}\wtd\Delta^{-1}
\nabla^2\wtd\Delta]\nonumber\\
&&+\wtd R_{\alpha\beta}-\half(\kappa+2a+2b-1)\wtd\Delta^{-1}
\wtd\nabla_\alpha\wtd\nabla_\beta\wtd\Delta
-\mu_i^{-1}\wtd\nabla_\alpha\wtd\nabla_\beta\mu_i\\
&&-\ft14((b-2)\kappa+a(a+b)d
+(b-2)(2a+2b-1))\wtd\Delta^{-2}\partial_\alpha\wtd\Delta\partial_\beta
\wtd\Delta \nonumber\\
&&+\wtd g_{\alpha\beta}[\ft{b}{2}\wtd\Delta^{-1}\wtd\nabla^2\wtd\Delta
+\ft{b}{4}(\kappa+2a+2b-3)\wtd\Delta^{-2}\partial^\gamma\wtd\Delta
\partial_\gamma\wtd\Delta+\ft{b}{2}\wtd\Delta^{-1}\mu_i^{-1}\partial^\gamma
\mu_i\partial_\gamma\wtd\Delta] \ .\nonumber
\end{eqnarray}
Note that $\wtd R_{\alpha\beta}$ as well as the covariant derivatives
$\wtd\nabla_\alpha$ are defined with respect to the $(k-1)$-dimensional metric
$d\wtd s^2 = \sum_{i=1}^kX_i^{-1}d\mu_i^2$.
While these expressions are rather unwieldy, they simplify considerably in
both the $S^5$ and the $S^7$ reductions, as many of the coefficients take
on simple values.

Finally, by taking the trace of the above, we find the expression for the
$D$-dimensional curvature scalar
\begin{eqnarray}
R&=& \wtd\Delta^{-a}[R^{(d)}
-(\kappa+a)\nabla^\rho(\wtd\Delta^{-1}\nabla_\rho\wtd\Delta)
+\ft14(\partial^\rho \wtd g_{\alpha\beta}\partial_\rho \wtd g^{\alpha\beta}
-X_i^{-2}\partial^\rho X_i\partial_\rho X_i)\nonumber\\
&&\qquad+\ft14(-(\kappa+a)(\kappa-1)+2b-(a+b)b(2k-1))
\wtd\Delta^{-2}\partial^\rho\wtd\Delta\partial_\rho\wtd\Delta] \nonumber\\
&&+\wtd\Delta^b[\wtd R
-(\kappa+2a+b-1)(\wtd\Delta^{-1}\wtd\nabla^2\wtd\Delta
+\wtd\Delta^{-1}\mu_i^{-1}\partial^\alpha\mu_i\partial_\alpha\wtd\Delta)
\nonumber\\
&&\qquad+\ft14(-(\kappa+2a+b-1)(\kappa+2a+2b-5)-a(a+b)d)
\wtd\Delta^{-2}\partial^\alpha\wtd\Delta \partial_\alpha\wtd\Delta
\nonumber\\
&&\qquad-2\mu_i^{-1}\wtd\nabla^2\mu_i+\wtd\Delta^{-1}(\sum X)^2
-\wtd\Delta^{-1}\sum X^2]\ .
\end{eqnarray}
To make contact with the Kaluza-Klein reductions, we note that
explicit computation of $\wtd R$ yields
\begin{equation}
\wtd R=\wtd\Delta^{-1}[2\wtd\Delta^{-1}\sum X^3\mu^2-2\wtd\Delta^{-1}\sum X
\sum X^2\mu^2+(\sum X)^2-\sum X^2]\ ,
\end{equation}
where we have followed a shorthand notation of removing indices so that,
$\sum X^3\mu^2\equiv\sum_{i=1}^kX_i^3\mu_i^2$, for example.  Note that for the
special case of $k=3$, corresponding to $S^5$, not all of the above
quantities are independent.  As a result we find that this expression
simplifies to yield $\wtd R_{(k=3)}=2\wtd\Delta^{-2} X_1X_2X_3
=2\wtd\Delta^{-2}$.  Additionally, we often find the following identities
useful:
\begin{eqnarray}
\partial^\alpha\wtd\Delta\partial_\alpha\wtd\Delta
&=&-4[\wtd\Delta^{-1}(\sum X^2\mu^2)^2-\sum X^3\mu^2]\ ,\nonumber\\
\wtd\nabla^2\wtd\Delta&=&2[\wtd\Delta^{-2}(\sum X^2\mu^2)^2-\wtd\Delta^{-1}
\sum X^3\mu^2-\wtd\Delta^{-1}\sum X\sum X^2\mu^2+\sum X^2]\ ,\nonumber\\
\mu_i^{-1}\partial^\alpha\mu_i\partial_\alpha\wtd\Delta&=&-2[\wtd\Delta^{-1}
\sum X\sum X^2\mu^2-\sum X^2]\ ,\nonumber\\
\mu_i^{-1}\wtd\nabla^2\mu_i&=&\wtd\Delta^{-1}[
\wtd\Delta^{-1}\sum X\sum X^2\mu^2-(\sum X)^2]\ .
\end{eqnarray}

The $S^5$ reduction of Type $IIB$ supergravity discussed in Section 
\ref{S5}
corresponds to the choice of $d=5$ and $a=b=\half$.  In this case we obtain
\begin{eqnarray}
\wtd\Delta^{1/2}R_{(k=3)}^{(D=10)}\!\!\!&=&\!\!R^{(5)}
-\half\nabla^\rho(\wtd\Delta^{-1}
\nabla_\rho\wtd\Delta) -\ft14(-\partial^\rho\wtd g_{\alpha\beta}
\partial_\rho\wtd g^{\alpha\beta}
+X_i^{-2}\partial^\rho X_i\partial_\rho X_i
+\wtd\Delta^{-2}\partial^\rho\wtd\Delta\partial_\rho\wtd\Delta)\nonumber\\
&&+2(\wtd\Delta^{-1}+3\sum X^{-1})\ ,
\end{eqnarray}
where we have used the simplified expression for $\wtd R_{(k=3)}$ given
above.
On the other hand, the $S^7$ reduction of eleven dimensional supergravity,
given by the line element (\ref{s7metred}), corresponds to the choice of
$d=4$ and $a={2\over3}$, $b={1\over3}$.  The eleven-dimensional curvature
scalar is
\begin{eqnarray}
\wtd\Delta^{2/3}R_{(k=4)}^{(D=11)}\!\!\!&=&\!\!R^{(4)}
-\ft23\nabla^\rho(\wtd\Delta^{-1}
\nabla_\rho\wtd\Delta) -\ft14(-\partial^\rho\wtd g_{\alpha\beta}
\partial_\rho\wtd g^{\alpha\beta}
+X_i^{-2}\partial^\rho X_i\partial_\rho X_i
+\wtd\Delta^{-2}\partial^\rho\wtd\Delta\partial_\rho\wtd\Delta)\nonumber\\
&&-\ft23\wtd\Delta^{-2}(\sum X^2\mu^2)^2+\ft83\wtd\Delta^{-1}\sum X^3\mu^2
-\ft43\wtd\Delta^{-1}\sum X\sum X^2\mu^2\nonumber\\
&&+4(\sum X)^2-\ft{14}{3}\sum X^2\ .
\end{eqnarray}
The last two lines involve undifferentiated scalars, and is used in
(\ref{ricci1}).  Curiously, the scalar kinetic terms in both cases have an
identical structure save for a total derivative, and take on a standard
Kaluza-Klein appearance (since $X_i^{-2}\partial^\rho X_i\partial_\rho
X_i = - \partial^\rho\hat g_{ij}\partial_\rho\hat g^{ij}$).  Finally note
that the implicitly defined term
$\partial^\rho\wtd g_{\alpha\beta}\partial_\rho\wtd g^{\alpha\beta}$
may be evaluated to give
\begin{equation}
-\partial^\rho\wtd g_{\alpha\beta}\partial_\rho\wtd g^{\alpha\beta}
= X_i^{-2}\partial^\rho X_i\partial_\rho X_i
+\wtd\Delta^{-2}\partial^\rho\wtd\Delta\partial_\rho\wtd\Delta
-2\wtd\Delta^{-1}X_i^{-1}\partial^\rho X_i\partial_\rho X_i\mu_i^2\ .
\end{equation}

\item

The general expression for a rotating $p$-brane carrying a single
charge is given in \ref{Solutions5}.  Following the procedure in the
previous sections, we may take the limit of large $p$-brane charge, by
performing the rescalings \bea &&m\longrightarrow \ep^{\tilde d} m\
,\qquad \sinh\b \longrightarrow \ep^{-{{\tilde d}\over 2}}\, \sinh\b\
,\nn\\ &&r\longrightarrow \ep\, r\ , \qquad x^\mu\longrightarrow
\ep^{1-\td d/2}\, x^\mu\ , \qquad \ell_i \rightarrow \ep\, \ell_i\ ,
\eea and then sending $\ep$ to zero.  We find that the metric becomes
$ds^2 =  \ep^{\alpha^2/2}\, d\td s^2$, where $\alpha$ is given by
(\ref{alpha}) and the metric $d\td s^2$ is given by
\bea
d\td s^2_{D} &=& \wtd \Delta^{\ft{\td d}{D-2}}\, e^{\td d\varphi}\,
\Big[-(H_1\cdots H_N)^{-\ft{d-2}{d-1}}\, f\, dt^2\nn\\
&& + (H_1\cdots H_N)^{\ft{1}{d-1}}\Big(
(\fft{\td d}{d}\,g\rho)^{\ft{(D-2)\alpha^2}{2\td d}}\, f^{-1}\,  d\rho^2 +
\rho^2\, d\vec y\cdot d\vec y\Big)\Big]\nn\\
&&+g^{-2}\, \wtd \Delta^{-\ft{d}{D-2}}\, e^{-d\varphi}\,
\sum_{i=1}^N X_i^{-1}\, \Big(d\mu_i^2 + \mu_i^2 (d\phi_i +
g\, A^i)^2\Big)\ ,
\eea
where
\bea
&&g\,\rho =(d/\td d)\, (g\, r)^{\td d/d}\ ,\qquad \vec y= g\, (\td
d/d)\, \vec x\ ,\nn\\
&& g^{-\td d}=2m\, \sinh^2\b\ ,\qquad \mu=2m\, (d/\td d)^{d-2}
\, g^{2+\td d-d}\ ,
\eea
and
\bea
&&f=-\fft{\mu}{\rho^{d-2}} + (\td d/d)^2 g^2\, \rho^2\, (H_1\cdots
H_N)\ ,\qquad
X_i=(H_1\cdots H_N)^{\ft{d}{(d-1)\td d}}\, H_i^{-1}\ ,\nn\\
&&\wtd \Delta = \fft{\sum_i X_i\,\mu_i^2}{(X_1\cdots X_N)^2}\ ,\qquad
e^{-\ft{2\td d}{\alpha^2}\varphi} = (\td d/d)\, g\, \rho\ .\qquad
A^i=\fft{1-H_i^{-1}}{g\, \ell_i\, \sinh\b}\, dt\ .
\eea
It follows that the $(d+1)$-dimensional metric becomes
\be
ds_{d+1}^2 =-(H_1\cdots H_N)^{-\ft{d-2}{d-1}}\, f\, dt^2 +
(H_1\cdots H_N)^{\ft{1}{d-1}}\Big(
e^{-(D-2)\, \varphi}\, f^{-1} \, d\rho^2 +
\rho^2\, d\vec y\cdot d\vec y\Big)\ .
\ee
The Einstein-frame metric is given by $ds_E^2 = e^{-\fft{(D-2)}{(d-1)}
\varphi}\, ds_{d+1}^2$. 
This is the metric of an $N$-charge black hole in a domain-wall background.
In the case when $a=0$, the domain wall specialises to $AdS_{d+1}$.

\end{enumerate}

\section{\bf ACKNOWLEDGEMENTS}

Much of the recent work on $AdS$ black holes described in these lectures was carried out in collaboration 
with my colleagues Miriam Cvetic, Parid Hoxha, Jim Liu, Hong L\"{u}, 
Jianxin Lu, Rene Martinez Acosta, Chris Pope, Hisham Sati and Tuan 
Tran. In writing these lectures, I would especially like to thank Jim Liu and 
Jianxin Lu for their help and Arthur Greenspoon for a careful reading 
of the manuscript. Conversations with Ergin Sezgin on singletons and related 
topics are also gratefully acknowledged. I would also like to thank the organizers of the TASI Summer School, 
K. T. Mahanthappa, Eva Silverstein, Shamit Kachru and Jeff Harvey and the 
organizers of the Banff Summer School, Yvan Saint-Aubin and Luc Vinet for their 
hospitality.

\section{\bf APPENDICES}

\appendix

\section{The Lagrangian, symmetries and transformation rules of D=11 
supergravity}
\la{D=11rules}

The unique supermultiplet in $D=11$ consists of a graviton described 
by the elfbein $e_{M}{}^{A}$, a gravitino described by the Rarita-Schwinger 
vector-spinor $\Psi_{M}$ and the $3$-form gauge field $A_{MNP}$.  Here 
letters at the beginning of the alphabet are local Lorentz indices and 
letters from the middle of the alphabet are spacetime world indices. The 
Lagrangian is given by 
\[
{\cal L}_{11}= 
\frac{1}{2\kappa_{11}е^{2}} e \left[ R-\frac{1}{2.4!}F^{MNPQ}F_{MNPQ}\right]
  -\frac{1}{12\kappa_{11}е^{2}} \frac{1}{3!4!^{2}}\epsilon^{M_{1}\ldots M_{11}}
 A_{M_{1}M_{2}M_{3}}F_{M_{4}M_{5}M_{6}M_{7}}F_{M_{8}M_{9}M_{10}M_{11}}
\]
\be
+\frac{1}{2\kappa_{11}е^{2}} e \left[-
 2i{\bar \Psi}_{M}\Gamma^{MNP}D_{N}(\frac{\omega+{\hat \omega}}{2})\Psi_{P} +
 \frac{i}{96}({\bar 
 \Psi}_{M}\Gamma^{MNPQRS}\Psi_{N}+12{\bar \Psi}^{P}е\Gamma^{QR}\Psi_{S})
 (F_{PQRS}+ {\hat F_{PQRS}}) \right]
\la{D11action}
\ee
where
\be
{\bar \Psi}= \Psi^{\dagger}е\Gamma^{0}е
\ee
and $\Psi$ is a Majorana spinor obeying
\be
{\bar \Psi}= \Psi^{T}еC^{-1}е
\ee
where $C$ is the charge conjugation matrix obeying
\be
C^{-1}е\Gamma_{A}еC=-\Gamma_{A}е{}^{T}е
\ee
and $\Gamma_{A}е$ are the $D=11$ Dirac matrices
\be
\{\Gamma^{A},\Gamma^{B}\}=2\eta^{AB}
\ee
\be
\eta^{AB}=diag(-,+,+,\ldots+)
\ee
Furthermore,
\be
e=det~{e_{M}{}^{A}}
\ee
\be
F_{MNPQ}=4\partial_{[M}A_{NPQ]}
\ee
\be
{\hat F}_{MNPQ}е=F_{MNPQ}е-3{\bar \Psi}_{[M}е\Gamma_{NP}е\Psi_{Q]}е
\ee
\be
D_{M}(\omega)\Psi_{N}=\partial_{M}\Psi_{N}+\frac{1}{4}\omega_{NAB}\Gamma^{AB}
\ee
\be
{\hat \omega}_{NAB}е=\omega_{NAB}е-\frac{i}{4}{\bar \Psi}_{P}\Gamma_{NAB}{}^{PQ}\Psi_{Q}е
\ee
\be
\omega_{NAB}е=\omega^{0}{}_{NAB}(e)+
\frac{i}{4}\left[{\bar \Psi}_{P}\Gamma_{NAB}{}^{PQ}\Psi_{Q}
-2({\bar \Psi}_{N}е\Gamma_{B}\Psi_{A}е
  -{\bar \Psi}_{N}е\Gamma_{A}\Psi_{B}е
  +{\bar \Psi}_{B}е\Gamma_{N}\Psi_{A}е)\right]
\ee
\be
\omega^{0}{}_{NAB}(e)=e^{N}е{}_{A}e^{P}е{}_{B}е(\Omega_{PMN}е+\Omega_{MNP}е
-\Omega_{NPM}е)
\ee
where $\Omega_{MNP}$ are the anholonomy coefficients
\be
[\partial_{A}е,\partial_{B}е]=
[e^{M}е{}_{A}е\partial_{M}е,e^{N}е{}_{B}е\partial_{N}е]=
\Omega_{AB}е{}^{C}е\partial_{C}е
\la{anholo}
\ee
and $e^{M}е{}_{A}$ is the inverse of $e_{M}е{}^{A}е$:
\be
e_{M}е{}^{A}еe^{N}е{}^{B}е\eta_{AB}е=\delta_{M}е{}^{N}е
\ee

The action is invariant under the following symmetry 
transformations:

a) $D=11$ general coordinate transformations with parameter 
$\xi^{M}е$:
\[
\delta 
e_{M}е{}^{A}е=e_{N}е{}^{A}е\partial_{M}е\xi^{N}е+\xi^{N}е\partial_{N}еe_{M}е{A}
\]
\[
\delta 
\Psi_{M}е=\Psi_{N}е\partial_{M}е\xi^{N}е+\xi^{N}е\partial_{N}е\Psi_{M}е
\]
\be
\delta 
A_{MNP}е=3A_{Q[MN}\partial_{P]}е\xi^{Q}е+\xi^{Q}е\partial_{Q}еA_{MNP}е
\la{susytransforms}
\ee
b) Local $SO(1,10)$ Lorentz transformations with parameter 
$\alpha_{AB}е=-\alpha_{BA}е$:
\[
\delta e_{M}е{}^{A}е=-e_{M}е{}^{B}е\alpha_{B}е{}^{A}е
\]
\[
\delta \Psi_{M}е=-\frac{1}{4}\alpha_{AB}е\Gamma^{AB}е\Psi_{M}е
\]
\be
\delta A_{MNP}е=0
\ee
c) $N=1$ supersymmetry transformations with anticommuting parameter 
$\epsilon$:
\[
\delta e_{M}{}^{A}=i{\bar \epsilon}\Gamma^{A}\Psi_{M}е
\]
\[
\delta \Psi_{M}е=D_{M}е({\hat \omega})\epsilon-
\frac{1}{288}(\Gamma_{M}е{}^{PQRS}е-8\delta_{M}е{}^{P}е\Gamma^{QRS}е)
{\hat F}_{PQRS}е\epsilon
\]
\be
\delta A_{MNP}е=3i{\bar \epsilon}\Gamma_{[MN}е\Psi_{P]}е
\la{susytransform}
\ee
d) Abelian gauge transformations with parameter 
$\Lambda_{MN}е=-\Lambda_{NM}е$:
\[
\delta e_{M}е{}^{A}е=0
\]
\[
\delta \Psi_{M}е=0
\]
\be
\delta A_{MNP}е=\partial_{[M}е\Lambda_{NP]}е
\ee
e) an odd number of space or time reflections together with
\be
A_{MNP}е \rightarrow -A_{MNP}е
\ee

\section{The field equations, symmetries and transformation rules 
of Type IIB supergravity}\la{TypeIIB}

Next we consider Type $IIB$ supergravity in $D = 10$ 
\cite{Schwarz,Howewest} 
which also describes 128 + 128 degrees of freedom, and corresponds to 
the field-theory limit of the Type $IIB$ superstring. The spectrum of 
the supergravity theory consists of a complex scalar $A$, a complex 
spinor $\lambda$, a complex 2-form $A_{MN}е$, a complex Weyl 
gravitino $\psi_{M}е$, a real graviton $e_{M}е^{R}$ and a real 
4-form $A_{MNPQ}$ whose 5-form field strength 
$F_{MNPQR}$ obeys a self-duality condition. The complex Weyl spinors 
$\psi_{M}е$ and $\lambda$ have opposite handedness
\be
\gamma_{11}е\psi_{M}е=-\psi_{M}е
\ee
\be
\gamma_{11}е\lambda=\lambda
\ee
Owing to the self-duality of $F_{MNPQR}$, there exists no covariant action principle.
It is therefore simplest to work directly with the field equations, which are:
\be
D^{M}еP_{M}е=\frac{1}{24}\kappa_{10}е^{2}еG_{MNP}еG^{MNP}е
+O(\psi^{2}е)
\ee
\be
\gamma^{M}еD_{M}е\lambda=\frac{1}{240}i\kappa_{10}е\gamma^{P_{1}е\ldots P_{5}е}
\lambda F_{P_{1}е\ldots P_{5}е}е+O(\psi^{2}е)
\ee
\be
D^{P}еG_{MNP}е=P^{P}еG^{*}е_{MNP}е-\frac{2}{3}i\kappa_{10}еF_{MNPQR}еG^{*}е{}^{PQR}е
+O(\psi^{2}е)
\ee
\be
\gamma^{MNP}еD_{M}е\psi_{P}е=-i\frac{1}{2}\gamma^{P}е\gamma^{M}е
\lambda^{*}еP_{P}е-\frac{1}{48}i\kappa_{10}е\gamma^{NPQ}е\gamma^{M}е\lambda 
G^{*}е{}_{NPQ}е+O(\psi^{3}е)
\ee
\[
R_{MP}е-\frac{1}{2}g_{MP}еR=
P_{M}еP^{*}е_{P}е+P^{*}е_{M}еP_{P}е-g_{MP}еP^{R}еP^{*}е_{R}е
+\frac{1}{6}\kappa_{10}е^{2}еF_{R_{1}е\ldots R_{4}еM}еF^{R_{1}е\ldots R_{4}е}е_{P}е
\]
\be
+\frac{1}{8}\kappa_{10}е^{2}е(G_{M}е^{RS}еG^{*}{}е_{PRS}е
+G^{*}{}е_{M}е^{RS}еG_{PRS}е)
-\frac{1}{24}\kappa_{10}е^{2}еg_{MP}еG^{RST}еG^{*}е{}_{RST}
+O(\psi^{2}е)
\ee
\be
F_{MNPQR}е=*F_{MNPQR}е
\ee
where
\be 
P_{M}е=f^{2}е\partial_{M}еB
\ee
\be
G_{MNP}е=f(F_{MNP}-BF^{*}е{}_{MNP}е)
\ee
\be
F_{MNP}е=3\partial_{[M}еA_{NP]}е
\ee
\be
f=(1-B^{*}еB)^{-1/2}е
\ee
\be
F_{MNPQR}е=5\partial_{[M}еA_{NPQR]}е-
\frac{5}{4}\kappa_{10}еIm(A_{[MN}еF^{*}е{}_{PQR]}е)
\ee
and where the supercovariant derivative $D_{M}е$ involves the 
composite $U(1)$ gauge field 
\be
Q_{M}е=f^{2}еIm(B\partial_{M}еB^{*}е).
\ee
Our notation is that $X^{*}е$ is the complex conjugate of $X$ while $*X$ 
is the Hodge dual of $X$.

We also have the identities
\be
D_{[M}еP_{P]}е=0
\ee
\be
D_{[M}еG_{NPQ]}е=-P_{[M}еG^{*}{}е^{NPQ]}е
\ee
\be
\partial_{[P}еQ_{M]}е=-iP_{[P}еP^{*}е_{M]}е
\ee
\be
\partial_{[N}еF_{M_{1}е\ldots M_{5}]е}=\frac{5}{12}i\kappa 
G_{[NM_{1}еM_{2}е}еG^{*}е_{M_{3}еM_{4}еM_{5}е]}е
\ee
The symmetries are $D=10$ general covariance, local $SO(1,9)$ Lorentz 
transformations, local supersymmetry with complex Weyl spinor 
parameter $\epsilon$
\be
\gamma_{11}е\epsilon=-\epsilon
\ee
and a global $SL(2,R)$ invariance.
The supersymmetry transformation rules are
\be
\delta B=\kappa_{10}еf^{-2}е{\bar \epsilon}^{*}е\lambda
\ee
\be
\delta A_{MN}е=f({\bar \epsilon}\gamma_{MN}е\lambda
+4i{\bar \epsilon}^{*}е\gamma_{[M}е\psi_{N]}е+B{\bar 
\epsilon}^{*}е\gamma_{MN}е\lambda+4iB{\bar 
\epsilon}\gamma_{[M}е\psi^{*}е_{N]}е)
\ee
\be
\delta \lambda=\frac{i}{\kappa_{10}е} \gamma^{M}е\epsilon^{*}е{\hat P}_{M}е
-\frac{1}{24}i\gamma^{MNP}е\epsilon {\hat G}_{MNP}е
\ee
\[
\delta \psi_{M}е= \frac{1}{\kappa_{10}}еD_{M}е\epsilon
+\frac{1}{480}i\gamma^{P_{1}е\ldots P_{5}е}е\gamma_{M}е\epsilon {\hat 
F}_{P_{1}е\ldots P_{5}е}е
+\frac{1}{96}(\gamma_{M}е{}^{NPQ}е{\hat G}_{NPQ}е-9\gamma^{PQ}е{\hat 
G}_{MPQ}е)\epsilon^{*}е
\]
\[
-\frac{7}{16}\kappa_{10}е\left(\gamma_{P}е\lambda {\bar 
\psi}_{M}е\gamma^{P}е\epsilon^{*}е-\frac{1}{1680}\gamma_{{P_{1}е\ldots 
P_{5}е}е}\lambda {\bar \psi}_{M}е\gamma^{P_{1}\ldots 
P_{5}е}е\epsilon^{*}е\right)
\]
\[
+\frac{1}{32}i\kappa_{10}е[\left 
(\frac{9}{4}\gamma_{M}е\gamma^{P}е
+3\gamma^{P}е\gamma_{M}е\right)\epsilon {\bar 
\lambda}\gamma_{P}е\lambda
\]
\be
-\left(\frac{1}{24}\gamma_{P}е\gamma^{P_{1}еP_{2}еP_{3}е}е
+\frac{1}{6}\gamma^{P_{1}еP_{2}еP_{3}е}е\gamma_{M}е\right)\epsilon 
{\bar \lambda}\gamma_{P_{1}еP_{2}еP_{3}е}е\lambda
+\frac{1}{960}\gamma_{M}е\gamma^{P_{1}е\ldots P_{5}е}е\epsilon 
{\bar \lambda}\gamma_{P_{1\ldots P_{5}е}е}е\lambda]
\ee
\be
\delta e_{M}е{}^{R}е=-2\kappa_{10}еIm({\bar 
\epsilon}\gamma^{R}е\psi_{N}е)
\ee
\be
\delta A_{MNPQ}е=2Re({\hat \epsilon}\gamma_{MNP}е\psi_{Q]е})
-\frac{3}{8}\kappa_{10}е(A_{[MN}е \delta A^{*}е_{PQ]}е-A^{*}е_{[MN}е \delta A_{PQ]}е
\ee
where
\be
{\hat P}_{M}е=P_{M}е-\kappa_{10}е{}^{2}е{\bar \psi}^{*}е_{M}е\lambda
\ee
\be
{\hat G}_{MNP}е=G_{MNP}е-3\kappa_{10}е{\bar 
\psi}_{[M}е\gamma_{NP]}е\lambda -6i\kappa_{10}е{\bar 
\psi}^{*}е_{[M}е\gamma_{N}е\psi_{P]}е
\ee
and
\be
{\hat F}_{MNPQR}е=F_{MNPQR}е-5\kappa_{10}е{\bar 
\psi}_{[M}е\gamma_{NPQ}е\psi_{R]}е-\frac{1}{16}\kappa_{10}е{\bar 
\lambda}\gamma_{MNPQR}е\lambda
\ee
and where 
\be
D_{M}е\epsilon=(\partial_{M}е + \frac{1}{4} 
\omega_{M}е^{RS}е\gamma_{RS}е-\frac{1}{2}iQ_{M}е)\epsilon
\ee
where
\be
\omega_{MNP}= \Omega_{NPM}е+\Omega_{NMP}е+\Omega_{MPN}е
\ee
and
\be
\Omega_{MNP}е=e^{R}е_{P}е\partial_{[M}еe_{N]}{}_{R}е
+\kappa_{10}е^{2}еIm({\bar \psi}_{M}е\gamma_{P}е\gamma_{N}е) 
\ee
The fields transform under $SL(2,R)$ with parameters $\alpha$ and 
$\gamma$ as
\be
\delta B=\alpha +2i\gamma B -\alpha^{*}еB^{2}е
\ee
\be
\delta \lambda =\frac{3}{2}i[\gamma +Im(\alpha B^{*}е)]\lambda
\ee
\be
\delta \psi_{M}е= \frac{1}{2}i[\gamma +Im(\alpha B^{*}е)]\psi_{M}
\ee
\be
\delta A_{MN}е=i\gamma A_{MN}е + \alpha A^{*}е_{MN}е
\ee
\be
\delta e_{M}е^{R}е=0
\ee
\be
\delta F_{MNPQR}е=0
\ee

\section{The Lagrangian, symmetries and transformation rules of the 
M2-brane}
\la{M2rules}

Let us introduce the coordinates $Z^M$ of a curved superspace
\begin{equation}
Z^M=(x^{\mu}, \theta^{\alpha})
\end{equation}
and the supervielbein $E_M{}^A (Z)$ where $M = \mu,\alpha$ are world
indices and $A = a,\alpha$ are tangent space indices.  We also define the
pull-back
\begin{equation}
E_i{}^A = \partial_iZ^ME_M{}^A~~.
\end{equation}
We also need the super $d$-form $A_{CBA}(Z)$. 
Then the supermembrane action is \cite{BST1,BST2} 
\begin{eqnarray}
S= T_2\int
d^3\xi\biggl\lbrack &-&\frac{1}{2}{\sqrt -\gamma}\gamma^{ij}
E_i{}^aE_j{}^b \eta_{ab}+\frac{1}{2}{\sqrt -\gamma}
\nonumber \\
&+&\frac{1}{3!}\epsilon^{i_1\ldots
i_d}E_{i}{}^{A}E_{j}{}^{B}E_{k}{}^{C}A_{CBA}
\biggr\rbrack.
\la{supermembrane}
\end{eqnarray}
Note that there is a kinetic term, a worldvolume cosmological term, and
a Wess-Zumino term.  
The target-space symmetries are superdiffeomorphisms, Lorentz invariance and
$d$-form gauge invariance.  The worldvolume symmetries are ordinary
diffeomorphisms and kappa invariance which we now examine in
more detail.  The transformation rules are

\begin{equation}
\delta Z^M E^a{}_M = 0, ~~~\delta Z^ME^{\alpha}{}_M =
\kappa^{\beta}(1+\Gamma)^{\alpha}{}_{\beta}
\la{kappatransf}
\end{equation}
where $\kappa^{\beta}(\xi)$ is an anticommuting spacetime spinor but
worldvolume scalar, and where

\begin{equation}
\Gamma^{\alpha}{}_{\beta} = \frac{1}{3!{\sqrt
-\gamma}}\epsilon^{ijk}E_{i}{}^{a}
E_{j}{}^{b}\ldots E_{k}{}^{c}\Gamma_{abc}~~.
\la{Gamma2}
\end{equation}
Here $\Gamma_a$ are the Dirac matrices in spacetime and 
\begin{equation}
\Gamma_{a_{1}..a_{d}}=\Gamma_{[a_1{\cdots}a_{d}]}~~.
\end{equation}
The matrix $\Gamma$ of (\ref{Gamma2}) is traceless and satisfies

\begin{equation}
\Gamma^2 = 1
\end{equation}
\noindent
when the equations of motion are satisfied and hence the matrices
$(1\pm\Gamma)/2$ act as projection operators.  The transformation rule
(\ref{kappatransf}) therefore permits us to gauge away one half of the fermion 
degrees of freedom.  As desribed below, this gives rise to a matching of physical 
boson and fermion degrees of freedom on the worldvolume.  

The kappa symmetry is achieved only if certain constraints on the antisymmetric
tensor field strength $F_{MNP..Q}$(Z) and the supertorsion are
satisfied.  One can show that the constraints on the background fields 
$E_M{}^A$ and $A_{MNP}$ are nothing but the equations of motion of 
eleven-dimensional supergravity.  

\section{The field equations, symmetries and transformation rules of 
the M5-brane}
\la{M5rules}

The $M5$-brane equations are elegantly derived from the superembedding 
formalism which treats the brane as a supermanifold $M$ embedded in a 
larger spacetime supermanifold $\unM$. In this subsection, therefore, we use the 
notations and conventions of \cite{hs1,Sezginsundell}. In
particular, we denote by $z^{\unM}=(x^{\unm},\theta^{\um})$ the local
coordinates on $\unM$, and $A=(a,\a)$ is the target tangent space
index. We use the ununderlined version of these indices to label the
corresponding quantities on the worldvolume. The embedded submanifold
$M$, with local coordinates $y^M$, is given as $z^{\unM}(y)$.

The suitable pullbacks of the super 3-form and the induced metric are 
\bea
A_{ijk} &=& \del_i z^{\unM}\del_j z^{\unN}\del_k z^{\unP}
A_{\unP\unN\unM}\ ,
\nn\w2
\gamma_{ij} &=& \left(\del_i z^{\unM}E_{\unM}{}^{\una}\right)
\left(\del_j z^{\unN}E_{\unN}{}^{\unb}\right)
\eta_{\una\unb}\ ,\la{cbg}
\eea
where $\eta_{\una\unb}$ is the Minkowski metric in eleven dimensions and
$E_{\unM}{}^{\unA}$ is the target space supervielbein. We define the basis 
one-forms $E^{\unA} =
d\xi^i\del_iz^{\unM}E_{\unM}{}^{\unA}$ and
$E^A = d\s^r \del_r y^M E_M{}^A$, where $E_M{}^A$ is the supervielbein on
$M_5$. The embedding matrix $E_A{}^{\unA}$ plays an important role in the
description of the model, and it is defined as
\be
E_A{}^{\unA} \equiv E_A{}^M\del_{M}z^{\unM}E_{\unM}{}^{\unA},
\ee
Here, we give the nonlinear field equations of the superfivebrane
equations, up to second order fermionic terms, that follow from the
superembedding condition $E_\a{}^{\una}=0$, which are proposed to arise
equally well from the ${\cF}$-constraint $\cF_{\a BC}=0$, where we have 
introduced the super $2$-form $A_{2}е$ and the following super $3$-form 
in $M_5$:
\be
\cF_3\equiv dA_2-f_5^{*} A_3\ .
\la{cf3a}
\ee
and where $f_5^*$ is the pullback associated with the embedding map
$f_5:M_5\hookrightarrow \unM$.  The details
of the procedures can be found in \cite{hsw1,Sezginsundell}. A key point is the
emergence of a super $3$-form $h$ in world superspace. This form arises
in the following component of the embedding matrix
\be
E_{\a}{}^{\ua} \se u_{\a}{}^{\ua}+h_{\a}{}^{\b'} u_{\b'}{}^{\ua}\ ,
\la{s}
\ee
where, upon the splitting of the indices to exhibit the $USp(4)$
R-symmetry group indices $i=1,...,4$, we have
\be
h_{\a}{}^{\b'}\ra h_{\a i \b}{}^{j}\se
{1\over 6} \d_i{}^j(\gamma^{abc})_{\a\b} h_{abc}\ ,
\ee
where $h_{abc}$ is a self-dual field defined on $M$. The pair
$(u_{\a}{}^{\ua},u_{\a'}{}^{\ua})$ make up an element of the group
$Spin(1,10)$.

The superembedding formalism was shown to give the following complete
$M5$-brane equations of motion:
\[
E_a{}^{\ua}E_{\ua}{}^{\b'}(\Gamma^a)_{\b'}{}^{\a}\se 0\ ,
\]
\[
\eta^{ab}\nabla_a E_b{}^{\una} E_{\una}{}^{b'}\se-\ft18 
(\Gamma^{b'a})_{\gamma'}{}^{\b}Z_{a\b}{}^{\gamma'}\ ,
\]
\be
\hat\nabla^c h_{abc}\se 
-\ft1{32}(\Gamma^c\Gamma_{ab})_{\gamma'}{}^{\b}Z_{c\b}{}^{\gamma'}\ ,
\la{se1}
\ee
where
\be
Z_{a\b}{}^{\gamma'}\se E_{\b}{}^{\ub}\left(E_{a}{}^{\una} T_{\una\ub}{}^{\uc}
-E_{a}{}^{\ud}E_{\ub}{}^{\gamma} (\nabla_{\gamma}E_{\ud}{}^{\d'})E_{\d'}{}^{\uc}
\right)E_{\uc}{}^{\gamma'} \ .
\la{z}
\ee
Recall that the inverse of the pair $(E_A{}^{\unA},E_{A'}{}^{\unA})$ is
denoted by $(E_{\unA}{}^A,E_{\unA}{}^{A'})$ and that $A=(a,\a)$ label
the tangential directions while $A'=(a',\a')$ label the normal
directions to the $M5$-brane worldvolume.

The non-vanishing parts of the target space torsion components 
$T_{\una\ub}{}^{\uc}$ are given by
\bea
T_{\ua\ub}{}^{\unc} &\se& -i(\Gamma^{\unc})_{\ua\ub}\ ,
\nn\w2
T_{\una\ub}{}^{\uc} &\se&-
{1\over36}(\Gamma^{\unb\unc\und})_{\ub}{}^{\uc}F_{\una\unb\unc\und}
-{1\over288}(\Gamma_{\una\unb\unc\und\une})_{\ub}{}^{\uc}
F^{\unb\unc\und\une}\ ,
\la{tt}
\eea
and $T_{\una\unb}{}^{\uc}$. The only other non-vanishing components of
$F_4$ are
\be
F_{\una\unb\uc\ud}\se -i(\Gamma_{\una\unb})_{\uc\ud}\ .
\la{h4}
\ee
The covariant derivative $\hat\nabla$ has an
additional, composite $SO(5,1)$ connection of the form $(\nabla
u)u^{-1}$ as explained in more detail in \cite{hsw1}.

The $M5$-brane equations of motion (\ref{se1}) live in superspace
\cite{hs1,hs2}. The component (i.e. Green-Schwarz) form of these
equations has also been worked out \cite{hsw1}. Up to fermionic
bilinears, the final result is:
\[
{\cE}_a(1-\Gamma)\gamma^b m_b{}^a \se 0\ ,
\]
\[
G^{mn}\nabla_m
{\cF}_{npq}\se Q^{-1}\left[4Y-2(mY+Ym)+mYm\right]_{pq}\ ,
\]
\be
G^{mn}\nabla_m{\cE}_n{}^{\unc} \se {Q\over \sqrt{-g}} \e^{m_1\cdots m_6
}\left(\ft1{6!}F^{\una}{}_{m_1\cdots m_6} + \ft1{(3!)^2}
F^{\una}{}_{m_1m_2m_3}\,{\cF}_{m_4m_5m_6}\, \right)P_{\una}{}^{\unc}\ .
\la{e1a}
\ee
Several definitions are in order. To begin with,
\bea
m_a{}^b \equiv \d_a{}^b-2 k_a{}^b\ ,\qquad
k_a{}^b \equiv h_{acd} h^{bcd}\ ,\qquad Q \equiv (1-\ft23 \tr\,k^2)\ ,
\nn\w2
Y_{ab} \equiv \left[4*F-2(m*F+*F m)+m*F m\right]_{ab}\ ,
\nn\w2
 P_{\una}{}^{\unc} \equiv
\d_{\una}{}^{\unc}-{\cE}_{\una}{}^m{\cE}_m{}^{\unc}\ ,
\quad\quad *F^{ab} \de \ft1{4!\sqrt{-g}}\e^{abcdef}F_{cdef}\ ,
\la{defsa}
\eea
The fields ${\cF}_{abc}$, $F_{\una_1\cdots\una_4}$ and its Hodge dual
$F_{\una_1\cdots\una_7}$ are the purely bosonic components of the
superforms
\be
{\cF}_3 \se dA_2-{\unC}_3\ , \qquad F_4 \se dA_3 \ , \qquad F_7 \se dA_6
+ \ft12\,A_3\wedge F_4\ .
\la{h7a}
\ee
The remaining nonvanishing component of $F_7$ is
\be
F_{\ua\ub\una\unb\unc\und\une}\se
-i(\Gamma _{\una\unb\unc\und\une})_{\ua\ub}\ .
\la{hh7}
\ee
The target space indices on $F_4$ and $F_7$ have been converted to
worldvolume indices with factors of ${\cE}_m{}^{\una}$ which are the
supersymmetric line elements defined as
\bea
{\cE}_m{}^{\una}(x) &\equiv& \del_m z^{\unM} E_{\unM}{}^{\una}\qquad
{\rm at}\ \theta=0\ ,\nn \\
{\cE}_m{}^{\ua}(x) &\equiv & \del_m z^{\unM}
E_{\unM}{}^{\ua}\qquad {\rm at}\ \theta=0\ .
\eea
The metric
\be
\gamma_{mn}(x) \equiv {\cE}_m{}^{\una}{\cE}_n{}^{\unb}\h_{\una\unb} \se
e_m{}^a e_n{}^b \eta_{ab}
\ee
is the standard $GS$ induced metric with determinant $g$, and $G^{mn}$
is another metric defined as
\be
G^{mn} \equiv (m^2)^{ab}e_a{}^m e_b{}^n\ .
 \la{gmna}
 \ee
Let us note that the connection in the covariant derivative $\nabla_m$
occurring in (\ref{e1}) is the Levi-Civita connection for the induced
metric $g_{mn}$ up to fermionic bilinears.

A key relation between $h_{abc}$ and ${\cF}_{abc}$ follows from the
dimension-$0$ components of the Bianchi identity $d{\cF}_3=-{\unF}_4$,
and is given by
\be
h_{abc}\se \ft14\,m_a{}^d {\cF}_{bcd}\ .
\la{hfa}
\ee
The matrix $\Gamma$ is the $\theta =0$ component of the matrix $\Gamma _{(5)}$
introduced above and it is given by
\be
\Gamma \se -{\bar {\Gamma}}+\ft13 h^{mnp}\Gamma _{mnp} \se
- \left[{\rm exp}~ (-\ft13\Gamma^{mnp}h_{mnp})\right] {\bar \Gamma}\ ,
\la{af}
\ee
where
\bea
&&{\bar \gamma} \equiv {1\over 6!\sqrt {-g}} \e^{m_1\cdots m_6}
\Gamma _{m_1\cdots m_6}\ ,
\la{c0}\w2
&& \Gamma _m\equiv {\cE}_m{}^{\una}\Gamma_{\una} \ ,
\qquad \Gamma^b\equiv\Gamma^m e_m{}^b\ , \qquad e_m{}^b \equiv E_m{}^a m_a{}^b\ .
\nn
\eea
The $\k$-symmetry transformation rules are
\bea
\d_{\k} z^{\una} &\se& 0 \ ,
\nn\w2
\d_{\k} z^{\ua} &\se&  \k^{\uc}\ft12 (1+\Gamma)_{\uc}{}^{\ua} \ ,
\nn\w2
\d_\k h_{abc} &\se& -\ft{i}{16} m_{d[a}\,{\cE}^d(1-\Gamma)\Gamma_{bc]} \k\ ,
\la{kt}
\eea
where $\Gamma$ is given by (\ref{af}). The $\k$-symmetry transformations are
the fermionic diffeomorphisms of the $M5$-brane worldvolume with
parameter $\k^\a=\k^{\ua}E_{\ua}{}^{\a}$. It
follows that
\be
\d_{\k} \cF_3\se \{d,i_\k\}\cF_3\se -i_\k \unF_4\ ,
\la{kcf}
\ee
which can also be verified by direct computation by combining (\ref{hf})
and (\ref{kt}).

The equations of motion (\ref{e1}) have been shown \cite{s} to be
equivalent to those which follow from an action with auxiliary scalar
field \cite{s2}.

We conclude this section by elucidating the consequences of the central
equation (\ref{hf}). To this end, we first note the useful identities
\bea
h_{abe}h^{cde}&\se&\d^{[c}_{[a}k_{b]}^{\ \;d]}\ ,
\nn\w2
k_{ac}k_b{}^c&\se&\ft16 \eta_{ab}\tr~k^2\ ,
\nn\w2 k_a{}^d
h_{bcd}&\se&k_{[a}{}^d h_{bc]d}\ ,
\la{hk1a}
\eea
which are consequences of the linear self-duality of $h_{abc}$. Taking
the Hodge dual of (\ref{hf}) one finds $*\cF_{abc} = -\cF_{abc} + 2
Q^{-1} m_a{}^d \cF_{bcd}$. Using the identity $m^2=2m-Q$, we readily
find the nonlinear self-duality equation
\be
*\cF_{mnp} \se Q^{-1}G_m{}^q \cF_{npq}\ .
\la{sda}
\ee
This equation can be expressed solely in terms of $\cF_3$. To do this, we
first insert (\ref{hf}) into (\ref{hk1}), which yields the identities
\bea
\cF_{abe}\cF^{cde} &\se& 2\delta_{[a}^{[c} X_{b]}^{\ \;d]} + \ft12
K^{-2} X_{[a}{}^{c} X_{b]}{}^{d} + 2(K^2-1) \delta_{[a}^{c}
\delta_{b]}^{d}\ ,
\nn\w2
X_{ac}X_b{}^c &\se& 4K^2(K^2-1)\eta_{ab}\ ,
\nn\w2
X_a{}^d \cF_{bcd} &\se& X_{[a}{}^d \cF_{bc]d}\ .
\la{xa}
\eea
where we have defined
\bea
K &\equiv & \sqrt{1+\ft1{24}\cF^{abc}\cF_{abc}}\ ,
\la{k}\w2
X_{ab} & \equiv & \ft12 K *\cF_a{}^{cd}\cF_{bcd}\ .
\la{xxa}
\eea
Next we derive the identities
\bea
Q(K+1) &\se& 2\ ,
\nn\w2
X_{ab} &\se&  \ft12 \cF_{acd}\cF_{b}{}^{cd} -
\ft1{12} \eta_{ab} \cF_{cde}\cF^{cde} \se 4 K(1+K) k_{ab}\ .
\eea
We can now express (\ref{sd}) entirely in terms of $\cF_3$
by deriving the identity
\be
Q^{-1}G_{mn} \se K \eta_{mn} - \ft12 K^{-1} X_{mn} \ .
\ee
Another way of writing (\ref{sd}) is
\be \cF^-_{abc}\se\ft12 (1+K)^{-2} \cF^+_{ade}\cF^{+def}\cF^+_{fbc}\ ,
\ee
where $K$ is a root of the quartic equation
\be
(K+1)^3(K-1)\se \ft1{24} \cF^{+ abc}\cF^{+}_{ade}\cF^{+def}\cF^+_{fbc}\ .
\ee

\section{The Lagrangian, symmetries and transformation rules of the 
D3-brane}
\la{D3rules}

Here we present a Lorentz invariant and supersymmetric worldvolume action for all Type $II$ Dirichlet 
$p$-branes, $p\le9$, in a general Type $II$ supergravity background 
\cite{Bergshoefftownsend}. The super 
Dirichlet $p$-brane action, in a general $N=2$ supergravity background
(for the string-frame metric) takes the form
\be
S_{d}е= S_{d}е_{DBI} + S_{d}е_{WZ} 
\la{introa}
\ee
where 
\be
S_{d}е_{DBI}= -T_{d}е\int d^{d}\xi\, e^{-\phi}\sqrt{-\det (\gamma_{ij} + 
{\cal F}_{ij})}
\la{introb}
\ee
is a Dirac-Born-Infeld type action and $I_{WZ}$ is a Wess-Zumino (WZ) type 
action to be discussed below; ${\cal F}_{ij}$ are the components of a
`modified' $2$-form field strength 
\be
{\cal F} = F-B\ ,
\la{introc}
\ee
where $F=dV$ is the usual field strength $2$-form of the BI field $V$ and $B$ is
the pullback to the worldvolume of a $2$-form potential $B$ on superspace, whose
leading component in a $\theta$-expansion is the $2$-form potential of
Neveu-Schwarz/Neveu-Schwarz (NS-NS) origin in Type $II$ superstring theory. We
use the same letter for superspace forms and their pullbacks to the
worldvolume since it should be clear from the context which is meant.
Superspace forms may be expanded in the coordinate basis of $1$-forms $dZ^M$ or
on the inertial frame basis $E^A= dZ^ME_M{}^A$, where $E_M{}^A$ is the
supervielbein. The basis $E^A$ decomposes under the action of the Lorentz group
into a Lorentz vector $E^a$ and a Lorentz spinor. The latter is a $32$-component
Majorana spinor for $IIA$ superspace and a pair of chiral Majorana spinors for
$IIB$ superspace. Thus
\be
(E^a, E^{\alpha\; I}), I=1,2
\la{introca}
\ee
for Type $IIB$.  We allow $\alpha$ to run from $1$ to $32$ but include a
chiral projector as appropriate. The worldvolume metric $g_{ij}$ appearing in 
(\ref{introb}) is defined in the standard way as 
\be
\gamma_{ij} = E_i^a E_j^b\eta_{ab}
\la{introd}
\ee
where $\eta$ is the $D=10$ Minkowski metric and
\be
E_i^A = \partial_i Z^M E_M{}^A\ .
\la{introe}
\ee

Thus $I_{DBI}$ is a straightforward extension to superspace of the corresponding
term in the bosonic action. The same is true for the WZ term. We introduce a 
R-R potential $A$ as a formal sum of $r$-form superspace potentials $A^{(r)}$, i.e.  
\be
A = \sum_{r=0}^{10} A^{(r)}\ .
\la{introf}
\ee
The even potentials are those of $IIB$ supergravity while the odd ones are those
of $IIA$ supergravity. In the bosonic case one could omit the $10$-form gauge
potential $F^{(10)}$ on the grounds that its $11$-form field strength is
identically zero. But an $11$-form field strength on {\it superspace} is not
identically zero; in fact we shall see that it is non-zero even
in a Minkowski background, a fact that is crucial to the $\kappa$-symmetry of
the super $9$-brane action.

The WZ term can now be written as 
\be
S_{d}е_{WZ} =T_{d}е\int_W A e^{\cal F}\  + \ mI_{CS}
\la{introg}
\ee
where, in the first term, the product is understood to be the exterior product of
forms and the form of appropriate degree is chosen in the `form-expansion'
of the integrand, i.e. $(p+1)$ for a Dirichlet $p$-brane. The $I_{CS}$ term is a
$(p+1)$-form Chern-Simons (CS) action that is present (for odd $p$) in a {\it
massive} $IIA$ background; its coefficient $m$ is the $IIA$ mass parameter.
This WZ term is formally the same as the known {\it bosonic} Dirichlet $p$-brane
WZ action, but  here the forms $C^{(r)}$ and $B$ are taken to be forms on {\it
superspace}, e.g.
\be
A^{(r)} = {1\over r!} dZ^{M_1}\cdots dZ^{M_r} A_{M_r\dots M_1}\ .
\la{introh}
\ee
This illustrates the standard normalization and the `reverse order' convention
for components of superspace forms. This convention goes hand in hand with the
convention for exterior differentiation of superspace forms in which the exterior
derivative acts `from the right'. Thus,
\be
dA^{(r)} = {1\over r!}dZ^{M_1}\cdots dZ^{M_r} dZ^N\partial_N A_{M_r\dots M_1}
\la{introi}
\ee

The field strength for the R-R field $A$ is
\be
F(A  ) = dA - HA + me^B
\la{introj}
\ee
where $m$ is the mass parameter of the $IIA$ theory. $F(A)$ can be
written as the formal sum
\be
F(A) = \sum_{n=1}^{11} F^{(n)}\ .
\la{introk}
\ee
Note that the top form is an $11$-form because we included a $10$-form 
$A^{(10)}$
in the definition of $A$. The field strengths $F^{(n)}$ will be subject to
superspace constraints, to be given below for bosonic backgrounds, in addition
to the constraint relating the bosonic components of
$F^{(n)}$ to the Hodge dual of the bosonic components of $F^{(10-n)}$. 

Each term in (\ref{introa}) is separately invariant under the global $IIA$ or
$IIB$ super-Poincar\'{e} group as well as under $(p + 1)$-dimensional general
coordinate transformations.  However, local kappa symmetry is achieved
by a subtle conspiracy between them, just as in the case of
super $p$-branes with scalar supermultiplets.

In these lectures, we are primarily interested in the $D3$-brane and so may 
set $p=3$ in 
the above formulae.

\section{D=4, N=2 gauged supergravity}
\la{D=4}

The $SO(8)$ gauged $N=8$ supergravity in four dimensions was obtained
in \cite{deWit1,deWit2} by gauging an $SO(8)$ subgroup of the global
$E_7$ symmetry group of \cite{Cremmer1,Cremmer2}.  To avoid some of
the complications of non-abelian gauge fields, one may consider a
truncation of this model to $N=2$, for which the bosonic sector
comprises the metric, four commuting $U(1)$ gauge potentials, three
dilatons and three axions.  In the absence of axions, this truncation
was obtained in \cite{Duffliu} by working in the symmetric gauge for
the 56-bein and incorporating three real scalars.  As was noted, there
is a straightforward generalization of the scalar ansatz to allow for
complex scalars.  Taking into account the $E_7$ self-duality condition
$\overline\phi^{ijkl}=\phi_{ijkl}={1\over4!}\epsilon_{ijklmnpq}\phi^{mnpq}$,
the scalar ansatz of \cite{Duffliu} may be generalized as:
\begin{equation}
\overline\phi^{ijkl}=\phi_{ijkl}=\sqrt{2}[
\Phi^{(1)}\epsilon^{(12)}+\Phi^{(2)}\epsilon^{(13)}+\Phi^{(3)}
\epsilon^{(14)}+
\overline\Phi^{(1)}\epsilon^{(34)}+\overline\Phi^{(2)}\epsilon^{(24)}+
\overline\Phi^{(3)}\epsilon^{(23)}]_{ijkl},
\end{equation}
where we follow the notation and conventions of \cite{Duffliu} (including the
definition of $SO(8)$ index pairs).  Note that the three complex scalars
may be parameterised in terms of their magnitudes and phases as
$\Phi^{(i)}=\phi^{(i)}e^{i\theta^{(i)}}$.

    Here, we shall consider the full $N=2$ truncation, where the three
axions are included as well as the other fields. In fact the structure
of the potential is little changed.  We find that the Lagrangian
including the axions may be written in the form
\begin{equation}
e^{-1}{\cal L}_4= R - \half\sum_i\left((\partial\phi^{(i)})^2
+\sinh^2\phi^{(i)}(\partial\theta^{(i)})^2\right)
- \half(F_{\mu\nu}^{(\alpha)+}{\cal M}_{\alpha\beta}
F^{(\beta)+\mu\nu}+{\rm h.c.})-g^2V\ ,\label{d44clag}
\end{equation}
where the potential is given simply by
\begin{equation}
V=-8(\cosh\phi^{(1)}+\cosh\phi^{(2)} +\cosh\phi^{(3)})\ .
\end{equation}
The complex symmetric scalar matrix ${\cal M}$ is quite complicated, and
incorporates all three complex scalars $\Phi^{(\alpha)}$ in a symmetric
manner; this is presented below.

In terms of the $N=2$ truncation, the three complex scalars each
parameterise an $SL(2;R)/SO(2)$ coset.  This may be made explicit by
performing the change of variables $(\phi^{(i)},\theta^{(i)})
\to(\varphi_i,\chi_i)$:
\begin{eqnarray}
\cosh\phi^{(i)}&=&\cosh\varphi_i+\half\chi_i^2\, e^{\varphi_i}\ , \nonumber\\
\cos\theta^{(i)}\sinh\phi^{(i)}&=&\sinh\varphi_i
-\half\chi_i^2\, e^{\varphi_i}\ ,\nonumber\\
\sin\theta^{(i)}\sinh\phi^{(i)}&=&\chi_i \, e^{\varphi_i}\ .
\end{eqnarray}
Defining the dilaton-axion combinations
\begin{equation}
A_i = 1 + \chi_i^2\,  e^{2\varphi_i}\ ,
\end{equation}
as well as
\begin{eqnarray}
B_1&=&\chi_2\, \chi_3 \, e^{\varphi_2+\varphi_3} +i\chi_1\,
e^{\varphi_1}\ ,\nonumber\\
B_2&=&\chi_1\, \chi_3 \, e^{\varphi_1+\varphi_3} +i\chi_2\,
e^{\varphi_2}\ ,\nonumber\\
B_3&=&\chi_1\, \chi_2 \, e^{\varphi_1+\varphi_2} +i\chi_3 \,
e^{\varphi_3}\ ,
\end{eqnarray}
we finally obtain the bosonic Lagrangian
\begin{equation}
e^{-1}{\cal L}_4=R - \half\sum_i\left((\partial\varphi_i)^2
+e^{2\varphi_i}(\partial\chi_i)^2\right)
- \half(F_{\mu\nu}^{(\alpha)+}{\cal M}_{\alpha\beta}
F^{(\beta)+\mu\nu}+{\rm h.c.})-g^2V\ .
\end{equation}
The potential $V$ is now given by
\begin{equation}
V=-8\sum_i\left(\cosh\varphi_i + \half\chi_i^2\, e^{\varphi_i}\right)
\ ,\label{eq:sl2pot}
\end{equation}
and the scalar matrix is
\begin{equation}
{\cal M}={1\over D}\left[\matrix{
e^{-\lambda_1} &e^{\varphi_1}B_1 &e^{\varphi_2}B_2 &e^{\varphi_3}B_3\cr
e^{\varphi_1}B_1 &e^{-\lambda_2}A_2A_3 &-e^{-\varphi_3}A_3 B_3
&-e^{-\varphi_2}A_2B_2\cr
e^{\varphi_2}B_2 &-e^{-\varphi_3}A_3B_3 &e^{-\lambda_3}A_1A_3
&-e^{-\varphi_1}A_1B_1\cr
e^{\varphi_3}B_3 &-e^{-\varphi_2}A_2B_2 &-e^{-\varphi_1}A_1B_1
&e^{-\lambda_4}A_1A_2}\right]\ ,
\end{equation}
where
\begin{equation}
D=1+\chi_1^2\, e^{2\varphi_1} +\chi_2^2\, e^{2\varphi_2} +\chi_3^2\,
e^{2\varphi_3}
-2i\, \chi_1\, \chi_2\, \chi_3\, e^{\varphi_1+\varphi_2+\varphi_3}\ .
\end{equation}
The scalar combinations $\{\lambda\}$ are defined as in \cite{Duffliu}:
\begin{eqnarray}
\lambda_1&=&           -\varphi_1-\varphi_2-\varphi_3\ ,\nonumber\\
\lambda_2&=&           -\varphi_1+\varphi_2+\varphi_3\ ,\nonumber\\
\lambda_3&=&\hphantom{-}\varphi_1-\varphi_2+\varphi_3\ ,\nonumber\\
\lambda_4&=&\hphantom{-}\varphi_1+\varphi_2-\varphi_3\ .
\label{eq:lambdas}
\end{eqnarray}

While this $N=2$ truncation of the $N=8$ theory essentially treats all
four $U(1)$ gauge fields equally, it was noted that one can make
contact with the theory obtained by reduction of a closed string on
$T^2$ through dualisation of two of the gauge fields.  To be specific,
we dualise $F_{\mu\nu}^{(2)}$ and $F_{\mu\nu}^{(4)}$, which singles
out the dilaton-axion pair $S=\chi_2 + i e^{-\varphi_2}$.  After an
additional field redefinition $S\to -1/S$, we obtain the bosonic
Lagrangian
\begin{eqnarray}
e^{-1}{\cal L}_4^{\rm dualized} &=&R -\half(\partial\varphi_2)^2
-\half e^{2\varphi_2}(\partial\chi_2)^2
+ \ft18\Tr (\partial ML \partial ML)
-g^2V\nonumber\\
&&\qquad
-\ft14 e^{-\varphi_2} F^T(LML)F - \ft14 \chi_2 \, F^TL{*F}\ ,
\label{eq:dlag}
\end{eqnarray}
where the potential is still given by (\ref{eq:sl2pot}).
The scalar matrix $M$ is given in terms of the
$SL(2;R)\times SL(2;R)$ vielbein
\begin{equation}
{\cal V}=e^{\varphi_3/2}\left[\matrix{1&-\chi_3\cr 0&e^{-\varphi_3}}\right]
\otimes e^{\varphi_1/2}\left[\matrix{1&-\chi_1\cr 0&e^{-\varphi_1}}\right],
\end{equation}
by $M={\cal V}^T{\cal V}$,
and the gauge fields have been arranged in the particular order
\begin{equation}
F_{\mu\nu}=[\matrix{F_{\mu\nu}^{(3)}&\widetilde F_{\mu\nu}^{(4)}&
\widetilde F_{\mu\nu}^{(2)}&-F_{\mu\nu}^{(1)}}]^T.
\end{equation}
Finally, $L=\sigma^2\otimes\sigma^2$ satisfies $L^2=I_4$, where
$\sigma^2$ is the standard Pauli matrix.  It is worth mentioning that
the pure scalar Lagrangian can be expressed as
\be
e^{-1} {\cal L}_{\rm scalar} = \sum_{i=1}^3 \Big[-\ft12 \tr \del {\cal M}_i
\del {\cal M}^{-1}_i + 4g^2 \tr {\cal M}_i \Big]\ ,
\ee
where ${\cal M}_i = {\cal V}_i^T\, {\cal V}_i$ and ${\cal V}_i$ is given by
\be
{\cal V}_i = e^{\varphi_i/2}\left[\matrix{1&-\chi_i\cr
0&e^{-\varphi_i}}\right]
\ .
\ee

We see that, save for the potential, the dualized Lagrangian is
indeed of the form obtained from $T^2$ compactification from six
dimensions.  In this case, two of the $SL(2;R)$'s now correspond to
$T$-dualities while the third corresponds to $S$-duality.  Note that
the initial choice of which two field strengths to dualise has
determined which of the three dilaton-axion pairs $(\varphi_i,\chi_i)$ is
to be identified with the strong-weak coupling $SL(2;R)$.

  Having shown that the bosonic Lagrangian is considerably simplified
by dualising to the field variables that arise in the $T^2$ reduction,
we may re-express the result (\ref{eq:dlag}) in the more explicit
notation of \cite{lpsol,cjlp1}.  Thus the bosonic sector of the gauged
$U(1)^4$ theory may be written as
\bea
e^{-1}{\cal L}_4 &=& R -\ft12 (\del\vec\varphi)^2 -\ft12 e^{-\vec
a\cdot\vec\varphi}\, (\del\chi)^2 - \ft12 e^{\vec
a_{12}\cdot\vec\varphi}\, (F_{\1 12})^2 - \ft12 e^{\vec
b_{12}\cdot\vec\varphi}\, (\cF^1_{\1 2})^2 -g^2 V\nn\\
&& -\ft14 \sum_{i=1}^2\Big( e^{\vec a_i\cdot\vec\varphi}\, (F_{\2i})^2 +
e^{\vec b_i\cdot\vec\varphi}\, (\cF_\2^i)^2\Big) - \ft12 \chi\,
\ep^{\mu\nu\rho\sigma}\, F_{\mu\nu\, i}\, \cF^i_{\rho\sigma}\ ,
\eea
where the field strengths are given by
\bea
F_{\2 1} &=& dA_{\1 1} + dA_{\0 12} {\cal A}_\1^2\ ,\qquad
\cF^1_{\2} = d\cA^1_{\2} - d\cA^1_{\0 2}\, \cA_\1^2 \ ,\nn\\
F_{\2 2} &=& dA_{\1 2} - \cA^1_{\0 2}\, dA_{\1 1} -
dA_{\0 12}\, A_\1^1\ ,\qquad
\cF_\2^2 = d\cA_\1^2 \ .
\eea
Here $\chi$, $A_{\0 12}$ and $\cA^1_{\0 2}$ are the axions $\chi_2$,
$\chi_1$ and $\chi_3$, and the potential is given by (\ref{eq:sl2pot}).

The inclusion of the potential term in the gauged supergravity theory
breaks all three $SL(2;R)$ symmetries to $O(2)$, acting as $\tau\to
(a\tau+b) /(c\tau+d)$, where
\begin{equation}
\pmatrix{a&b\cr c&d} =
\pmatrix{\cos\alpha/2&\sin\alpha/2\cr-\sin\alpha/2&\cos\alpha/2}.
\label{o2rot}
\end{equation}
In terms of the original $(\phi,\theta)$ scalar variables in
(\ref{d44clag}), this $O(2)$ subgroup corresponds to
$\theta\longrightarrow \theta+\alpha$.  The $O(2)$ symmetry is,
however, sufficient for generating dyonic solutions.  Nevertheless, we
note that the fermionic sector and in particular the supersymmetry
transformations are not invariant under this symmetry of the bosonic
sector.  One manifestation of this particular situation is the fact
that magnetic black holes of this theory are not supersymmetric, even
though they may be extremal.  Furthermore, in a related note, while it
is clear that the dualisation procedure performed above runs into
difficulty in the full $N=8$ theory with non-abelian $SO(8)$ gauging,
the fermionic sector does not admit such a straightforward
dualisation, even in the $N=2$ abelian truncation.  This is easily
seen by the fact that the gravitini are necessarily charged under the
gauge fields and hence couple to the bare gauge potentials themselves.


\begin{thebibliography}{99}

\bibitem{Wess}
J. Wess and J. Bagger,
{\sl Supersymmetry and supergravity},
Princeton University Press (1983).

\bibitem{Gates}
S. J. Gates, Jr., M. T. Grisaru, M. Rocek and W. Siegel,
{\sl Superspace {\it or} one thousand and one lessons in 
supersymmetry},
Benjamin Cummings (1983).

\bibitem{Srivastava}
Prem P. Srivastava,
{\sl Supersymmetry, superfields and supergravity},
Adam Hilger (1986).

\bibitem{West}
P. West,
{\sl Introduction to supersymmetry and supergravity},
World Scientific (1986).

\bibitem{Bailin}
D. Bailin and A. Love,
{\sl Supersymmetric gauge field theory and string theory},
I. O. P.

\bibitem{M}
M. J. Duff,
{\sl $M$-theory: the theory formerly known as strings},
Int. J. Mod. Phys. {\bf A11} (1996) 5623.

\bibitem{String}
M.~J. Duff, R.~R. Khuri and J.~X. Lu,
{\sl String solitons},
Phys. Rep. {\bf 259} (1995) 213, hep-th/9412184

\bibitem{Das}
D. Z. Freedman and A. Das,
\newblock{\sl Gauge internal symmetry in extended supergravity},
\newblock Phys. Lett. {\bf B74} (1977) 333.

\bibitem{CDGR}
S. M. Christensen, M. J. Duff,  G. W. Gibbons and M. Rocek,
{\sl Vanishing one-loop $\beta$-function in gauged $N > 4$ supergravity},
Phys. Rev. Lett. {\bf 45}, 161 (1980).

\bibitem{Supergravity81}
M. J. Duff,
{\sl Ultraviolet divergences in extended supergravity},
in Proceedings of the 1981 Trieste Conference ``Supergravity 81'', eds. Ferarra
and Taylor, C. U. P. 1982.

\bibitem{DN}
B. de Wit and H. Nicolai,
\newblock{\sl $N=8$ supergravity},
\newblock Nucl. Phys. {\bf B208} (1982) 323.

\bm{Cremmer} 
E. Cremmer, B. Julia and J. Scherk, 
{\sl Supergravity theory in eleven dimensions}, 
Phys. Lett. {\bf B76} (1978) 409.

\bibitem{DP}
 M. J. Duff and C. N. Pope,
{\sl Kaluza-Klein supergravity and the seven sphere},
in Proceedins of the 1982 Trieste conference 
``Supersymmetry and Supergravity 82'', eds Ferrara, Taylor and van 
Nieuwenhuizen, World Scientific 1983.

\bibitem{CS}
E. Cremmer and J. Scherk,
{\sl Spontaneous compactification of extra space dimensions},
Nucl. Phys. {\bf B118} (1977) 61.

\bibitem{Freundrubin}
P. G. O. Freund and M. A. Rubin,
{\sl Dynamics of dimensional reduction},
Phys. Lett. {\bf B97} (1980) 233.

\bibitem{DV}
M. J. Duff and P. van Nieuwenhuizen,
{\sl Quantum inequivalence of different field representations},
Phys. Lett. {\bf B94} (1980) 179.

\bibitem{PTV}
K. Pilch, P. K. Townsend and P. van Nieuwenhuizen,
{\sl Compactification of $d=11$ supergravity on $S^{4}$ (or $11=7+4$,
too)},
Nucl. Phys. {\bf B242} (1984) 377.

\bibitem{PPV}
M. Pernici, K. Pilch and P. van Nieuwenhuizen,
{\sl Gauged maximally extended supergravity in seven dimensions},
Phys. Lett. {\bf B143} (1984) 103.

\bibitem{TV}
P. K. Townsend and P. van Nieuwenhuizen,
{\sl Gauged seven-dimensional supergravity},
Phys. Lett. {\bf B125} (1983) 41-6.

\bibitem{DNP}
 M. J. Duff, B. E. W. Nilsson and C. N. Pope,
{\sl Kaluza-Klein supergravity},
Phys. Rep. {\bf 130} (1986) 1.

\bibitem{Schwarz}
J. H. Schwarz,
{\sl Covariant field equations of chiral $N=2$ $D=10$ supergravity},
Nucl. Phys. {\bf B226} (1983) 269.

\bibitem{Howewest}
P. Howe and P.C. West,
{\sl The complete $N=2$, $d=10$ supergravity},
Nucl. Phys. {\bf B238} (1984) 181.

\bibitem{Schwarzcompact}
J. H. Schwarz,
{\sl Spontaneous compactification of extended supergravity in ten 
dimensions},
Physica {\bf A124} (1984) 543.

\bibitem{GM}
M. G\"{u}naydin and N. Marcus,
{\sl The spectrum of the $S^{5}$ compactification of the $N=2$,
$D=10$ supergravity and the unitary supermultiplet},
Class. Quant. Grav. {\bf 2} (1985) L11.

\bibitem{KRV}
H. J. Kim, L. J. Romans and P. van Nieuwenhuizen,
{\sl Mass spectrum of chiral ten-dimensional $N=2$ supergravity on
$S^{5}$},
Phys. Rev. {\bf D32} (1985) 389.

\bibitem{PPV2}
M. Pernici, K. Pilch and P. van Nieuwenhuizen,
{\sl Gauged $N=8$, $d=5$ supergravity},
Nucl. Phys. {\bf B259} (1985) 460.

\bibitem{GRW}
M. G\"{u}naydin, L. J. Romans and N. P. Warner,
{\sl Gauged $N=8$ supergravity in five dimensions},
Phys. Lett. {\bf B154} (1985) 268.

\bibitem{Gunaydin1}
M. G\"{u}naydin,
{\sl Singleton and doubleton supermultiplets of space-time supergroups 
and infinite spin superalgebras}, in Proceedings of the 1989 Trieste 
Conference ``Supermembranes and Physics in $2+1$ Dimensions'', 
eds Duff, Pope and Sezgin, World Scientific 1990.

\bibitem{Dirac}
P. A. M. Dirac,
{\sl A remarkable representation of the 3+2 de Sitter group},
J. Math. Phys. {\bf 4} (1963) 901.

\bibitem{Fronsdal}
C. Fronsdal,
{\sl Dirac supermultiplet},
Phys. Rev. {\bf D26} (1982) 1988.

\bibitem{Flato}
M. Flato and C. Fronsdal,
{\sl Quantum field theory of singletons. The rac},
J. Math. Phys. {\bf 22} (1981) 1100. 

\bibitem{Heidenreich}
W. Heidenreich,
{\sl All linear, unitary, irreducible representations of de Sitter 
supersymmetry with positive energy},
Phys. Lett. {\bf B110} (1982) 461.

\bibitem{FN}
D. Z. Freedman and H. Nicolai,
{\sl Multiplet shortening in $OSp(4|8)$},
Nucl. Phys. {\bf B237} (1984) 342.

\bibitem{Sezgin}
E. Sezgin,
{\sl The spectrum of the eleven dimensional supergravity compactified
on the round seven sphere}, Trieste preprint, 1983, in Supergravity in
Diverse Dimensions, vol. 2, 1367, (eds A. Salam and E. Sezgin World 
Scientific, 1989);
Fortschr. Phys. {\bf 34} (1986) 217.

\bibitem{NS}
H. Nicolai and E. Sezgin,
{\sl Singleton representations of $OSp(N,4)$},
Phys. Lett. {\bf B143} (1984) 389.

\bm{GRW2} M. G\"{u}naydin, L.J. Romans and N.P. Warner, {\sl Spectrum
generating algebras in Kaluza-Klein theories}, Phys. Lett. {\bf B146}
(1984) 401.

\bibitem{Nahm}
W. Nahm,
{\sl Supersymmetries and their representations},
Nucl. Phys. {\bf B135} (1978) 149.

\bibitem{Fifteen}
M. J. Duff,
{\sl Supermembranes:  The first fifteen weeks},
 Class. Quant. Grav. {\bf 5}, 189 (1988).

\bibitem{BD}
M. P. Blencowe and M. J. Duff,
{\sl Supersingletons},
Phys. Lett. {\bf B203}, 229 (1988).

\bibitem{NST}
H. Nicolai, E. Sezgin and Y. Tanii,
{\sl Conformally invariant supersymmetric field theories on $S^{p}
\times S^{1}$ and super $p$-branes},
Nucl. Phys. {\bf B305} (1988) 483.

\bibitem{BST1}
E.~Bergshoeff, E.~Sezgin and P.~Townsend,
\newblock {\sl Supermembranes and eleven-dimensional supergravity},
\newblock Phys. Lett. {\bf B189} (1987) 75.

\bibitem{BST2}
E.~Bergshoeff, E.~Sezgin and P.~Townsend,
\newblock {\sl Properties of the eleven-dimensional supermembrane theory},
\newblock Ann. Phys. {\bf 185} (1988) 330.

\bibitem{BSTan}
E. Bergshoeff, E. Sezgin and Y. Tanii.
{\sl Stress tensor commutators and Schwinger terms in singleton
theories},
Int. J. Mod. Phys. {\bf A5} (1990) 3599.

\bibitem{BDPS}
E. Bergshoeff, M. J. Duff, C. N. Pope and E. Sezgin,
{\sl Supersymmetric supermembrane vacua and singletons},
 Phys. Lett. {\bf B199}, 69 (1988).

\bibitem{BDPS2}
E. Bergshoeff, M. J. Duff, C. N. Pope and E. Sezgin,
{\sl Compactifications of the eleven-dimensional supermembrane},
Phys. Lett. {\bf B224}, 71 (1989).

\bibitem{Sutton}
M. J. Duff and C. Sutton,
{\sl The membrane at the end of the universe}
New. Sci. {\bf 118} (1988) 67.

\bibitem{Classical}
M. J. Duff,
{\sl Classical and quantum supermembranes},
Class. Quant. Grav. {\bf 6}, 1577 (1989).

\bibitem{BD2}
M. P. Blencowe and M. J. Duff,
{\sl Supermembranes and the signature of spacetime},
 Nucl. Phys. {\bf B310} 387 (1988).

\bibitem{BSS}
E. Bergshoeff, A. Salam, E. Sezgin, Y. Tanii,
{\sl $N=8$ supersingleton quantum field theory},
Nucl.Phys. {\bf B305} (1988) 497.

\bibitem{DPS}
M. J. Duff, C. N. Pope and E. Sezgin,
{\sl A stable supermembrane vacuum with a discrete spectrum},
Phys. Lett. {\bf B225}, 319 (1989).

\bibitem{dust} 
M.J. Duff and K.S. Stelle, 
{\sl Multi-membrane solutions of $D=11$ supergravity}, 
Phys. Lett. {\bf B253} (1991) 113. 

\bibitem{AETW}
A.~Achucarro, J.~Evans, P.~Townsend and D.~Wiltshire,
\newblock {\sl Super p-branes},
\newblock Phys. Lett. {\bf B198} (1987) 441.

\bibitem{GNST}
M. G\"{u}naydin, B. E. W. Nilsson, G. Sierra and P. K. Townsend,
{\sl Singletons and superstrings},
Phys. Lett. {\bf B176} (1986) 45.

\bibitem{DTV}
M. J. Duff, P. K. Townsend and P. van Nieuwenhuizen,
{\sl Spontaneous compactification of supergravity on the three-sphere},
Phys. Lett. {\bf B122} (1983) 232.

\bibitem{super}  
M. J. Duff, 
{\sl Supermembranes}, lectures given at 
the T. A. S. I. Summer School, University of Colorado, Boulder, June 1996;
the Topical Meeting, Imperial College, London, July 1996 and the 26th
British Universities Summer School in Theoretical Elementary Particle
Physics, University of Swansea, September 1996.
Int. J. Mod. Phys. {\bf A11} (1996) 5623, hep-th/9611203. 

\bibitem{Hull}
C. M. Hull,
{\sl Duality and the signature of spacetime},
hep-th/9807127.

\bibitem{HS}
G.~T. Horowitz and A.~Strominger,
\newblock {\sl Black strings and p-branes},
\newblock Nucl. Phys. {\bf B360} (1991) 197.

\bibitem{DLgauge}
M. J. Duff and J. X. Lu,
\newblock {\sl The self-dual Type $IIB$ superthreebrane},
\newblock Phys. Lett. {\bf B273} (1991) 409.

\bibitem{CHS1}
C. G.~Callan, J.~A. Harvey and A.~Strominger,
\newblock {\sl World sheet approach to heterotic instantons and solitons},
\newblock Nucl. Phys. {\bf  B359} (1991) 611.

\bibitem{CHS2}
C. G.~Callan, J.~A. Harvey and A.~Strominger,
\newblock {\sl Worldbrane actions for string solitons},
\newblock Nucl. Phys. {\bf  B367} (1991) 60.

\bibitem{Luscan}
M. J. Duff and J. X. Lu,
\newblock{\sl Type $II$ $p$-branes: the brane scan revisited},
\newblock Nucl. Phys. {\bf B390} (1993) 276.

\bm{Gueven}
R. Gueven,
{\sl Black $p$-brane solutions of $D = 11$ supergravity theory},
Phys. Lett. {\bf B276} (1992) 49.

\bm{lm} J.T. Liu and R. Minasian, {\sl Black holes and membranes in 
$AdS_7$}, hep-th/9903269.

\bibitem{GT}
G. W. Gibbons and P. K. Townsend,
{\sl Vacuum interpolation in supergravity via super $p$-branes},
Phys. Rev. Lett. {\bf 71} (1993) 3754.

 \bibitem{DGT}
M. J. Duff, G. W. Gibbons and P. K. Townsend,
{\sl Macroscopic superstrings as interpolating solitons},
Phys. Lett. B. {\bf 332} (1994) 321.

\bibitem{GHT}
G. W. Gibbons, G. T. Horowitz and P. K. Townsend,
{\sl Higher-dimensional resolution of dilatonic black hole
singularities},
Class. Quant. Grav. {\bf 12} (1995) 297.

\bibitem{DLPS}
M.J. Duff, H. L\"u, C.N. Pope and E. Sezgin,
{\sl Supermembranes with fewer supersymmetries},
Phys. Lett. {\bf B371} (1996) 206, hep-th/9511162.

\bm{ccdfft} L. Castellani, A. Ceresole, R. D'Auria, S. Ferrara,
P. Fr\'e and M. Trigiante, {\sl $G/H$ M-branes and $AdS_{p+2}$ geometries},
hep-th/9803039.

\bibitem{ADP}
M. A. Awada, M. J. Duff and C. N. Pope,
{\sl $N=8$ supergravity breaks down to $N=1$},
Phys. Rev. Lett. {\bf 50} (1983) 294.

\bibitem{DNP2}
 M. J. Duff, B. E. W. Nilsson and C. N. Pope,
{\sl Spontaneous supersymmetry breaking via the squashed seven sphere},
Phys. Rev. Lett. {\bf 50} (1983) 2043.

\bibitem{Yano}
T. Nagano and K. Yano,
Ann. Math. {\bf 69} (1959) 451.

\bm{lublack} 
M.J. Duff and J.X. Lu, 
{\sl Black and super $p$-branes in diverse dimensions}, 
Nucl. Phys. {\bf B416} (1994) 301. 

\bibitem{DR}
M.J. Duff and J. Rahmfeld, 
{\sl Bound states of black holes and other $p$-branes},
Nucl. Phys. {\bf B481} (1996) 332.
    
\bibitem{KKLP}
N. Khviengia, Z. Khviengia, H. L\"{u} and C.N. Pope,
{\sl Interecting $M$-branes and bound states},
Phys. Lett. {\bf B388} (1996) 21.  

\bm{Rahmfeld} M. J. Duff, S. Ferrara, R. Khuri and J. Rahmfeld,
{\sl Supersymmetry and dual string solitons}, Phys. Lett. {\bf B356} 
(1995) 479.

\bibitem{DR2}
M.~J. Duff and J.~Rahmfeld,
\newblock {\sl Massive string states as extreme black holes},
\newblock Phys. Lett. {\bf B345} (1995) 441.

\bibitem{FG}
D. Z. Freedman and G. W. Gibbons,
{\sl Electrovac ground state in gauged $SU(2) \times SU(2)$ 
supergravity},
Nucl. Phys. {\bf B233} (1984) 24.
 
\bibitem{CKKTV}
P. Claus, R. Kallosh, J. Kumar, P.K. Townsend and A. Van
Proeyen, {\sl Supergravity and the large $N$ limit of theories with
sixteen supercharges},
hep-th/9801206.

\bibitem{DFFFTT}
G. Dall'Agata, D. Fabbri, C. Fraser, P. Fr\'{e}, P. Termonia and M. 
Trigiante,
{\sl The $OSp(4|8)$ singleton action from the supermembrane}
hep-th/9807115.

\bm{T} A. A. Tseytlin, {\sl Extreme dyonic black holes in string theory}, 
Mod. Phys. Lett. {\bf A11} (1996) 689, hep-th/9601177.

\bm{CT1} M. Cvetic and A. A. Tseytlin, {\sl General class of BPS 
saturated black holes as exact superstring solutions}, Phys. Lett. 
{\bf B366} (1996) 95, hep-th/9510097.

\bm{CT2} M. Cvetic and A. A. Tseytlin, {\sl Solitonic strings and BPS 
saturated dyonic black holes}, Phys. Rev. {\bf D53} (1996) 5619, 
hep-th/9512031.

\bibitem{DLP}
M. J. Duff, H. L\"u and C. N. Pope,
{\sl Supersymmetry without supersymmetry},
Phys. Lett. {\bf B409} (1997) 136, hep-th/9704186.

\bibitem{Kachru}
S. Kachru and E. Silverstein,
{\sl $4d$ conformal field theories and strings on orbifolds},
hep-th/9802183.

\bm{DNP3} M.J. Duff, B.E.W. Nilsson and C.N. Pope, {\sl
Compactification of $D=11$ supergravity on K3$\times T^3$},
Phys. Lett. {\bf B129} (1983) 39.

\bibitem{DLP1}
M. J. Duff, H. L\"u and C. N. Pope,
{\sl $AdS_{5} \times S^{5}$ untwisted},
hep-th/9803061.

\bibitem{DLP2}
M. J. Duff, H. L\"u and C. N. Pope,
{\sl $AdS_{3} \times S^{3}$ (un)twisted and squashed},
hep-th/9807173.

\bm{Halyo}
E. Halyo,
{\sl Supergravity on $AdS_{5/4}$ x Hopf fibrations and conformal field 
theories},
hep-th/9803193.

\bm{Page} 
D.N. Page and C.N. Pope, 
{\sl Which compactifications of $D=11$ supergravity are stable?},
Phys. Lett. {\bf B144} (1984) 346.

\bm{Oz} Y. Oz and J. Terning, {\sl Orbifolds of $AdS_{5} \times S^{5}$
and 4-D conformal field theories}, hep-th/9803167.

\bm{Romans}
L. Romans,
{\sl New compactifications of chiral $N=2$, $d=10$ supergravity},
Phys. Lett. {\bf B153} (1985) 392.

\bm{klebanov} 
I. Klebanov and E. Witten, 
{\sl Superconformal field theory on three-branes at a Calabi-Yau singularity}, 
hep-th/9807080.

\bm{Romans2} L.J. Romans, {\sl Gauged $N=4$ supergravities in five
dimensions and their magnetovac backgrounds}, Nucl. Phys. {\bf B267}
(1986) 443.

\bibitem{Howe}
M. J. Duff, P.~Howe, T.~Inami and K. S. Stelle,
\newblock {\sl Superstrings in $D = 10$ from supermembranes in $D = 11$},
\newblock Phys. Lett. {\bf B191} (1987) 70.

\bibitem{Luduality}
M. J. Duff and J. X. Lu,
\newblock {\sl Duality rotations in membrane theory},
\newblock Nucl.Phys. {\bf B347} (1990) 394.

\bibitem{Hulltownsend}
C. M. Hull and P. K. Townsend,
\newblock {\sl Unity of superstring dualities},
\newblock Nucl. Phys. {\bf B438} (1995) 109.

\bibitem{Townsendeleven}
P. K. Townsend,
\newblock {\sl The eleven-dimensional supermembrane revisited},
\newblock Phys.Lett.{\bf B350} (1995) 184.

\bibitem{Wittenvarious}
E.~Witten,
\newblock {\sl String theory dynamics in various dimensions},
\newblock Nucl. Phys. {\bf B443} (1995) 85.

\bibitem{Duffliuminasian}
M. J. Duff, J.~T. Liu and R.~Minasian,
\newblock {\sl Eleven-dimensional origin of string/string duality: A one 
loop test},
\newblock Nucl.Phys. {\bf B452} (1995) 261.

\bibitem{Becker1}
K.~Becker, M.~Becker and A.~Strominger,
\newblock {\sl Fivebranes, membranes and nonperturbative string theory},
\newblock Nucl.Phys. {\bf B456} (1995) 130.

\bibitem{Schwarzpower}
J. H. Schwarz,
\newblock {\sl The power of $M$-theory},
\newblock Phys. Lett. {\bf B360} (1995) 13.

\bibitem{Horava}
P. Horava and E. Witten,
\newblock {\sl Heterotic and Type $I$ string dynamics from eleven 
dimensions},
\newblock Nucl. Phys. {\bf B460} (1996) 506.

\bibitem{TownsendM}
P. Townsend,
\newblock {\sl $D$-Branes From $M$-Branes},
\newblock Phys. Lett. {\bf B373} (1996) 68.

\bibitem{Aharony}
O. Aharony, J. Sonnenschein and S. Yankielowicz,
\newblock {\sl Interactions of strings and $D$-branes from $M$-theory},
\newblock  Nucl. Phys. {\bf B474} (1996) 309.

\bibitem{DuffM}
M. J. Duff,
\newblock {\sl  $M$-theory (the theory formerly known as strings)},
\newblock  I. J. M. P {\bf A11} (1996) 5623. 

\bibitem{BanksM}
T. Banks, W. Fischler, S. H. Shenker and L. Susskind, 
\newblock {\sl $M$ theory as a matrix model: a conjecture},
\newblock Phys. Rev. {\bf D55} (1997) 5112.

\bibitem{slan}
R. Slansky,
\newblock {\sl Group theory for unified model building},
\newblock Phys. Rep. {\bf 79} (1981) 1.

\bm{Nilssonpope} 
B.E.W. Nilsson and C.N. Pope, 
{\sl Hopf fibration of $D=11$ supergravity}, 
Class. Quantum Grav. {\bf 1} (1984) 499.

\bibitem{Wittenworld}
E. Witten, 
\newblock {\sl Strong coupling and the cosmological constant}, 
\newblock Mod. Phys. Lett. {\bf A10} (1995) 2153.

\bibitem{Orzalesi}
M. J. Duff and C. Orzalesi, 
\newblock {\sl The cosmological constant in
spontaneously compactified $D = 11$ supergravity}, 
\newblock Phys. Lett. {\bf B122} (1983) 37. 

\bibitem{Maldacena1}
J. Maldacena,
{\sl The large $N$ limit of superconformal field theories and
supergravity},
hep-th/9711200.

\bibitem{Polchinski}
J. Polchinski,
{\sl Dirichlet branes and Ramond-Ramond charges},
Phys. Rev. Lett. {\bf 75} (1995) 4724

\bm{Wittenbound} 
E. Witten, 
{\sl Bound states of strings and $p$-branes}, 
Nucl. Phys. {\bf B460} (1996) 335, hep-th/9510135.

\bm{Gubser1} S. S. Gubser, I. R. Klebanov and A. W. Peet, {\sl Entropy
and temperature of black 3-branes}, Phys. Rev. {\bf D54} (1996) 3915, 
hep-th/9602135.

\bm{Klebanov2} I. R. Klebanov, {\sl World volume approach to 
absorption by non-dilatonic branes}, Nucl. Phys. {\bf B496} 231, 
hep-th/9702076.

\bm{Gubser2} S. S. Gubser, I. R. Klebanov and A. A. Tseytlin, {\sl 
String theory and classical absorption by three-branes}, Nucl. 
Phys. {\bf B499} (1997) 217, hep-th/9703040.

\bm{Gubser3} S. S. Gubser and I. R. Klebanov, {\sl Absorption by 
branes and Schwinger terms in the world volume theory}, Phys. Lett. 
{\bf B413} (1997) 41, hep-th/9708005.

\bibitem{IMSY}
N. Itzhak, J. M. Maldacena, J. Sonnenschein and S. Yankielowicz,
{\sl Supergravity and the large $N$ limit of theories with sixteen
supercharges},
hep-th/9802042.

\bibitem{Ferrara2}
S. Ferrara and C. Fronsdal,
{\sl Conformal Maxwell theory as a singleton field theory on
$AdS_{5}$, $IIB$ three-branes and duality},
hep-th/9712239.

\bibitem{KKR}
R. Kallosh, J. Kumar and A. Rajaraman,
{\sl Special conformal symmetry of worldvolume actions},
hep-th/9712073.

\bibitem{Boonstra}
H. J. Boonstra, B. Peters, K. Skenderis,
{\sl Branes and anti-de Sitter space-time},
hep-th/9801076.

\bibitem{Gunaydin}
M. G\"{u}naydin and D. Minic,
{\sl Singletons, doubletons and $M$-theory},
hep-th/9802047.

\bibitem{Gubserklebanovpolyakov}
S. S. Gubser, I. R. Klebanov and A. M. Polyakov,
{\sl Gauge theory correlators from non-critical string theory},
hep-th/9802109.

\bibitem{Horowitz}
G. T. Horowitz and H. Ooguri,
{\sl Spectrum of large $N$ gauge theory from supergravity},
hep-th/9802116.

\bibitem{Wittenads}
E. Witten,
{\sl Anti-de Sitter space and holography},
hep-th/9802150.

\bibitem{Berkooz}
M. Berkooz,
{\sl A supergravity dual of a $(1,0)$ field theory in six dimensions},
hep-th/9802195.

\bibitem{Ferrara}
S. Ferrara and C. Fronsdal,
{\sl On $N=8$ supergravity on $AdS_{5}$ and $N=4$ superconformal
Yang-Mills theory},
hep-th/9802203.

\bibitem{Rey}
Soo-Jong Rey and Jungtay Lee,
{\sl Macroscopic strings as heavy quarks in large $N$ gauge theory and
anti-de Sitter supergravity},
hep-th/9803001.

\bibitem{Maldacena2}
J. Maldacena,
{\sl Wilson loops in large $N$ field theories},
hep-th/9803002.

\bm{lnv} A. Lawrence, N. Nekrasov and C. Vafa, {\sl On conformal field
theories in four dimensions}, hep-th/9803015.

\bm{thooft}
G. t'Hooft,
{\sl Dimensional reduction in quantum gravity},
gr-qc/9310026.

\bm{Susskind1}
L. Susskind,
{\sl The world as a hologram},
hep-th/9409089.

\bm{hawkingpope} 
S.W. Hawking and C.N. Pope, 
{\sl Generalised spin structures in quantum gravity}, 
Phys. Lett. {\bf B73} (1978) 42.

\bm{pope} 
C.N. Pope, 
{\sl Eigenfunctions and spin$^c$ structures in $CP^2$}, 
Phys. Lett. {\bf B97} (1980) 417.

\bm{DLPhet} 
M.J. Duff, H. L\"u, C.N. Pope, 
{\sl Heterotic phase transitions and singularities of the gauge dyonic string}, 
hep-th/9603037. 

\bm{DLPgauge} 
M.J. Duff, James T. Liu, H. L\"u, C.N. Pope, 
{\sl Gauge dyonic strings and their global limit}, 
hep-th/9711089. 

\bm{Boonstra1} 
H.J. Boonstra, B. Peeters and K. Skenderis, 
{\sl Duality and asymptotic geometries}, 
Phys. Lett. {\bf B411} (1997) 59, hep-th/9706192.

\bm{Boonstra2} 
H.J. Boonstra, B. Peeters and K. Skenderis, 
{\sl Brane intersections, anti-de Sitter spacetimes and dual superconformal 
theories}, 
hep-th/9803231.

\bibitem{Horowitz2}
G.T. Horowitz, J. Maldacena and A. Strominger,
{\sl Nonextremal black hole microstates and U duality},
hep-th/9603109.

\bibitem{Horowitz3}
G.T. Horowitz and A. Strominger,
{\sl Counting states of near extremal black holes},
hep-th/9602051.

\bibitem{Seibergwitten}
N. Seiberg and E. Witten,
{\sl The D1/D5 system and singular CFT},
hep-th/9903224.

\bibitem{Maldacenamichelsonstrominger}
J. Maldacena, J. Michelson and A. Strominger, 
{\sl Anti-de Sitter Fragmentation},
 hep-th/9812073.

\bibitem{Maldacena3}
J. Maldacena and A. Strominger,
{\sl $AdS_{3}$ black holes and a stringy exclusion principle},
hep-th/9804085.

\bm{romans} L.J. Romans, {\sl Supersymmetric, cold and lukewarm black
holes in cosmological Einstein-Maxwell theory}, Nucl. Phys. {\bf B383}
(1992) 395; {\sl Black holes in cosmological 
Einstein-Maxwell theory}, in ``Trieste 1992, Proceedings,
High energy physics and cosmology'' 416.

\bibitem{Behrndt1}
K. Behrndt, A.H. Chamseddine and W.A. Sabra,
{\sl BPS black holes in $N=2$ five-dimensional $AdS$ supergravity},
Phys. Lett. {\bf B442} (1998) 97, hep-th/9807187.

\bm{birm} D. Birmingham, {\sl Topological black holes in ante-de
Sitter space}, hep-th/9808032.

\bibitem{Caldarelliklemm1} M. M. Caldarelli and D. Klemm, {\sl
Supersymmetry of anti-de Sitter black holes}, hep-th/9808097.

\bibitem{Klemm}
D. Klemm,
{\sl BPS black holes in gauged $N=4$ $D=4$ supergravity},  hep-th/9810090.

\bibitem{Behrndt2}
K. Behrndt, M. Cvetic and W.A. Sabra,
{\sl Non-extreme black holes five-dimensional $N=2$ $AdS$
supergravity}, hep-th/9810227.

\bibitem{Duffliu}
M.J. Duff and J.T. Liu,
{\sl Anti-de Sitter black holes in gauged N=8 supergravity},
hep-th/9901149.

\bibitem{Chamblin}
A. Chamblin, R. Emparan, C.V. Johnson and R.C. Myers,
{\sl Charged $AdS$ black holes and catastrophic holography},
hep-th/9902170.

\bibitem{Cveticgubser1} M. Cvetic and S.S. Gubser,
{\sl Phases of R charged black holes, spinning
branes and strongly coupled gauge theories}, hep-th/9902195.

\bibitem{Cveticgubser2} M. Cvetic and S.S. Gubser, {\sl Thermodynamic
stability and phases of general spinning branes}, hep-th/9903132.

\bibitem{Sabra} W. Sabra,
{\sl Anti-de Sitter black holes in N=2 gauged supergravity},
hep-th/9903143.

\bibitem{Caldarelliklemm2} M.M. Caldarelli and D. Klemm, {\sl M-theory
and stringy corrections to anti-de Sitter black holes and conformal
field theories}, hep-th/9903078.

\bibitem{Wittenads2} E. Witten, {\sl Anti-de Sitter space, thermal phase
transition, and confinement in gauge theories},
Adv. Theor. Math. Phys. {\bf 2} (1998) 505, hep-th/9803131.

\bibitem{deWit1}
B. de Wit and H. Nicolai,
{\sl $N=8$ supergravity with local $SO(8) \times SU(8)$ invariance},
Phys. Lett. {\bf B108} (1982) 285.

\bibitem{deWit2}
B. de Wit and H. Nicolai,
{\sl N=8 supergravity},
Nucl. Phys. {\bf B208} (1982) 323.

\bm{duffpope3} 
M.J. Duff and C.N. Pope, 
{\sl Kaluza-Klein supergravity and the seven-sphere}, 
in: Supersymmetry and supergravity 82, eds. S. Ferrara, J.G. Taylor 
and P. van Nieuwenhuizen (World Scientific, Singapore, 1983).

\bibitem{Duffads} 
M.J. Duff, 
{\sl Anti-de Sitter space, branes, singletons, superconformal field theories 
and all that},
hep-th/9808100.

\bibitem{Duffpope} 
M.J. Duff and C.N. Pope, 
{\sl Consistent truncations in Kaluza-Klein theories}, 
Nucl. Phys. {\bf B255} (1985) 355.

\bibitem{Pope} 
C.N. Pope, 
{\sl Consistency of truncations in Kaluza-Klein}, 
published in the Proceedings of the 1984 Santa Fe meeting.

\bibitem{deWitnicolaiwarner}
B. de Wit, H. Nicolai and N.P. Warner,
{\sl The embedding of gauged $N=8$ supergravity into $D=11$ 
supergravity}, 
Nucl. Phys. {\bf B255} (1985) 29.

\bibitem{DNPW}
M. J. Duff, B. E. W Nilsson, C. N. Pope and N. Warner,
{\sl On the consistency of the Kaluza-Klein ansatz},
Phys. Lett. {\bf B149} (1984) 90. 

\bibitem{deWitnicolai}
B. de Wit and H. Nicolai,
{\sl The consistency of the $S^{7}е$ truncation in $D=11$ 
supergravity}, 
Nucl. Phys. {\bf B281} (1987) 211. 

\bibitem{Duffliurahmfeld1}
M.J. Duff, J.T. Liu and J. Rahmfeld,
{\sl Four-dimensional string-string-string triality},
Nucl. Phys. {\bf B459} (1996) 125, hep-th/9508094.

\bm{cvyod} M. Cvetic and D. Youm, {\sl Dyonic BPS saturated black
holes of heterotic string on a six torus}, Phys. Rev. {\bf D53} (1996)
584, hep-th/9507090. 

\bm{lpsol} H. L\"u and C.N. Pope, {\sl $p$-brane solitons in maximal
supergravities}, Nucl. Phys. {\bf B465} (1996) 127, hep-th/9512012.

\bm{khor} R.R. Khuri and T. Ortin, {\sl Supersymmetric black holes in
$N=8$ supergravity}, Nucl. Phys. {\bf B467} (1996) 355.

\bm{cveticts} M. Cvetic and A.A. Tseytlin,
{\sl General class of BPS saturated dyonic black holes as exact
superstring solutions}, Phys. Lett. {\bf B366} (1996) 95,
hep-th/9510097.

\bm{tsey} A.A. Tseytlin, {\sl Harmonic superpositions 
of M-branes},
Nucl. Phys. {\bf B475} (1996) 149, hep-th/9604035.

\bibitem{Cveticyoum3} M. Cvetic and D. Youm, {\sl Near BPS saturated
rotating electrically charged black holes as string states},
Nucl. Phys. {\bf B477} (1996) 449, hep-th/9605051.

\bibitem{Cveticyoum2} M. Cvetic and D. Youm, 
{\sl Rotating intersecting M-branes},
Nucl. Phys. {\bf B499} (1997) 253, hep-th/9612229.

\bibitem{Gubserthermo}
S. Gubser,
{\sl Thermodynamics of spinning $D3$-branes},
hep-th/9810225.

\bibitem{Csaki}
C. Csaki, Y. Oz, J. Russo and J. Terning,
{\sl Large N QCD from rotating branes},
hep-th/9810186.

\bibitem{Kraus}
P. Kraus, F. Larsen and S.P. Trivedi,
{\sl The Coulomb branch of gauge theory from rotating branes},
hep-th/9901056.

\bibitem{Cai}
R. Cai and K. Soh,
{\sl Critical behavior in the rotating D-branes},
hep-th/9812121.

\bm{klt} P. Kraus, F. Larsen and S.P. Trivedi, {\sl The Coulomb branch
of gauge theory from rotating branes}, hep-th/9811120.

\bm{rosf} J.G. Russo and K. Sfetsos, {\sl Rotating $D3$-branes and QCD
in three dimensions}, hep-th/9901056.

\bibitem{Csaki2} C. Csaki, J. Russo, K. Sfetsos and J. Terning,
{\sl Supergravity Models for $3+1$ Dimensional QCD}, hep-th/9902067.

\bm{sfetsos} K. Sfetsos,
{\sl Rotating NS5-brane solution and its exact string theoretical 
description}, hep-th/9903201.

\bibitem{thorne} K.S. Thorne, {\sl Black holes: the membrane
paradigm}, New Haven, USA: Yale Univ. Pr. (1986) 367p.

\bibitem{mype}
R.C. Myers and M.J. Perry,
{\sl Black holes in higher dimensional spacetimes},
Ann. Phys. {\bf 172} (1986) 304.

\bm{clp1} M. Cvetic, H. L\"u and C.N. Pope, {\sl Spacetimes of
boosted $p$-branes and CFT in infinite momentum frame},
hep-th/9810123, to appear in Nucl. Phys. {\bf B}.

\bm{nilsson} B.E.W. Nilsson, {\sl On the embedding of $d=4$, $N=8$
gauged supergravity in $d=11$, $N=1$ supergravity}, Phys. Lett.
{\bf B155} (1985) 54.

\bibitem{Pope2} C.N. Pope, {\sl The embedding of the
Einstein-Yang-Mills equations in $D=11$ supergravity}, Class. Quantum
Grav. {\bf 2} (1985) L77.

\bm{ppv} M. Pernici, K. Pilch and P. van Nieuwenhuizen, {\sl
Noncompact gaugings and critical points of maximal supergravity in
seven dimensions}, Nucl. Phys. {\bf B249} (1985) 381.

\bm{lpss} H. L\"u, C.N. Pope, E. Sezgin and K.S. Stelle, {\sl
Dilatonic $p$-brane solitons}, Phys. Lett. {\bf B371} (1996) 46,
hep-th/9511203.

\bibitem{Cremmer1}
E. Cremmer and B. Julia,
{\sl The $N=8$ supergravity theory. 1. The Lagrangian},
Phys. Lett. {\bf B80} (1978) 48.

\bibitem{Cremmer2}
E. Cremmer and B. Julia,
{\sl The $SO(8)$ supergravity},
Nucl. Phys. {\bf B159} (1979) 141.

\bm{cjlp1} E. Cremmer, B. Julia, H. L\"u and C.N. Pope, {\sl
  Dualisation of dualities}, Nucl. Phys. {\bf B523} (1998) 73,
  hep-th/9710119.

\bibitem{Rahmfeld3}
M.~J. Duff and J.~Rahmfeld,
\newblock {\sl Bound states of black holes and other $p$-branes},
\newblock  Nucl. Phys. {\bf B481} (1996) 332.

\bibitem{Aichelburg}
P.~C. Aichelburg and F.~Embacher,
\newblock {\sl Exact superpartners of $N=2$ supergravity solitons},
\newblock Phys. Rev. {\bf D34} (1986) 3006.

\bibitem{Duffliurahmfeld2}
M.~J. Duff, J.~T. Liu and J.~Rahmfeld,
\newblock {\sl Dipole moments of black holes and string states},
\newblock Nucl. Phys. {\bf B494} (1997) 161.

\bibitem{Freedmannicolai}
D. Z. Freedman and H. Nicolai,
{\sl Multiplet shortening in $OSp(N,4)$},
Nucl. Phys. {\bf B237} (1984) 342.

\bibitem{Rahmfeld1}
M.~J. Duff and J.~Rahmfeld,
\newblock {\sl Massive string states as extreme black holes},
\newblock Phys. Lett. {\bf B345} (1995) 441.


\bibitem{Caldarelli}
M. M. Caldarelli and D. Klemm,
{\sl Supersymmetry of anti-de Sitter black holes},
{\tt hep-th/9808097}.

\bibitem{Rahmfeld2}
J. Rahmfeld,
\newblock {\sl Extremal black holes as bound states},
\newblock  Phys. Lett. {\bf B372} (1996) 198.

\bibitem{graham}
C. R. Graham and R. Lee, 
{\sl Einstein metrics with prescribed conformal infinity on the ball},
Adv. Math. {\bf 87} (1991) 186. 

\bibitem{Supergravity}
E. Cremmer,
{\sl Symmetries in extended supergravity},
in Supergravity 81, Eds S. Ferrara and J. G. Taylor, (C. U. P. 1982).

\bibitem{Duffsupermembranes}
M. J. Duff, 
\newblock {\sl Supermembranes},
\newblock Lectures given at the Theoretical Advanced Study Institute in
Elementary Particle Physics (TASI 96), Boulder, 1996, hep-th/9611203.

 \bibitem{Salamsezgin}
Abdus Salam and Ergin Sezgin,
{\sl Supergravities in diverse dimensions},
World Scientific (1989).

\bibitem{Wittenfermion}
E. Witten,
{\sl Fermion quantum numbers in Kaluza-Klein theory},
in Proceedings of the Shelter Island II conference (1983) 227, eds Khuri,
Jackiw, Weinberg and Witten (M. I. T. Press, 1985).

\bibitem{Kaluza}
M.~J. Duff, B.~E.~W. Nilsson and C.~N. Pope,
\newblock {\sl {K}aluza-{K}lein supergravity},
\newblock Phys. Rep. {130} (1986) 1.

\bibitem{Klein}
 M. J. Duff, 
\newblock {\sl Kaluza-Klein theory in perspective},
\newblock in the Proceedings of the Nobel Symposium {\it Oskar Klein
Centenary}, Stockholm, September 1994, (Ed Lindstrom, World Scientific
1995), hep-th/9410046. 

\bibitem{Townsend}
P. K. Townsend,
{\sl M(embrane) theory on $T^{9}е$},
in the Strings 97 Proceedings, hep-th/9708054.

\bibitem{Cvetic10} 
M. Cvetic, M.J. Duff, P. Hoxha, James T. Liu, 
H. Lu, J.X. L\"{u}, R. Martinez-Acosta, C.N. Pope, H. Sati
and T.A. Tran, 
\newblock {\sl Embedding $AdS$ black holes in ten and eleven dimensions}, 
\newblock  hep-th/9903214.

\bibitem{Akama}
K. Akama,
{\sl Pregeometry}
in {\it Lecture Notes in Physics, 176, Gauge Theory and Gravitation, 
Proceedings, Nara, 1982}, edited by K. Kikkawa, N. Nakanishi and H. 
Nariai, (Springer-Verlag) 267-271,
hep-th/000113.

\bibitem{Rubakov}
V. A. Rubakov and M. E. Shaposhnikov,
{\sl Do we live inside a domain wall?},
Phys. Lett. {\bf B125} (1983) 136.

\bibitem{Gibbonswiltshire}
G.W.~Gibbons and D.L.~Wiltshire,
{\sl Space-time as a membrane in higher dimensions},
Nucl. Phys. {\bf B287} (1987) 717.

\bibitem{Horowitzstrominger}
G. T. Horowitz and A. Strominger,
\newblock {\sl Black strings and $p$-branes},
\newblock Nucl. Phys. {\bf B360} (1991) 197.

\bibitem{Antoniadis}
I. Antoniadis,
{\sl A possible new dimension at a few TeV},
Phys. Lett. {\bf B246} (1990) 377.

\bibitem{Wittencalabi}
E. Witten,
\newblock {\sl Strong coupling expansion of Calabi-Yau 
compactification},
\newblock Nucl. Phys. {\bf B471} (1996) 135.

\bibitem{Lykken}
J. Lykken,
{\sl Weak sclae superstrings}
Phys. Rev. {\bf D54} (1996) 3693. 

\bibitem{Arkani}
N. Arkani-Hamed, S. Dimopoulos and G. Dvali,
{\sl The hierarchy problem and new dimensions at a millimeter},
Phys. Lett. {\bf B429} (1998) 263.

\bibitem{Dienes}
K. R.~Dienes, E.~Dudas and T.~Gherghetta,
{\sl Extra space-time dimensions and unification},
Phys. Lett. {\bf B436} (1998) 55.

\bibitem{Wittensearch}
E. Witten,
{\sl Search for a realistic Kaluza-Klein theory},
Nucl. Phys. {\bf B186} (1981) 412-428.

\bibitem{Castellani}
L. Castellani, R. D'Auria and P. Fr\'{e},
{\sl Supergravity and superstrings: a geometric perspective}, 
(in 3 volumes), World Scientific Press, 1991.

\bibitem{Stelle}
K. S. Stelle,
{\sl An introduction to $p$-branes},
In *Seoul/Sokcho 1997, Dualities in gauge and string theories 39-104.
 
\bibitem{Dealwis}
S. P. De Alwis,
\newblock {\sl A note on brane tension and $M$-theory}, 
\newblock {\tt hep-th/9607011}.

\bibitem{Aharony1}
O. Aharony, S Gubser, J. Maldacena, H. Ooguri and Y. Oz,
{\sl Large N field theories , string theory and gravity},
hep-th/9905111.

\bibitem{Han}
S. K. Han and I. G. Koh,
{\sl $N=4$ supersymmetry remaining in Kaluza-Klein monopole 
background in $D=11$ supergravity},
Phys. Rev. {\bf D31} (1985) 2503.

\bibitem{Nastase1}
H. Nastase, D. Vaman, P. van Nieuwenhuizen,
{\sl Consistent nonlinear KK reduction of 11-D supergravity on 
$AdS_{7}е \times S^{4}е$ and selfduality in odd dimensions},
hep-th/9905075.

\bibitem{Nastase2}
H. Nastase, D. Vaman, P. van Nieuwenhuizen,
{\sl Consistency of the $AdS_{7}е \times S^{4}е$ reduction and the 
origin of selfduality in odd dimensions},
hep-th/9911238.

\bibitem{Lupope1}
H. L\"{u} and C. N. Pope,
{\sl Exact embedding of $N=1$, $D=7$ gauged supergravity in $D=11$},
hep-th/9906168.

\bibitem{Lupope2}
H. L\"{u}, C. N. Pope and T.A. Tran,
{\sl Five-dimensional $N=4$, $SU(2) \times U(1)$ gauged supergravity 
from Type $IIB$},
hep-th/9909203.

\bibitem{Lupope3}
M. Cvetic, H. L\"u and C. N. Pope,
{\sl Four-dimensional $N=4$ $SO(4)$ gauged supergravity from $D=11$},
hep-th/9910252.

\bibitem{Hawkingreall}
S.W. Hawking and H.S. Reall,
{\sl Charged and rotating AdS black holes and their CFT duals},
hep-th/9908109.

\bibitem{Gunaydintakemae}
M. G\"{u}naydin and S. Takemae,
{\sl Unitary supermultiplets of $OSp(8*|4)$ and their 
$AdS_{7}е/CFT_{6}е$ duality},
hep-th/9910110.
 
\bibitem{hs1} 
P.S. Howe and E. Sezgin, 
{\it Superbranes}, 
Phys. Lett. {\bf B390} (1997) 133, hep-th/9607227.

\bibitem{Sezginsundell}
E. Sezgin and P. Sundell,
{\sl Aspects of the $M5$-brane},
hep-th/9902171. 

\bibitem{hsw1} 
P.S. Howe, E. Sezgin and P.C. West, 
{\sl Covariant field equations of the $M$-theory five-brane}, 
Phys. Lett. {\bf 399B} (1997) 49.

\bibitem{hs2} 
P.S. Howe and E. Sezgin, 
{\sl D=11,p=5}, 
Phys. Lett. {\bf B394} (1997) 62.

\bibitem{s} 
I. Bandos, K. Lechner, A. Nurmagambetov, P. Pasti, D. Sorokin and M. Tonin, 
{\sl Covariant action for the superfivebrane of M-theory},
Phys. Rev. Lett. {\bf 78} (1997) 4332.

\bibitem{s2} 
I. Bandos, K. Lechner, A. Nurmagambetov, P. Pasti, D. Sorokin and M. Tonin, 
{\sl On the equivalence of different formulations of the M theory five-brane}, 
hep-th/9703127.
 
\bibitem{Bergshoefftownsend}
E.~Bergshoeff and P.K.~Townsend,
{\sl Super D-branes},
Nucl. Phys. {\bf B490}, 145 (1997)

\bibitem{Randallsundrum1}
L. Randall and R. Sundrum,
{\sl A large mass hierarchy from a small extra dimension},
hep-th/9905221.

\bibitem{Randallsundrum2}
L. Randall and R. Sundrum,
{\sl An alternative to compactification},
hep-th/9906064.

\bibitem{Petersen}
J. L. Petersen,
{\sl Introduction to the Maldacena conjecture on $AdS/CFT$},
hep-th/9902131.

\bibitem{Yau}
E. Witten and S. T. Yau,
{\sl Connectedness of the boundary in the AdS/CFT correspondence},
hep-th/9910245.         

\bibitem{Banks}
T. Banks,
{\sl $M$ theory and cosmology},
hep-th/9911067

\bibitem{Eleven} 
M. J. Duff, 
{\sl The world in eleven dimensions: supergravity, 
supermembranes and M-theory}, a reprint volume with commentaries,
I.O.P Publishing, 1999.

\bibitem{vanN}
P. Van Nieuwenhuizen,
{\sl Supergravity},
Phys. Rept. {\bf 68} (1981) 189.

\bibitem{Ahn}
C. Ahn and S-J. Rey,
{\sl Three-dimensional CFTs and RG flow from squashing M2-brane 
horizon},
hep-th/9908110

\bibitem{Landsteiner}
K. Landsteiner and E. Lopez,
{\sl The thermodynamic potentials of Kerr-AdS black holes and their 
CFT duals},
hep-th/9911124.

\bibitem{Deser}
S. Deser, D. Seminara, {\sl Counterterms/M-theory corrections to 
$D=11$ supergravity},
Phys. Rev. Lett. {\bf 82} (1999) 2435.

\bibitem{Harmark1}
T. Harmark and N. A. Obers,
{\sl Thermodynamics of spinning branes and their dual field theories},
hep-th/9910036.

\bibitem{Harmark2}
T. Harmark and N. A. Obers,
{\sl Phase structure of non-commutative field spinning branes and their 
dual field theories},
hep-th/9911169.



  
\end{thebibliography}
\end{document}